%% file: ANA-SIMU-2018-04-PAPER.tex
\documentclass[cernpreprint, atlasdraft=false, texlive=2020, USenglish, texmf]{atlasdoc}
 
\usepackage{atlaspackage}
\usepackage{atlasbiblatex}
 
\usepackage{atlasphysics}
\usepackage{atlasjetetmiss}
\usepackage{atlasprocess}
\usepackage{atlasmisc}
 
\usepackage{subfig}
\usepackage{multirow}
 
\graphicspath{{logos/}{figures/}}
 
\usepackage{ANA-SIMU-2018-04-PAPER-defs}

\AtlasTitle{AtlFast3: the next generation of fast simulation in ATLAS}
 
\AtlasAbstract{
The ATLAS experiment at the Large Hadron Collider has a broad physics
programme ranging from precision measurements to direct searches for
new particles and  new interactions, requiring ever larger and ever
more accurate datasets of simulated Monte Carlo events. Detector
simulation with \textsc{Geant4} is accurate but requires significant
CPU resources. Over the past decade, ATLAS has developed and utilized
tools that replace the most CPU-intensive component of the simulation
-- the calorimeter shower simulation -- with faster simulation
methods. Here, AtlFast3, the next generation of high-accuracy fast
simulation in ATLAS, is introduced. AtlFast3 combines parameterized
approaches with machine-learning techniques and is deployed to meet
current and future computing challenges and simulation needs of the
ATLAS experiment. With highly accurate performance and significantly
improved modelling of substructure within jets, AtlFast3 can simulate large numbers of events for a wide range of physics processes.
}
 
\author{The ATLAS Collaboration}
\AtlasJournalRef{Comput Softw Big Sci 6 (2022) 7}
\AtlasDOI{10.1007/s41781-021-00079-7}
\PreprintIdNumber{CERN-EP-2021-174}

\hypersetup{pdftitle={ATLAS document},pdfauthor={The ATLAS Collaboration}}
 
\begin{document}
 
\maketitle
 
\tableofcontents

\section{Introduction}
\label{sec:intro}
The physics programme of the ATLAS experiment~\cite{PERF-2007-01} at the Large Hadron Collider~\cite{Evans:2008zzb} (LHC) relies on the accurate simulation of billions of Monte Carlo (MC) events to complement the data delivered by the LHC. Simulated events are prepared by generating the appropriate physics process, simulating the passage of particles through the detectors, digitizing the detector response, and then reconstructing those events using the same reconstruction algorithms that are applied to the recorded data. The simulation of the passage of particles through the detectors for analyses of data taken during LHC Run~2 requires approximately 40\% of the computing resources of the ATLAS experiment~\cite{Calafiura:2729668,foundation2018hep}. Run~2 was the second data-taking run of the LHC and lasted from 2015 to 2018.
 
The complex accordion geometry of the ATLAS electromagnetic calorimeter (see Section~\ref{sec:calorimeters}) makes the simulation of shower development particularly CPU intensive when using the \GEANT~\cite{Agostinelli:2002hh} toolkit (G4). In fact, 80\% of the total simulation time for a typical sample of the production of top and anti-top quark pair (\ttbar) is devoted to modelling this shower development~\cite{SOFT-2010-01}. Therefore, fast approaches for calorimeter simulation are crucial in reducing the CPU needs of the ATLAS experiment and to enable the production of the required numbers of simulated events for precision physics analyses.
 
The ATLAS Collaboration has already developed and deployed a fast simulation tool, called \AFII~\cite{SOFT-2010-01} (AF2), that performs the simulation of the entire ATLAS detector by combining different tools to simulate different sub-detectors and particles. In particular, \AFII relies on a fast simulation of the calorimeters called FastCaloSim~\cite{ATL-PHYS-PUB-2010-013}. \AFII has been used to produce approximately 32 billion of the 52 billion events simulated for physics analyses of the Run~2 data. \AFII is known to have certain limitations, particularly in the modelling of jets of particles reconstructed with large-radius clustering algorithms and the detailed description of their substructure. In this paper, the ATLAS Collaboration introduces a new fast simulation tool, \AFIII or AF3, which has the same CPU performance as \AFII, but better accuracy in reproducing \GEANT. The ATLAS Collaboration is using \AFIII for a large resimulation campaign of Run~2 MC events and plans to use \AFIII extensively for Run~3 and beyond.
 
\AFII and \AFIII perform the simulation of the entire ATLAS detector by combining a number of components. Key components of \AFIII are two parametric calorimeter simulations: the new version of FastCaloSim, referred to as \FCaloSII, and \FGAN. Parametric simulations of the calorimeter response simulate the energy of a particle shower as a single step based on an underlying parametrization instead of simulating how every particle propagates and interacts inside the calorimeter volume. This makes the CPU performance of these tools essentially independent of the particle energy and scales linearly with the number of particles entering the calorimeter volume.
 
\AFIII is designed to simulate particle showers to a level of precision such that no sizeable differences from the \GEANT simulation can be resolved by the reconstruction algorithms, including those for electron, photon, and $\tau$-lepton reconstruction and identification, and jet reconstruction and clustering.
 
The fast simulation of showers in the calorimeter can be factorized into several components: the total shower energy, the energy sharing between calorimeter layers, the average lateral shower development within a layer, the uncorrelated energy fluctuations in individual showers compared to average showers and, for hadronic showers, the correlated fluctuations between the longitudinal and lateral energy distributions. The energy that is deposited in the calorimeter depends on the kinetic energy (\Ekin) of the particle and is the energy used for parameterization unless otherwise specified. The \Ekin is defined as the particle energy minus its mass. For antiprotons and antineutrons, the rest mass is added instead of being subtracted as their annihilation will result in additional energy deposited in the calorimeter.
 
The simulation of the total shower energy and its longitudinal distribution between layers, including correlations, provides an approximate simulation of jets, electrons, photons and $\tau$-leptons, albeit with overestimated reconstruction and identification efficiencies. The simulation of the average lateral energy spread plays an important role in the reconstruction and identification of objects for physics analysis. For speed, a simplified geometry of the calorimeter cells (see Section~\ref{sec:detector}) is used in \AFIII, where each cell belongs to a longitudinal sampling layer of the  calorimeter and is either a cuboid in $\eta$,\footnote{ATLAS uses a right-handed Cartesian coordinate system with its origin at the nominal interaction point (IP) in the centre of the detector. The $z$-axis is along the beam pipe, and the $x$-axis points from the IP to the centre of the LHC ring. Cylindrical coordinates ($r$, $\phi$) are used in the transverse plane, $\phi$ being the azimuthal angle around the beam pipe. The rapidity is defined as $y=(1/2)\ln[(E+p_z)/(E-p_z)]$, while the pseudorapidity is defined in terms of the polar angle $\theta$ as $\eta = -\ln \tan(\theta/2)$.} $\phi$, and $r$ for the layers in the detector barrel, or a cuboid in $\eta$, $\phi$, and $z$ for the layers in the detector endcaps up to $|\eta|<3.2$, or a cuboid in $x$, $y$, and $z$ for the forward calorimeter layers. This means that the accordion structure of the real ATLAS liquid-argon electromagnetic calorimeter must be emulated. The improved average shower energy distribution and a correction for the accordion structure in \AFIII reproduces the reconstruction and identification efficiencies of the \GEANT simulation, especially for electrons and photons.
 
Fluctuations also play an important role in the calorimeter simulation. The simulation of independent and uncorrelated energy fluctuations in calorimeter cells in individual showers relative to average showers is required for an accurate description of electrons and photons of all energies. Correct modelling of the fluctuations is also crucial for hadrons, where at low energy the fluctuations in the lateral energy distribution are dominated by sampling fluctuations, noise and additional proton--proton interactions. For medium-energy hadrons, correlated fluctuations between the longitudinal and lateral energy distributions in hadronic showers play an important role and are accurately simulated by \FGAN (see Section~\ref{sec:fcsgan}).
 
\AFIII combines the strengths of the \FCaloSII and \FGAN calorimeter simulation approaches. The updated calorimeter parameterization in \FCaloSII is used to simulate electromagnetic showers and hadronic showers with low and high energies while Generative Adversarial Networks (GANs)~\cite{GAN} trained for \FGAN~\cite{ATL-SOFT-PUB-2020-006} are introduced for medium-energy hadrons because of their ability to model correlated fluctuations. The two tools are combined to optimize the performance of the reconstruction and a smooth interpolation between them is performed. Both calorimeter simulation approaches are derived using single particles simulated with \GEANT to model the shower development in the ATLAS electromagnetic and hadronic calorimeters. As the calorimeters do not necessarily contain the full showers, the rate at which secondary particles punch through to the muon spectrometer is parameterized and those particles are simulated with \GEANT. The parameterization of  muon punch-through is a new feature in \AFIII. \GEANT is also used to simulate particles in the inner detector and hadrons with energies below a few hundred \MeV\ in the calorimeters.
 
Section~\ref{sec:detector} introduces the ATLAS detector and provides details about the detectors used in \AFIII. Datasets and samples are presented  in Section~\ref{sec:datasets}. Sections~\ref{sec:fcsv2} and ~\ref{sec:fcsgan} discuss the \FCaloSII and \FGAN parameterizations, respectively. The simulation of muon punch-through is discussed in Section~\ref{sec:muonPunchThrough}. Section~\ref{sec:af3} discusses how \AFIII is constructed by combining \FCaloSII and \FGAN. The performance of \AFIII in the reconstruction and in physics analysis is discussed in Section~\ref{sec:af3perf}. Section~\ref{sec:conclusion} concludes the paper.

\section{The ATLAS detector}
\label{sec:detector}
The ATLAS detector~\cite{PERF-2007-01,ATLAS-TDR-19} at the LHC covers nearly the entire solid angle around the collision point.
It consists of an inner tracking detector surrounded by a thin superconducting solenoid, electromagnetic and hadron calorimeters,
and a muon spectrometer incorporating three large superconducting air-core toroidal magnets. An extensive software suite~\cite{ATL-SOFT-PUB-2021-001} is used in the reconstruction and analysis of real and simulated data, in detector operations, and in the trigger and data acquisition systems of the experiment.
 
\subsection{Inner detector}
The inner detector is immersed in a \SI{2}{\tesla} axial magnetic field
and provides charged-particle tracking in the range \(|\eta| < 2.5\).
The high-granularity silicon pixel detector is closest to the collision point and typically provides four measurements per track,
the first hit normally being in the innermost layer.
It is followed by the silicon microstrip tracker, which usually provides eight measurements per track.
These silicon detectors are complemented by the transition radiation tracker (TRT),
which enables radially extended track reconstruction up to \(|\eta| = 2.0\).
The TRT also provides electron identification information
based on the fraction of hits (typically 30 in total) above a higher energy-deposit threshold corresponding to transition radiation.
 
\subsection{Calorimeters}
\label{sec:calorimeters}
\FloatBarrier
The calorimeter system covers the pseudorapidity range \(|\eta| < 4.9\) and exploits several technologies to measure the energy deposited by different types of particles. In Table~\ref{tab:cal_layers} the different calorimeter modules, different layers in the radial direction, the acronyms used to refer to each layer, the coverage in $\eta$ by each layer and the transition regions with gaps between different layers are summarized.  In the transition regions of the calorimeter, the detector response deteriorates relative to the rest of the acceptance.
 
\begin{table}[]
\centering
\caption{The different calorimeter modules, their number of layers in the radial direction, the acronyms used to refer to each layer and their coverage in $\eta$. The first layers of the electromagnetic calorimeter are Presampler layers and have finer granularity than the subsequent layers.}
\vspace{0.2cm}
\resizebox*{0.75\textwidth}{!}{
\noindent\makebox[\textwidth]{
\begin{tabular}{lllll}
\hline
Calorimeter   & Layers & Module Name   & $\eta$-coverage  & Sampling Layer                                                            \\
\hline
\multirow{2}{*}{Electromagnetic calorimeters} & 4 & Electromagnetic Barrel (EMB)  & $|\eta| < 1.5$     & PreSamplerB, EMB1, EMB2, EMB3 \\
& 4 & Electromagnetic Endcap (EMEC) & $1.5 < |\eta| < 1.8$   & PreSamplerE \\
& & & $1.5 < |\eta| < 3.2$   & EME1, EME2 \\
& & & $1.5 < |\eta| < 2.5$ & EME3  \\
\hline
\multirow{4}{*}{Hadronic calorimeters}  & 4   & Hadronic Endcap (HEC)   & $1.5 < |\eta| < 3.2$  & HEC0, HEC1, HEC2, HEC3  \\
& 3  & Tile Barrel (TileBar) & $|\eta| < 1.0$ & TileBar0, TileBar1, TileBar2         \\
& 3  & Tile Extended Barrel (TileExt)  & $0.8 < |\eta| < 1.7$   & TileExt0, TileExt1, TileExt2 \\
& 3  & Tile Gap (TileGap)  & $1.0 < |\eta| < 1.6$ & TileGap1, TileGap2, TileGap3         \\
\hline
Forward calorimeter                          & 3                & FCal    & $3.1 < |\eta| < 4.9$                 & FCal0, FCal1, FCal2                  \\
\hline
\multirow{3}{*}{Transition regions} & - & between barrel and endcap & $|\eta| \approx 1.45$ & - \\
& - & between outer and inner wheel of endcap & $|\eta| = 2.5$ & - \\
& - & between endcap and FCal & $|\eta| \approx 3.2$ & - \\
\hline
\end{tabular}}}
\label{tab:cal_layers}
\end{table}
 
\begin{figure}[htb]
\centering
\includegraphics[width=.7\linewidth]{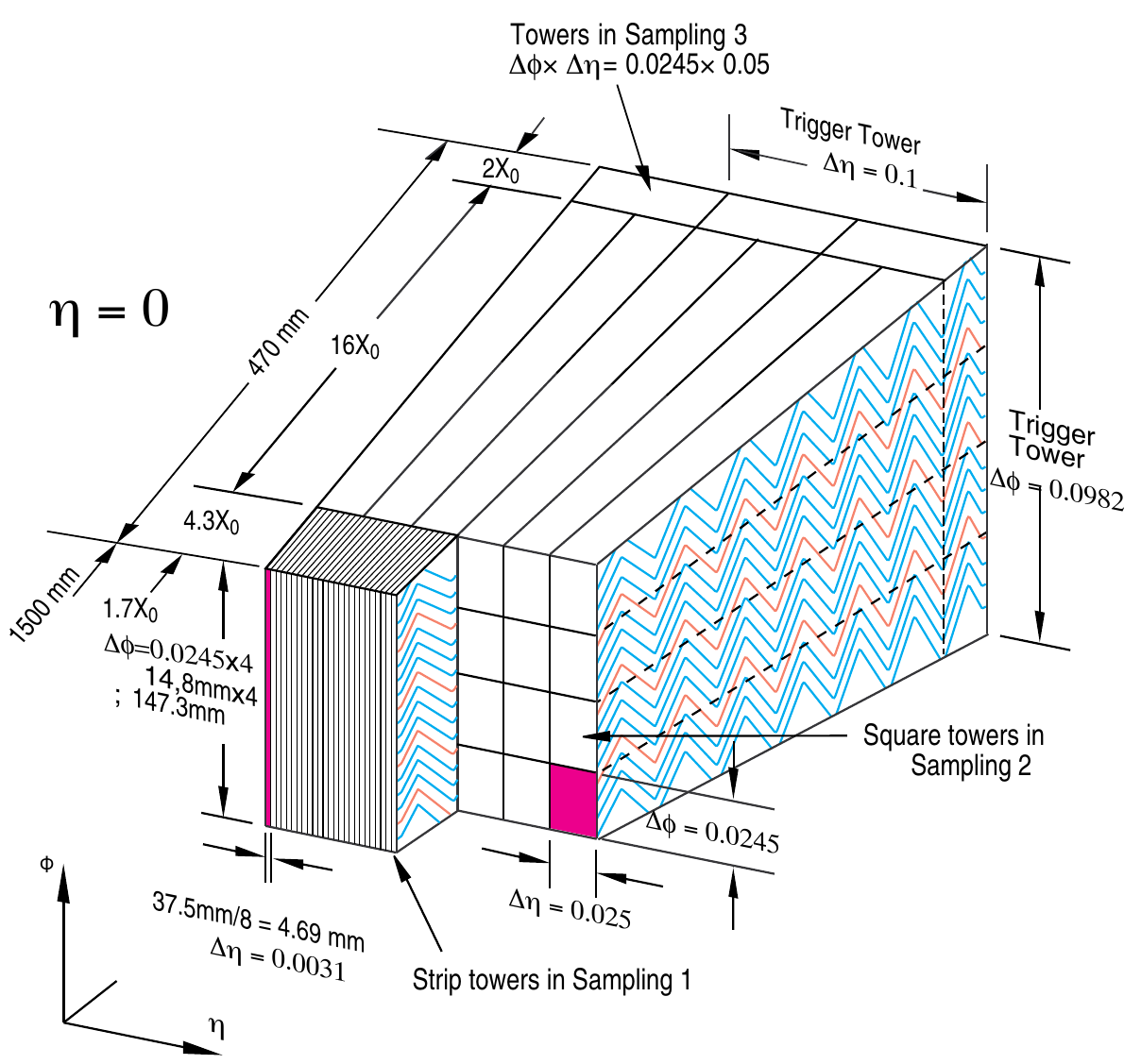}
\caption{Granularity of the electromagnetic barrel LAr calorimeters~\cite{CERN-LHCC-96-040}. The accordion geometry is indicated with blue and orange lines on the side of the tower.}
\label{fig:det_granularity}
\end{figure}

Electromagnetic showers are measured by high-granularity lead/liquid-argon (LAr) calorimeters. The electromagnetic barrel (EMB) and electromagnetic endcap calorimeters (EMEC) provide coverage within the region \(|\eta|< 3.2\).  There are three EMEC sampling layers in the precision-measurement region ($1.5 < |\eta| < 2.5$) and two layers in the higher-$|\eta|$ region ($2.5 < |\eta| < 3.2$). An additional thin LAr Presampler covering \(|\eta| < 1.8\) corrects for energy loss in material upstream of the calorimeters. The electromagnetic calorimeters have an accordion-shape geometry, which provides several active layers in a compact design without any gaps. An illustration of this structure and of the segmentation in each layer is shown in Figure~\ref{fig:det_granularity}. Each layer consists of a number of cells, and groups of cells are referred to as towers, which are used by the trigger. The thickness of the calorimeter is given in units of radiation length, $X_0$.
 
Hadronic showers are measured in a steel/scintillator-tile calorimeter segmented into three barrel structures within \(|\eta| < 1.7\) (one TileBar and two TileExt), and two copper/LAr hadron endcap calorimeters (HEC) for $1.5 < |\eta| < 3.2$. The granularity of HEC3 decreases for $|\eta| > 2.5$. The transition region between the barrel and the endcap has additional detectors, the Tile Gap layers.
 
Coverage at higher $|\eta|$ ($3.1 < |\eta| < 4.9$) is provided by the forward calorimeter (FCal). Two different technologies, copper/LAr and tungsten/LAr, are used in the FCal, which provides both for electromagnetic and hadronic energy measurements.
 
The energy deposited in the calorimeters is read out from cells which define the granularity of the detector. The granularity varies significantly depending on the layer and can vary with $\eta$ and $\phi$.
 
\subsection{Muon spectrometer}
The muon spectrometer (MS) comprises separate trigger and
high-precision tracking chambers and measures the deflection of muons in a magnetic field generated by the superconducting air-core toroidal magnets.
The field integral of the toroids ranges between \num{2.0} and 6.0~Tm across most of the detector. A set of precision chambers covers the region \(|\eta| < 2.7\) with three layers of monitored drift tubes (MDTs),
complemented by cathode-strip chambers (CSCs) in the forward region, where the background is highest. In the barrel region ($|\eta| \leq 1.05$), the MDT chambers are located in and around eight coils of superconducting toroid magnets.
In the endcap ($1.05 \leq |\eta| \leq 2.7$) sections of the MS, the MDTs are located both in front of and behind the endcap toroid magnets. The innermost detector in the endcap region is instrumented with CSCs instead of MDTs to withstand higher rate and background conditions.
 
Secondary particles, created in showers in the calorimeters, leaking into the MS can have a significant impact on muon reconstruction and depends on the calorimeter simulation.
This effect is called muon punch-through and the technique used to simulate it is described in Section~\ref{sec:muonPunchThrough}.

\section{Dataset simulation and preprocessing}
\label{sec:datasets}
This section discusses the simulation of the datasets used to derive the calorimeter parameterizations and the datasets used to validate the performance of \AFIII.
 
\subsection{Simulation of reference samples with \GEANT}
\label{sec:refsamples}
The reference samples used to generate the calorimeter parameterizations discussed in this paper were produced using \GEANT version 10.1.3~\cite{Agostinelli:2002hh}, which was released in 2016 and use the FTFP\_BERT\_ATL physics list~\cite{Allison:2016lfl}.  FTFP\_BERT\_ATL uses the Bertini intra-nuclear cascade model below 9~\GeV{} and transitions to the Fritiof model~\cite{Andersson:1986gw,Andersson:1996xi,Ganhuyag:1997gz} with a pre-compound model for 12~\GeV{} and higher. The default \GEANT electromagnetic physics list is used. The standard configuration for ATLAS simulation in physics analyses, referred to as the MC16 campaign, uses a fast simulation technique known as Frozen Showers~\cite{SOFT-2010-01} to simulate electromagnetic showers in the FCal. This latter configuration has been used for full simulation samples for all papers on Run~2 data published by ATLAS to date. The reference samples do not use the Frozen Shower technique, but it is used for the validation samples discussed in Section~\ref{sec:validation}.
 
The reference datasets consist of single-particle events produced on a cylinder with $r=1148$\,mm and $|z|<3550$\,mm located just outside the TRT because only particles reaching the calorimeters have to be parameterized. This means that the impact of the cryostat and solenoid material is included in the reference datasets. The particles are produced with directions of propagation consistent with production at the interaction point and simulated without the spread of the LHC beam bunches, for positive and negative values of $\eta$ and with a uniform distribution in $\phi$. The impact of the spread of the beam bunches is negligible because particles are parameterized according to where they enter the calorimeter. Photons ($\gamma$) and electrons $(e^{\pm})$ are used to parameterize electromagnetic showers, and positively and negatively charged pions ($\pi^{\pm}$) are used to parameterize hadronic showers. The positive and negative charge states of electrons and pions  are combined since the difference in shower development due to different charge is negligible.  Charged pions are used to model the simulation of hadrons because the dependence of the hadronic showers on particle type is very small. The \GEANT simulation is run in a special configuration with simulation steps which are smaller than
those in the default configuration in the calorimeter so that details of the spatial position of the deposited energy or hits are saved. These detailed hits are used to parameterize lateral shower shapes in \FCaloSII, discussed in Section~\ref{sec:shapepara}, and for the training of \FGAN. The sum of the energies of these detailed hits is lower than that of the standard \GEANT hits due to a small bug in \GEANT. This introduces a small bias in the energy distributions for \FGAN and also for \FCaloSII standard hits in the derivation of the longitudinal energy parameterization of \FCaloSII. The bias is non-negligible for photons and electrons, and this is corrected for in \FCaloSII. After the correction, no significant impact on the physics performance of \AFIII is observed. Deposited energies in simulation are digitized without considering electronic noise and cross-talk between calorimeter cells in the readout electronics.
 
The calorimeter parameterization is obtained for 100 uniform $\eta$ slices to provide coverage up to $|\eta|=5$. This range slightly exceeds the calorimeter $\eta$ range to include particles depositing only a fraction of their energy in the calorimeters. In each slice, up to 19 single-particle samples were produced, starting from a minimum momentum of 16~\MeV{} and up to 4.2~\TeV{} with the momentum doubling for each sample. The \FGAN and \FCaloSII parameterizations for pions  are derived from the samples with energies of 256~\MeV{} and higher as discussed in Section~\ref{af3:energy:low}. For photons, the samples with energies of 64~\MeV{} and higher are used for \FCaloSII and those with energies of 128~\MeV{} and higher are used for \FGAN. All samples are used to derive the energy interpolation splines discussed in Section~\ref{sec:energy_interpol}. Ten thousand events were produced for each of the samples with energies up to 256~\GeV{}. With increasing energies above 256~\GeV, the number of events was progressively reduced, reaching 1000 events for the highest-energy sample, due to the significant increase in the time required to simulate the events in \GEANT.
 
Single-(anti)proton, single-(anti)neutron and single-(anti)kaon samples were produced in the same $\eta$ and momentum slices to derive the corrections described in Section~\ref{sec:hadron_correction}. These corrections are sufficient to provide good performance for these stable hadrons, thus avoiding a dedicated parameterization and thereby reducing the memory footprint of \AFIII.

\subsection{Voxelization}
\label{sec:datasets:voxels}
The spatial energy deposits in each layer in the \GEANT datasets are grouped into volumes called `voxels' for parameterization of \FCaloSII and \FGAN.
Only layers with a significant amount of energy, referred to as `relevant layers', are considered in the parameterization. Relevant layers are defined using criteria on the fraction of energy deposited in the layers with respect to the total energy deposited in the calorimeters. The criteria used by the two simulators to determine the relevant layers are slightly different:
 
\begin{itemize}
\item In \FCaloSII, only layers with energy fractions larger than 0.1\% are used; this procedure is performed for each sample independently.
\item In \FGAN all samples in the same $\eta$ slice are processed with the same number of voxels. The relevant layers are determined using only the 1~\TeV{} energy point and have an energy fraction larger than 0.1\%.
In addition, a layer with less energy is considered relevant if it is  in front of a relevant layer. For example, PreSamplerB is always included in \FGAN in the barrel region of the detector even if the energy deposited there is below the threshold.
\end{itemize}
 
The coordinates of hits in relevant layers, $\eta^{\text{hit}}$ and $\phi^{\text{hit}}$, are calculated relative to the extrapolated position of the particle in that layer, $\eta^{\text{extr}}$ and $\phi^{\text{extr}}$. The extrapolation is calculated from the momentum of the particle at the point where it enters the calorimeter and propagated through the calorimeter, taking into account the magnetic field for charged particles. Equation~(\ref{eqn:shape:deta_dphi}) shows the relative angular coordinates, $\Delta\phi$ and $\Delta\eta$, of the showers. The coordinates of the calorimeter cell associated with the hit,  $z_{\text{cell}}$ and $r_{\text{cell}}$, are used to transform the relative hit coordinates into millimeters, $\Delta\phi^{\SI{}{\milli\meter}}$ and $\Delta\eta^{\SI{}{\milli\meter}}$, which provides a more convenient description of the shower.
 
\begin{eqnarray}
\label{eqn:shape:deta_dphi}
\Delta\eta & = & \eta^{\text{hit}} - \eta^{\text{extr}}, \\ \nonumber
\Delta\phi & = & \phi^{\text{hit}} - \phi^{\text{extr}}, \\ \nonumber
\Delta\eta^{\SI{}{\milli\meter}} & = & \Delta\eta \times \eta_{\text{Jacobi}} \times \sqrt{r^2_{\text{cell}} + z^2_{\text{cell}}}, \\ \nonumber
\Delta\phi^{\SI{}{\milli\meter}} & = & \Delta\phi \times r_{\text{cell}},
\end{eqnarray}
 
where $\eta_{\text{Jacobi}} = | 2 \times \exp({-\eta_{\text{extr.}}}) / \left(1+\exp(-2\eta_{\text{extr.}})\right)|$.
 
\begin{eqnarray}
\label{eqn:shape:r_alpha}
\Delta R^{\SI{}{\milli\meter}} & = & \sqrt{\left(\Delta\eta^{\SI{}{\milli\meter}}\right)^2 + \left(\Delta\phi^{\SI{}{\milli\meter}}\right)^2}, \\ \nonumber
\alpha & = & \text{arctan2}\left(\Delta\phi^{\SI{}{\milli\meter}} , \Delta\eta^{\SI{}{\milli\meter}}\right).
\end{eqnarray}

The hit positions are then transformed to polar coordinates defined in Eq.~(\ref{eqn:shape:r_alpha}) and grouped into voxels of different size:
 
\begin{itemize}
\item In \FCaloSII, the shower symmetry along $\phi$ with respect to the centre of the shower in each layer is exploited. A binning of \SI{1}{\milli\meter} in the radial direction is used in the high-granularity EMB1 and EME1 layers while \SI{5}{\milli\meter} is used in the other layers. Along the angular direction, $\alpha$, eight uniform bins are used in all relevant layers. These settings are used for all particles. The size of each voxel is much smaller than the calorimeter cell dimensions.
 
\item In \FGAN, the size of the voxel is optimized for each particle type and detector $\eta$ slice. In the radial direction, a variable bin width is used with increasingly wider bins. An example is shown in Table~\ref{tab:voxel} for pions in the barrel. Only layers with a large fraction of the total energy are binned along the angular direction. The angular positions of the showers in the other layers are neglected and simulated uniformly. Ten bins of equal size are used for layers binned in the angular direction. Due to the variable-width bins in the radial direction, the size of the voxels can be significantly larger than the cell dimensions. This voxel definition is optimized for an accurate training of the GANs since using as many bins as \FCaloSII would significantly increase the training time and instability of the GANs, ultimately reducing the performance of \FGAN.
\end{itemize}
 
\begin{table}[]
\centering
\caption{The binning used for the voxelization of pions in the different calorimeter layers for \FGAN in the $0 < |\eta| < 0.8$ range. Each ellipsis indicates when the same binning continues until the subsequent listed number.}
\vspace{0.2cm}
\resizebox*{\textwidth}{!}{
\begin{tabular}{llc}
\hline
Layer & Bin boundaries in $\Delta R^\text{mm}$ [mm] & Number of bins in $\phi$\\
\hline
PreSamplerB    &  5, 10, 30, 50, 100, 200, 400, 600   &   ~~1                         \\
EMB1           &  1, 4, 7, 10, 15, 30, 50, 90, 150, 200        &              10 \\
EMB2           &  5, 10, 20, 30, 50, 80, 130, 200, 300, 400   &          10                 \\
EMB3            &  50, 100, 200, 400, 600   &                      ~~1      \\
TileBar0        &  10, 20, 30, $\cdots$ 100, 130, 160, 200, 250 $\cdots$ 400, 1000, 2000   &            10           \\
TileBar1        &  10, 20, 30, $\cdots$  100, 130, 160, 200, 250, $\cdots$ 400, 600, 1000, 2000   &         10               \\
TileBar2     &  0, 50, 100, $\cdots$ 300, 400, 600, 1000, 2000   &       ~~1               \\
\hline
\end{tabular}}
\label{tab:voxel}
\end{table}
 
\subsection{Validation datasets}
\label{sec:validation}
A range of different Monte Carlo samples commonly used for physics performance studies and physics analysis are used to validate the performance of \AFIII.
Table~\ref{tab:samples} summarizes the key samples, which are also discussed in this section. The matrix element (ME) order describes the precision at which the process is produced by the generator in perturbative quantum chromodynamics (QCD); this can be leading order (LO), next-to-leading order (NLO) or next-to-next-to-leading order (NNLO).
 
The production of \ttbar{} events was performed with \POWHEGBOX[r2330]~\cite{Jezo:2018yaf} interfaced with \PYTHIA[6.427]~\cite{Sjostrand:2006za} for the parton shower and hadronization modelling with the CT10~\cite{Gao:2013xoa} set of parton distribution functions (PDFs) and the Perugia2012 set of tuned parameters (P2012 tune)~\cite{Skands:2010ak}. At least one of the top quarks is required to produce a lepton when decaying. This sample can be used to validate small-radius jets, leptons and $b$-jets.
 
Events containing $Z$ bosons decaying into a pair of electrons, muons or $\tau$-leptons were generated with \POWHEGBOX[r2856]~\cite{Frixione:2002ik} at \NLO{} in QCD using the \CTEQ[6L1]~\cite{Pumplin:2002vw} PDF set. The events were interfaced with \PYTHIA[8.186]~\cite{Sjostrand:2007gs} for the parton shower and hadronization modelling using the AZNLO tune~\cite{STDM-2012-23}. The samples were generated with $\pT(Z) = 0$. These samples are used to validate electrons, muons and $\tau$-leptons.
 
Events containing a new hypothetical spin-1 boson, $W^\prime$, decaying into a $WZ$ pair, which subsequently decay into hadrons, were generated using \PYTHIA[8.235] with the \NNPDF[23LO]~\cite{carrazza2013parton} PDF set. The $W^\prime$ bosons were generated with a mass of 13~\TeV{}, and the differential cross-section is reweighted to have a flat distribution of jet \pT{} from $200~\GeV$ to $3~\TeV{}$. Similarly, a sample of $Z^\prime$ bosons with a mass of 4~\TeV{} was generated using \PYTHIA[8.235], and the $Z^\prime$ boson was subsequently decayed into a top and anti-top quark pair. The top quarks were forced to decay into hadrons in the samples, which allows the substructure of jets to be validated. Similar to the $W^\prime$ sample, the differential cross-section is reweighted to have a flat distribution of jet \pT{} from $200~\GeV$ to $3~\TeV{}$ to better populate kinematic regions with higher jet \pT. These samples are used to validate the substructures of various jets with very high transverse momentum.
 
Samples of multijet events were simulated with the \PYTHIA[8.186] general-purpose event generator interfaced to \EVTGEN[1.2.0]~\cite{Lange:2001uf} for decay of heavy-flavour mesons.  The \NNPDF[23] PDF set~\cite{Ball:2012cx} and the A14 tune~\cite{ATL-PHYS-PUB-2014-021} were used.  The rapidly falling spectrum of leading-jet momenta requires this simulation to be filtered in leading-jet \pT{}. These samples are used to validate jets in a range of \pT{} regimes.
 
Higgs boson production via gluon--gluon fusion (ggF) was simulated at \NNLO accuracy in the strong coupling constant \alphas using \POWHEG NNLOPS \cite{Nason:2004rx,Frixione:2007vw,Alioli:2010xd,Hamilton:2013fea,Hamilton:2015nsa}, which achieves NNLO accuracy for arbitrary inclusive $gg\to H$ observables by reweighting the Higgs boson rapidity spectrum of MJ-\MINLO ~\cite{Hamilton:2012np,Campbell:2012am,Hamilton:2012rf} to that of HNNLO~\cite{Catani:2007vq}. The \textsc{PDF4LHC15} PDF set \cite{Butterworth:2015oua} and the \AZNLO tune of \PYTHIA[8] were used. This simulation was interfaced with \PYTHIA[8.230] for parton shower and non-perturbative hadronization effects. The Higgs boson was decayed into a pair of photons.
 
The impact of \pileup in the same and neighbouring proton-bunch crossings was modelled by combining detector signals from simulated inelastic \pp events with the hard-scattering (HS) event~\cite{SIMU-2020-01}. These \pileup events were generated with \PYTHIA[8.186]~\cite{Sjostrand:2007gs} using the \NNPDF[23] set of PDFs and the A3 tune~\cite{ATL-PHYS-PUB-2016-017}. The \pileup events were simulated using \GEANT even for samples produced with \AFII or \AFIII, and the same \pileup events were reused for all samples. An average number of \pileup interactions per \pp bunch crossing of 38 with a standard deviation of 12 was used, similar to the \pileup distribution recorded by the ATLAS experiment during Run~2.
 
The validation samples are reconstructed using the standard algorithms for the ATLAS experiment~\cite{PERF-2015-01, EGAM-2018-01, MUON-2018-03,Cacciari:2008gp, PERF-2015-09, PERF-2016-07}.
The energy scale and resolution of reconstructed leptons and jets, as well as their reconstruction and identification efficiencies in the simulation are corrected to match those measured in data using the standard procedures of the ATLAS experiment~\cite{JETM-2018-05}. Unless stated otherwise, the same reconstruction code and corrections are applied to samples simulated with \GEANT, \AFII and \AFIII.
 
\begin{table}[!htbp]
\caption{Summary of the Monte Carlo generator settings for the simulation of samples for  validation of \AFIII. See text for details.
}
\label{tab:samples}
\vspace{0.2cm}
\centering
\resizebox{\textwidth}{!}{
\def\arraystretch{1.5}
\newcolumntype{L}[1]{>{\raggedright\let\newline\\\arraybackslash}p{#1}}
\begin{tabular}{L{0.20\textwidth}L{0.29\textwidth}L{0.18\textwidth}L{0.17\textwidth}L{0.16\textwidth}L{0.1\textwidth}}
\toprule
Process & Generator & ME Order & PDF & Parton Shower & Tune \\
\midrule
\multicolumn{6}{c}{SM process samples}\\
\midrule
\ttbar & \POWHEGBOX[r2330.3] & NLO & \CT[10]  & \PYTHIA[6.427] & P2012 \\
$Z^\star/\gamma^\star \rightarrow \tau\tau$ & \POWHEGBOX[r2856] & NLO & \CTEQ[6L1]& \PYTHIA[8.186] & AZNLO \\
$Z \rightarrow \mumu$ & \POWHEGBOX[r2856] & NLO & \CTEQ[6L1]  & \PYTHIA[8.186] & AZNLO \\
$Z \rightarrow ee$ & \POWHEGBOX[r2856] & NLO & \CTEQ[6L1]  & \PYTHIA[8.186] & AZNLO \\
$W^\prime(13~\TeV)\rightarrow WZ\rightarrow 4q$ & \PYTHIA[8.235] & LO &\NNPDF[23LO] & \PYTHIA[8.235] & A14 \\
$Z^\prime(4~\TeV)\rightarrow\ttbar$ & \PYTHIA[8.235] & LO & \NNPDF[23LO] & \PYTHIA[8.235] & A14 \\
Dijet: leading jet $\pT=\text{140--400}$~\GeV{} & \PYTHIA[8.186] & NLO & \NNPDF[23] & \PYTHIA[8.186] & A14 \\
Dijet: leading jet $\pT=\text{1.8--2.5}$~\TeV{} & \PYTHIA[8.186]  & NLO & \NNPDF[23] & \PYTHIA[8.186] & A14 \\
ggF $\text{Higgs}\to\gamma\gamma$ & \POWHEG[v2~NNLOPS] & NNLO & PDF4LHC15 & \PYTHIA[8.230] & AZNLO \\
\bottomrule
\end{tabular}
}
\end{table}
\section{Calorimeter simulation with \FCaloSII}
\label{sec:fcsv2}
 
\FCaloSII parameterizes the longitudinal and lateral
development of showers in the calorimeter.  During \AFIII simulation,
energy is directly deposited in calorimeter cells using the parameterized responses. The longitudinal parameterization along with a correction to the energy resolution is discussed in Section~\ref{sec:epara}. Parameterization of the average lateral shower distribution is discussed in Section~\ref{sec:shapepara}. Finally, the simulation of hits using longitudinal and average lateral shower parameterization is described in Section~\ref{sec:NumberOfHits}.

\subsection{Longitudinal shower development }
\label{sec:epara}
As particles shower in the calorimeter, they deposit energy in the various layers. The amount of energy deposited in each layer depends on how deep in the calorimeter the shower was initiated. The amount of energy deposited is highly correlated between layers, making it difficult to independently parameterize the response for each layer.
 
Principal Component Analysis (PCA)~\cite{Jolliffe2011} is used to classify showers from the samples introduced in Section~\ref{sec:datasets} for each slice of energy, $\eta$ bin, and particle type. The PCA transformation is performed twice. The initial PCA, referred to as the `\PCAI', is used to classify showers into bins referred to as `PCA bins'. A second PCA transformation, referred to as the `\PCAII', is performed in each bin of the \PCAI to generate uncorrelated and approximately Gaussian distributions. These Gaussian distributions from each PCA bin are used in the \FCaloSII simulation. The steps of this PCA chain are discussed in detail below.
 
\begin{figure}[hbtp!]
\centering
\subfloat[]{ \label{fig:epara:pcachain_hist}
\includegraphics[width=0.47\textwidth]{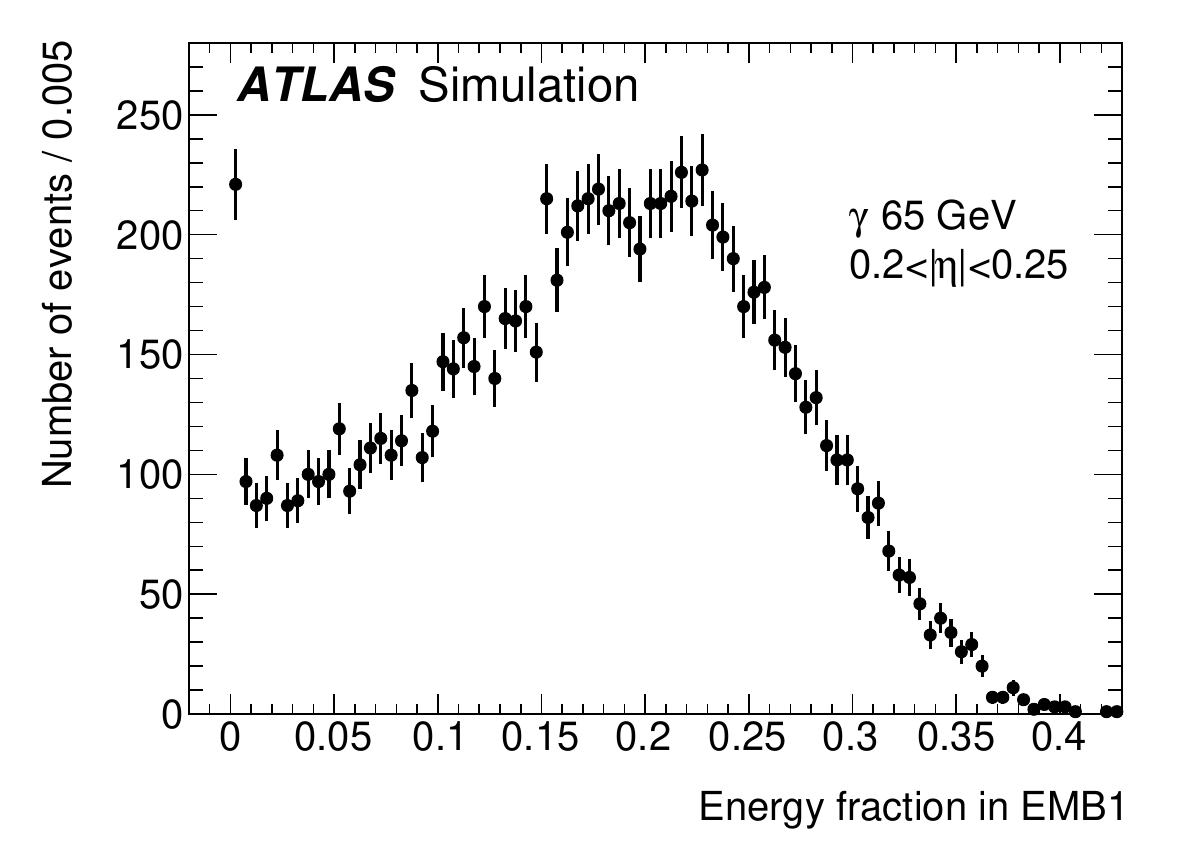}  }
\subfloat[]{ \label{fig:epara:pcachain_gauss}
\includegraphics[width=0.47\textwidth]{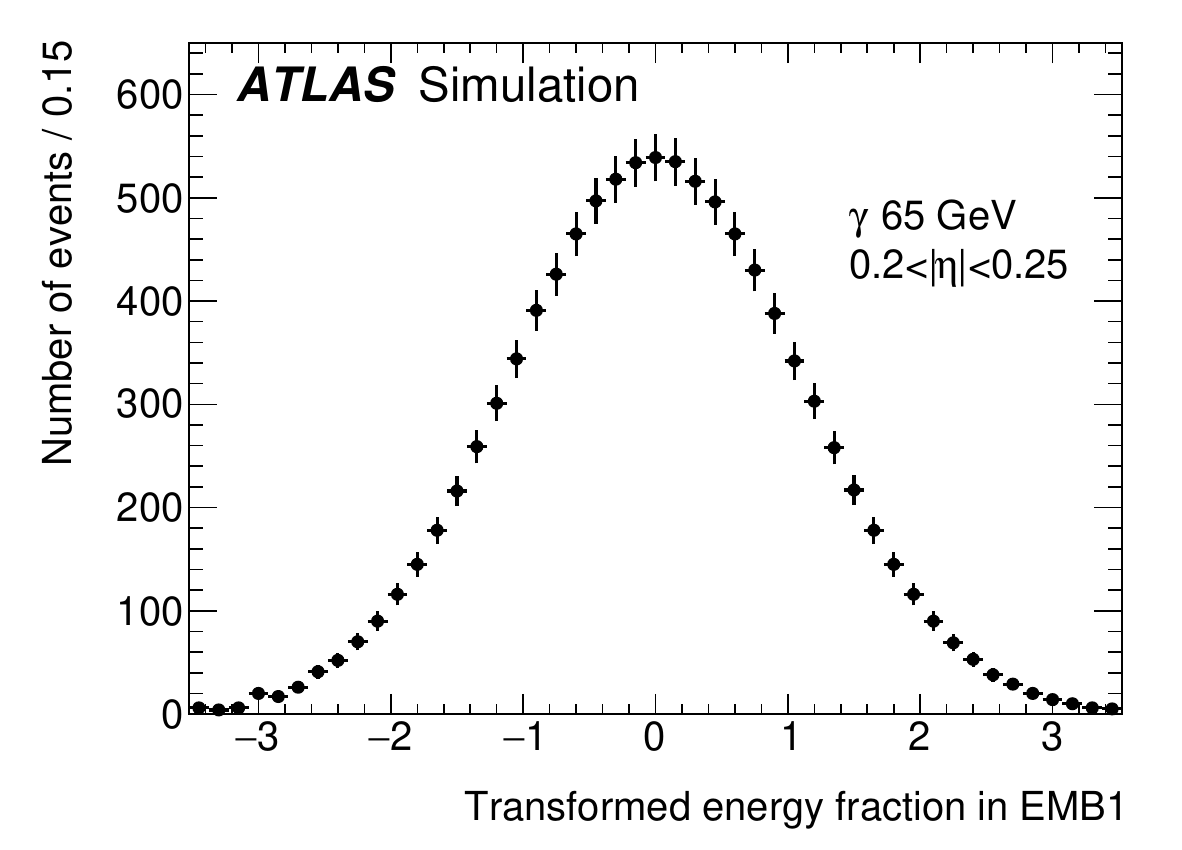}  }\\
\subfloat[]{ \label{fig:epara:pcachain_PC0}
\includegraphics[width=0.47\textwidth]{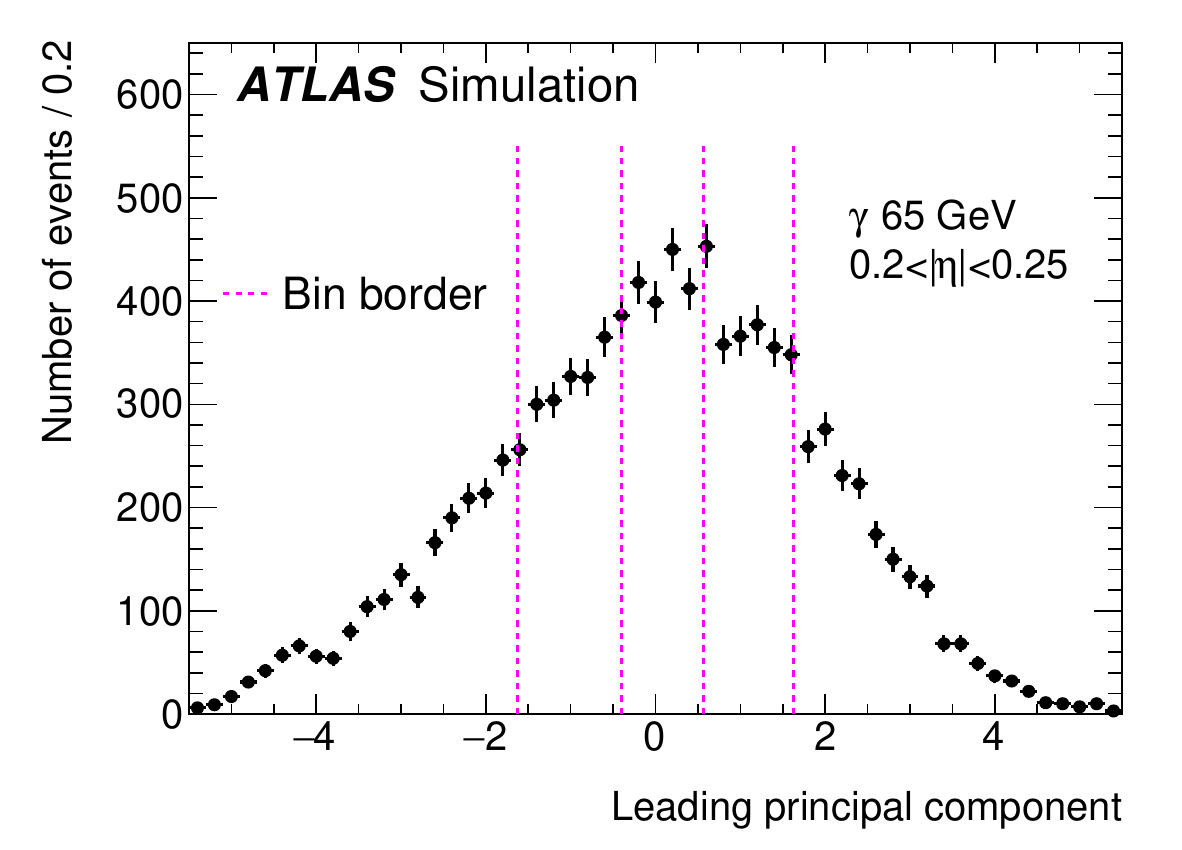}  }
\caption{ \label{fig:epara:pcachain}
Example of the steps in the \PCAI transformation for 65~\GeV\ photons with  $0.2 < |\eta| < 0.25$: (a) shows the distribution of energy fractions in EMB1, (b) the Gaussian distribution, and (c) is the leading principal component of the \PCAI  with bin borders (dotted pink lines) showing five PCA bins. The steps of the \PCAII are identical to those of the \PCAI but performed in each PCA bin separately to generate uncorrelated Gaussian distributions using all principal components of the \PCAII. The errors bars indicate the size of the statistical uncertainty.
}
\end{figure}
 
The distribution of the fraction of energy deposited (see Figure~\ref{fig:epara:pcachain_hist}) in each calorimeter layer and the total energy deposited (summed over all layers) are used to classify the showers. Only the relevant layers as defined in Section~\ref{sec:datasets:voxels} are considered. The energy fraction in each layer is integrated and transformed into a Gaussian distribution using a cumulative distribution function transformation (see Figure~\ref{fig:epara:pcachain_gauss}). These Gaussian distributions from each layer and each event are used to construct a PCA matrix to perform the \PCAI.
 
The \PCAI converts the set of correlated energies into a set of linearly uncorrelated quantities by an orthogonal transformation of the coordinate system. The transformation is calculated using the covariance and the eigenvectors of the PCA matrix. The principal components with highest and second-highest variance are referred to as the leading and sub-leading principal components of the \PCAI. Figure~\ref{fig:epara:pcachain_PC0} shows the leading principal component of the \PCAI. To classify the shower, the leading and in some cases the sub-leading principal component of the \PCAI is binned in equally populated PCA bins (covering equal ranges of cumulative PCA bin probability).
A bin with zero deposited energy is included, as this improves the modelling of low-energy particles. Typically, five PCA bins in the leading principal component of the \PCAI are used (see Figure~\ref{fig:epara:pcachain_PC0}.) However, within the transition regions of the calorimeter layers, given in Table~\ref{tab:cal_layers}, the sub-leading principal component is also used in order to determine PCA bins in  two dimensions. The exact number of bins in each region is determined from a $\chi^2$ test giving the best modelling of all energy fractions. The \PCAI removes non-linear correlations between layers and roughly classifies the showers according to their depth.
 
\begin{figure}[hbtp!]
\centering
\subfloat[]{
\includegraphics[width=0.47\textwidth]{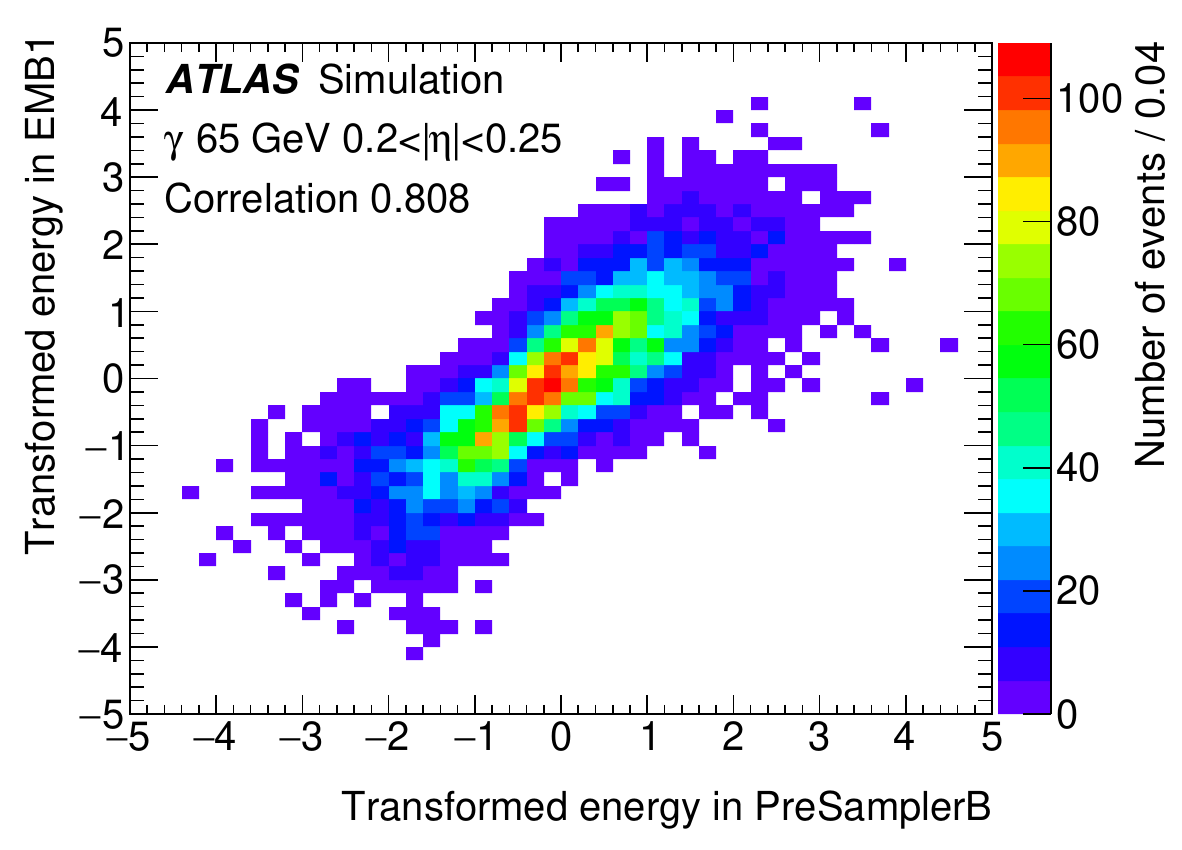}  }
\subfloat[]{
\includegraphics[width=0.47\textwidth]{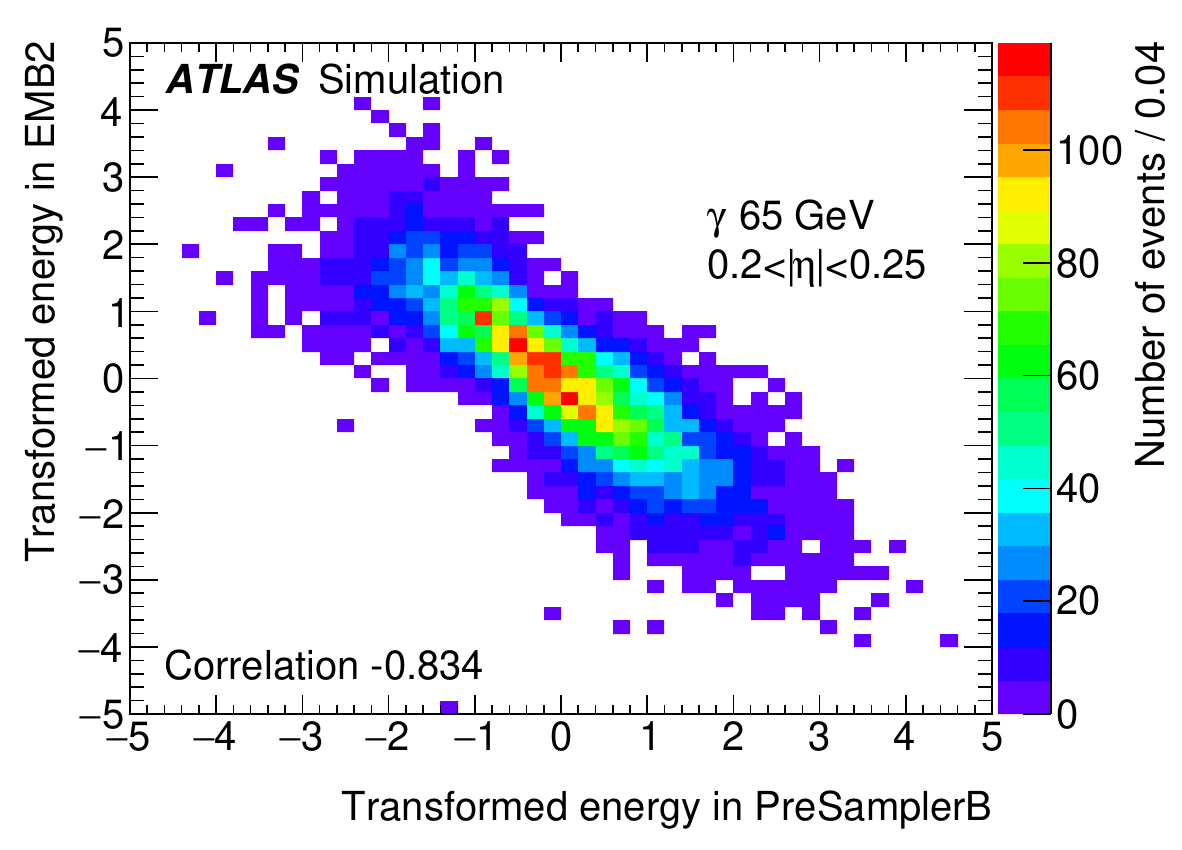}  }\\
\subfloat[]{
\includegraphics[width=0.47\textwidth]{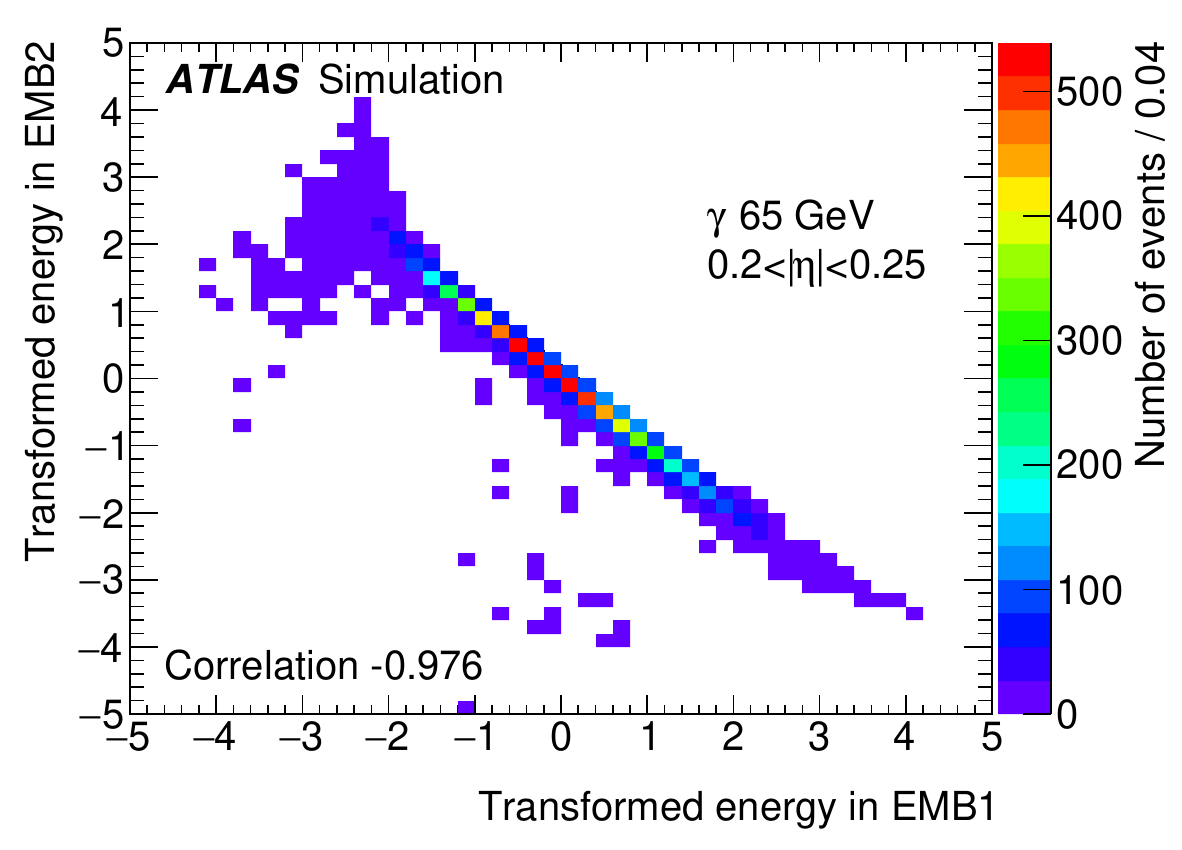}  }
\caption{ \label{fig:epara:correlations_before}
Correlations between the transformed energies deposited in several layers, before PCA rotation, showing (a) Presampler barrel vs EM barrel 1, (b) Presampler vs EM barrel 2 and (c) EM barrel 1 vs EM barrel 2. The energies were transformed into Gaussian distributions. The correlation factors obtained from these 2D histograms are displayed.
}
\end{figure}

\begin{figure}[hbtp!]
\centering
\subfloat[]{
\includegraphics[width=0.47\textwidth]{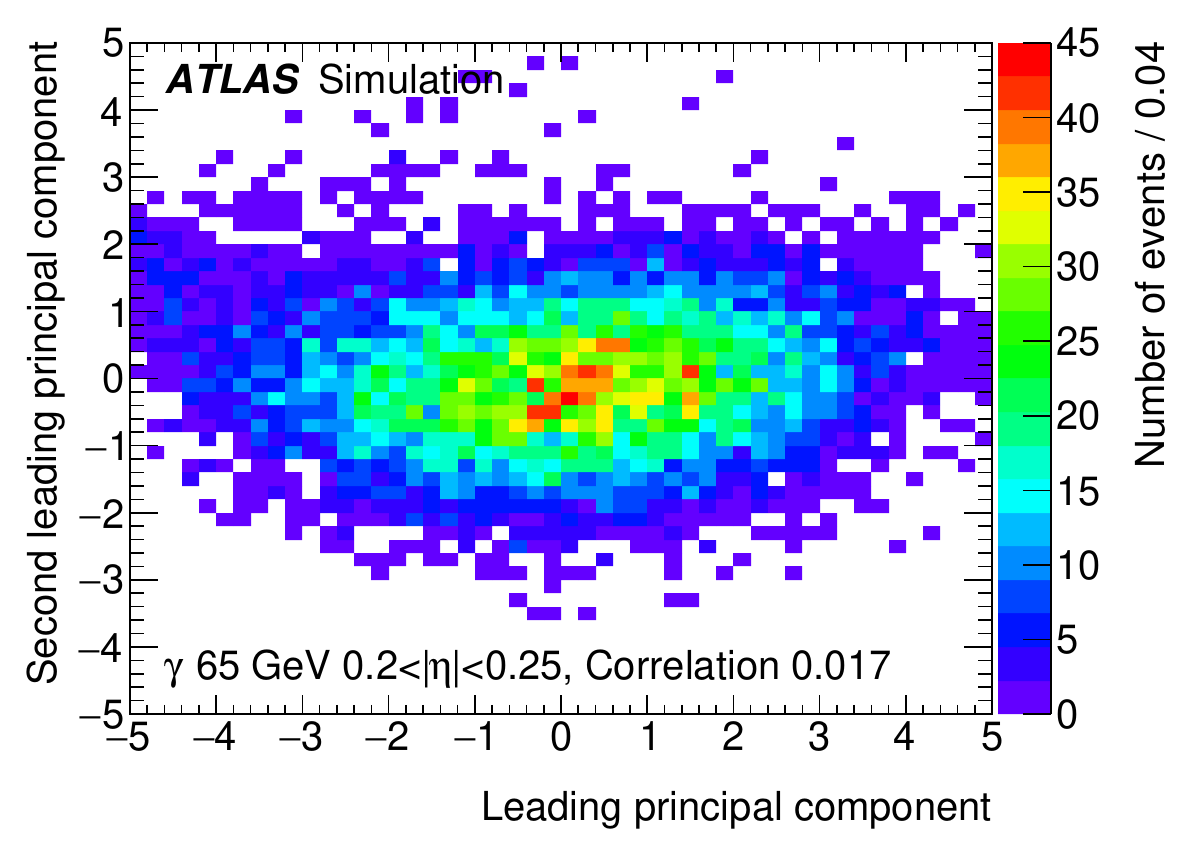}  }
\subfloat[]{
\includegraphics[width=0.47\textwidth]{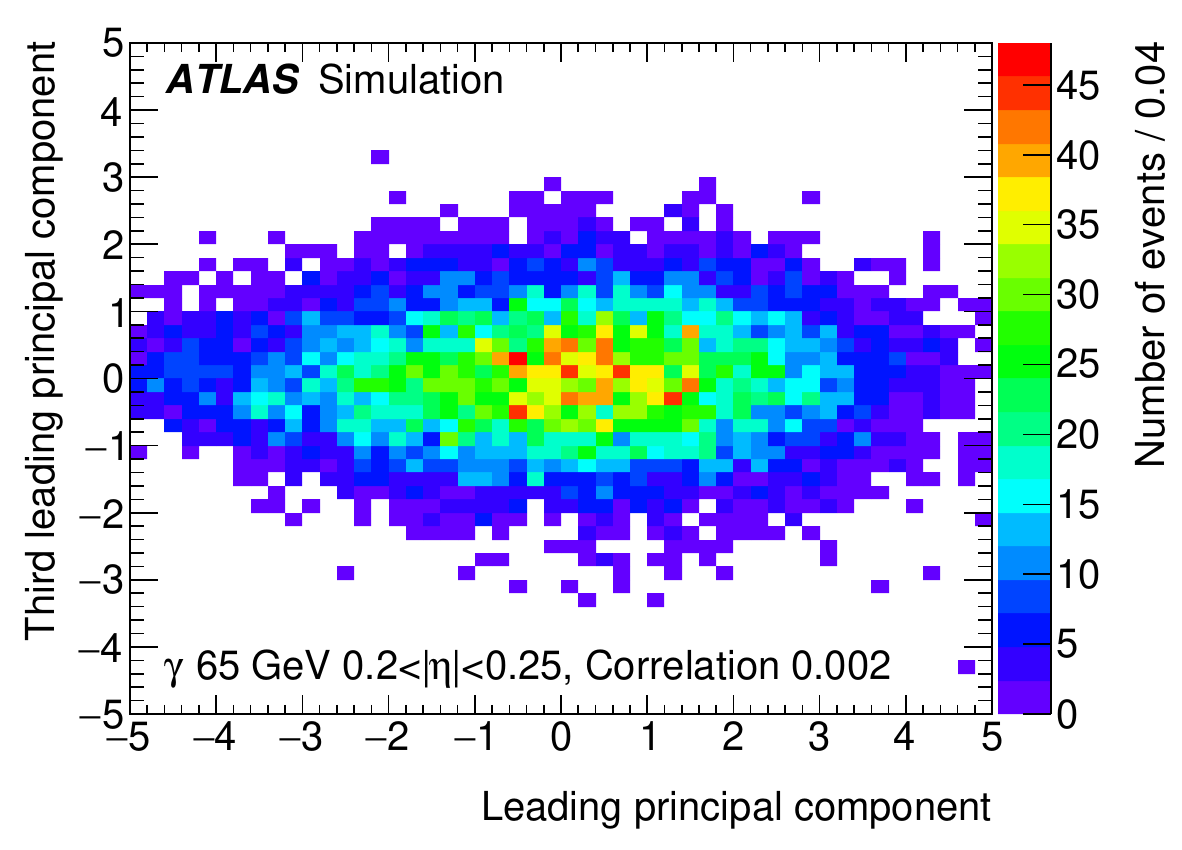}  }\\
\subfloat[]{
\includegraphics[width=0.47\textwidth]{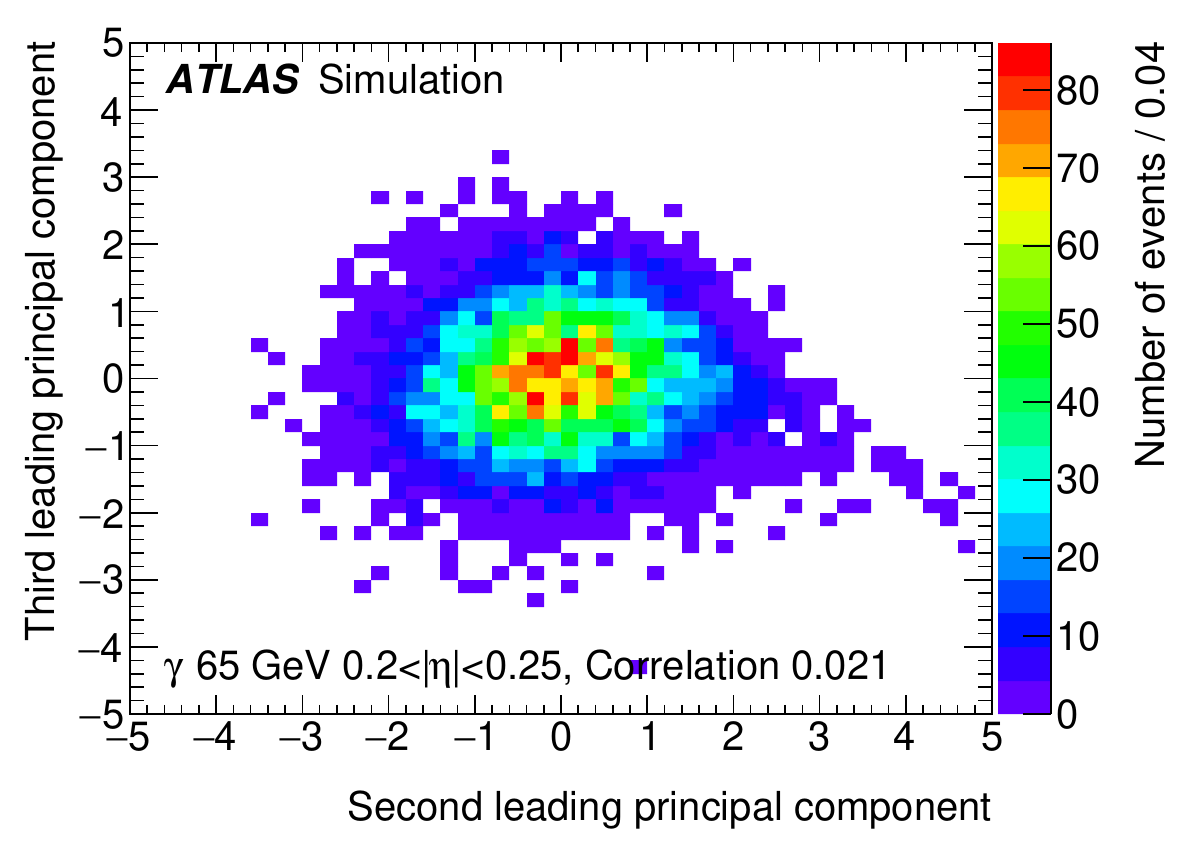}  }
\caption{ \label{fig:epara:correlations_after}
Correlations between the \PCAIh components after the PCA rotation. The individual components are approximately Gaussian distributed.
}
\end{figure}

The effectiveness of the PCA transformation is demonstrated in Figures~\ref{fig:epara:correlations_before} and~\ref{fig:epara:correlations_after}, which show the correlations before and after the \PCAI. The correlations between different layers are calculated from the Gaussian inputs of the PCA matrix. After the \PCAI, the correlations are calculated using a subset of the principal components and are strongly reduced.
 
The total energy and the energy fractions in each \PCAIh bin are transformed into Gaussian distributions following the same method and then the \PCAII is performed. The steps of the \PCAII are identical to those of the \PCAI except that only the events in a given \PCAIh bin are used. The \PCAIIh rotation removes any further correlations in each \PCAIh bin to produce uncorrelated Gaussian outputs using all principal components.  The mean and RMS of these Gaussian distributions, the PCA matrices, the PCA bin probabilities and the inverse cumulative distributions are stored and used in the simulation.
 
During simulation, the steps of the PCA chain are executed in reverse. For each simulated particle, a PCA bin is selected using random numbers distributed according to the PCA bin probabilities. The uncorrelated Gaussian distributions, stored in the parameterization, in the selected PCA bin are used to generate uncorrelated random numbers. These random numbers are rotated using the inverse PCA matrix of the \PCAII to generate correlated random numbers. The correlated random numbers  are then mapped back to the total energy and the energy fractions deposited in each layer using the error function and the inverse cumulative distributions of the \PCAI.
 
\begin{figure}[hbtp!]
\centering
\subfloat[]{\includegraphics[width=0.47\textwidth]{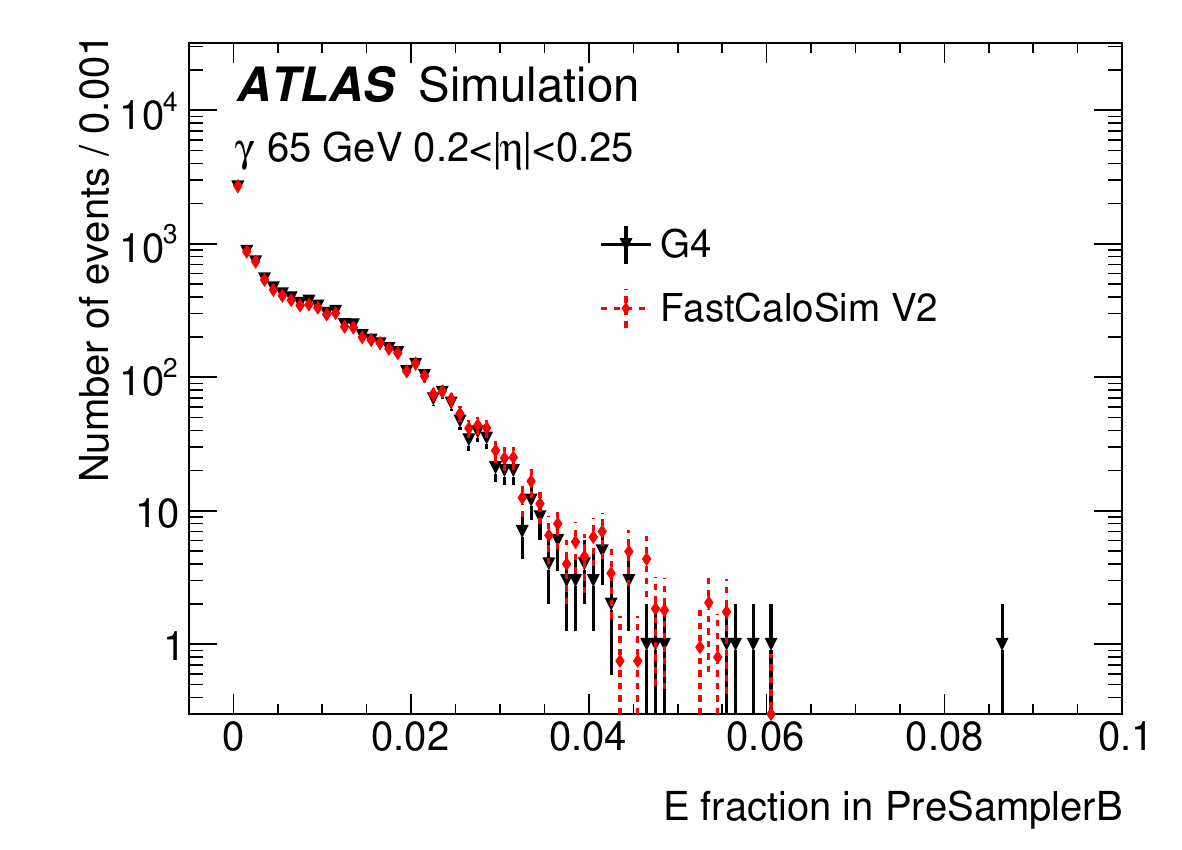}  }
\subfloat[]{\includegraphics[width=0.47\textwidth]{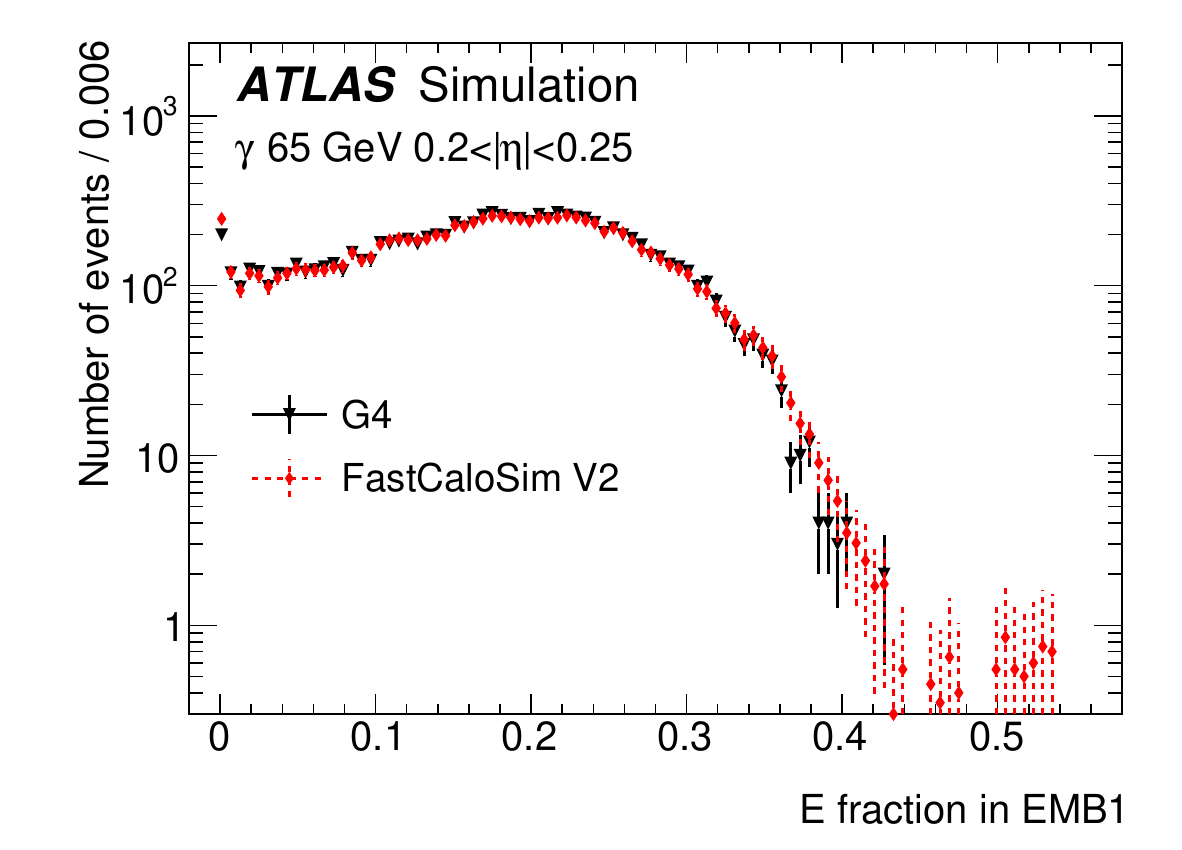}  } \\
\subfloat[]{\includegraphics[width=0.47\textwidth]{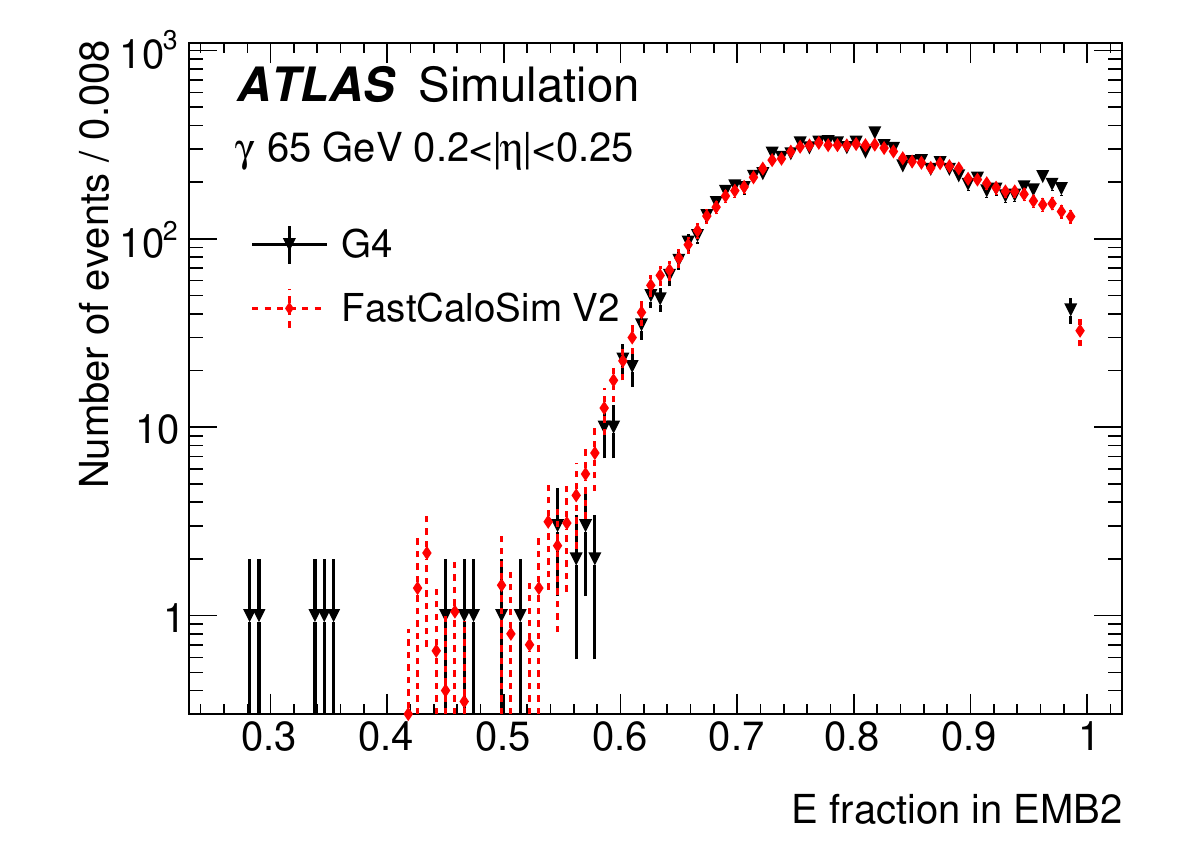}  }
\subfloat[]{\includegraphics[width=0.47\textwidth]{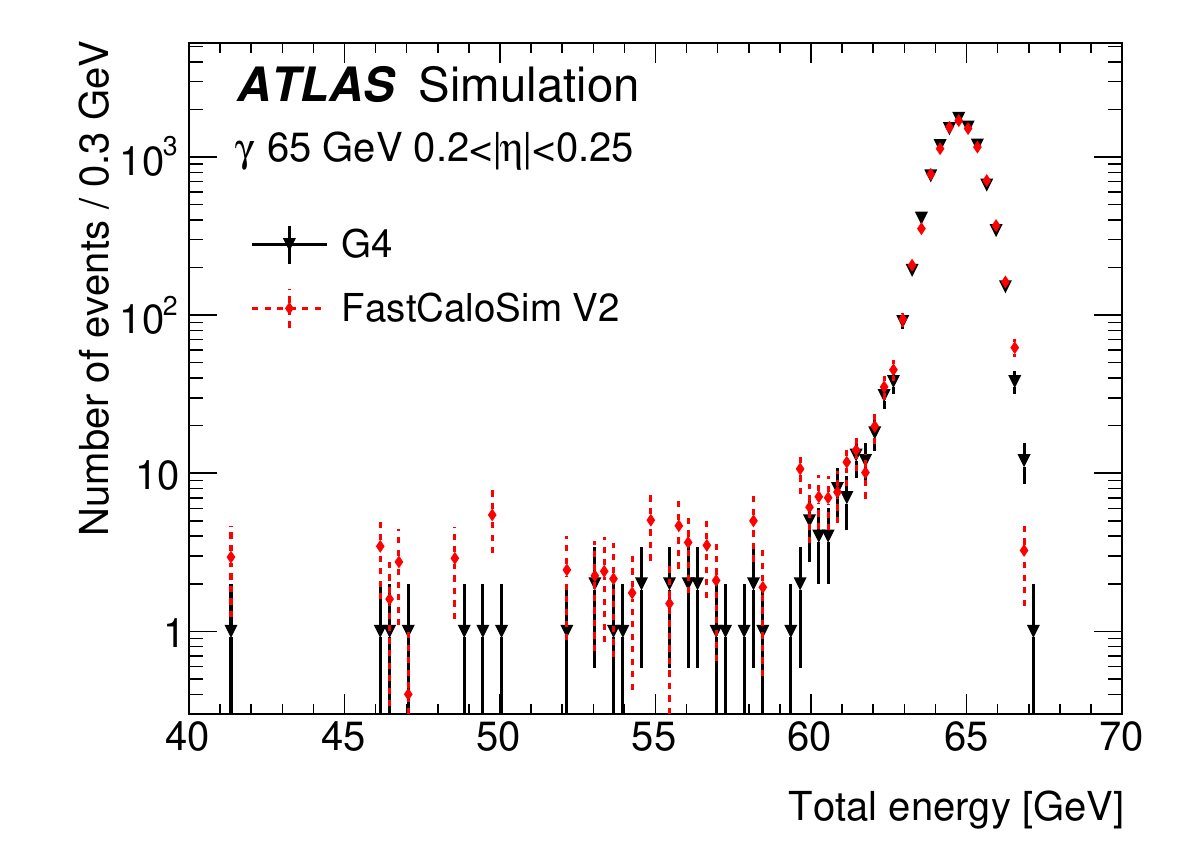}  }
\caption{ \label{fig:epara:vali431004}
Validation of the energy, $E$, parameterization is shown for 65~\GeV\ photons with $0.2 < |\eta| < 0.25$, comparing the input \GEANT sample (black triangles) with \FCaloSII (red dots). Good agreement is observed for all layers and the total energy. The errors bars indicate the size of the statistical uncertainty.}
\end{figure}

The validation of the longitudinal energy parameterization is shown in Figure~\ref{fig:epara:vali431004} for 65~\GeV\ photons with $0.2<|\eta|<0.25$. Incoming photons with an energy of 65~\GeV\ are simulated using the \FCaloSII parameterization and compared with \GEANT simulation. In this case, no digitization or reconstruction algorithms are applied, but the energy deposited in the active regions of the calorimeter has been scaled by the sampling fraction. This simulation without digitization and reconstruction is referred to as `stand-alone simulation'.
 
In general, the energy fractions deposited in each calorimeter layer using the \FCaloSII parameterization are observed to be in good agreement with \GEANT.
However, for the total energy distribution, a residual difference in the mean and a larger RMS are observed. These small differences can impact the modelling of complex quantities, e.g.\ the Higgs boson invariant mass distribution reconstructed from two photons. Additional corrections are therefore applied to improve the modelling of both the resolution and the mean, as discussed in Sections~\ref{sec:energy:res_corr} and~\ref{sec:hadron_correction}.
 
\subsection{Average lateral shower shape }
\label{sec:shapepara}
\begin{figure}[hbtp!]
\subfloat[photon]{\includegraphics[width=0.45\textwidth]{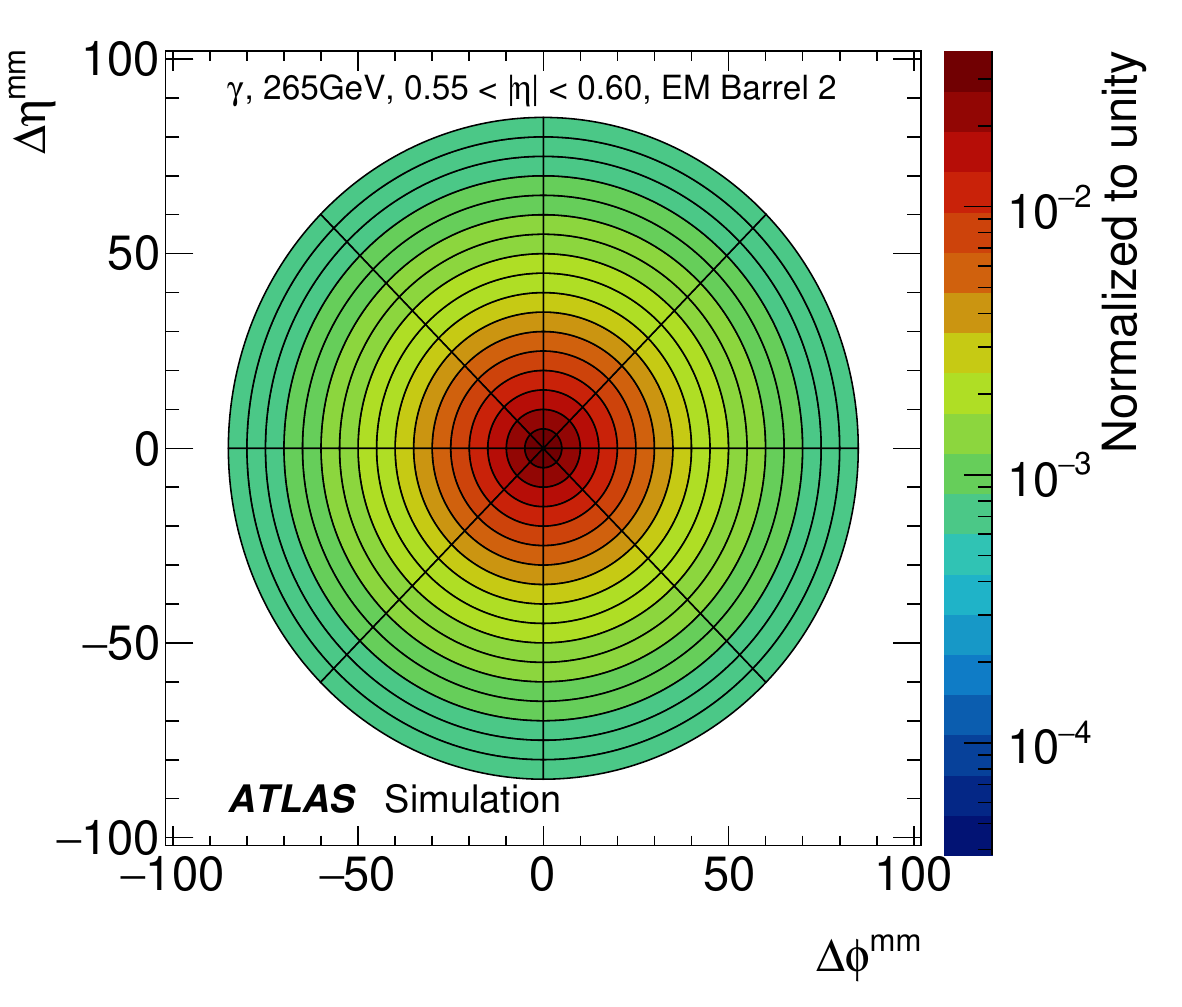}}
\label{fig:shape:param:photon_full}
\subfloat[pion]{\includegraphics[width=0.45\textwidth]{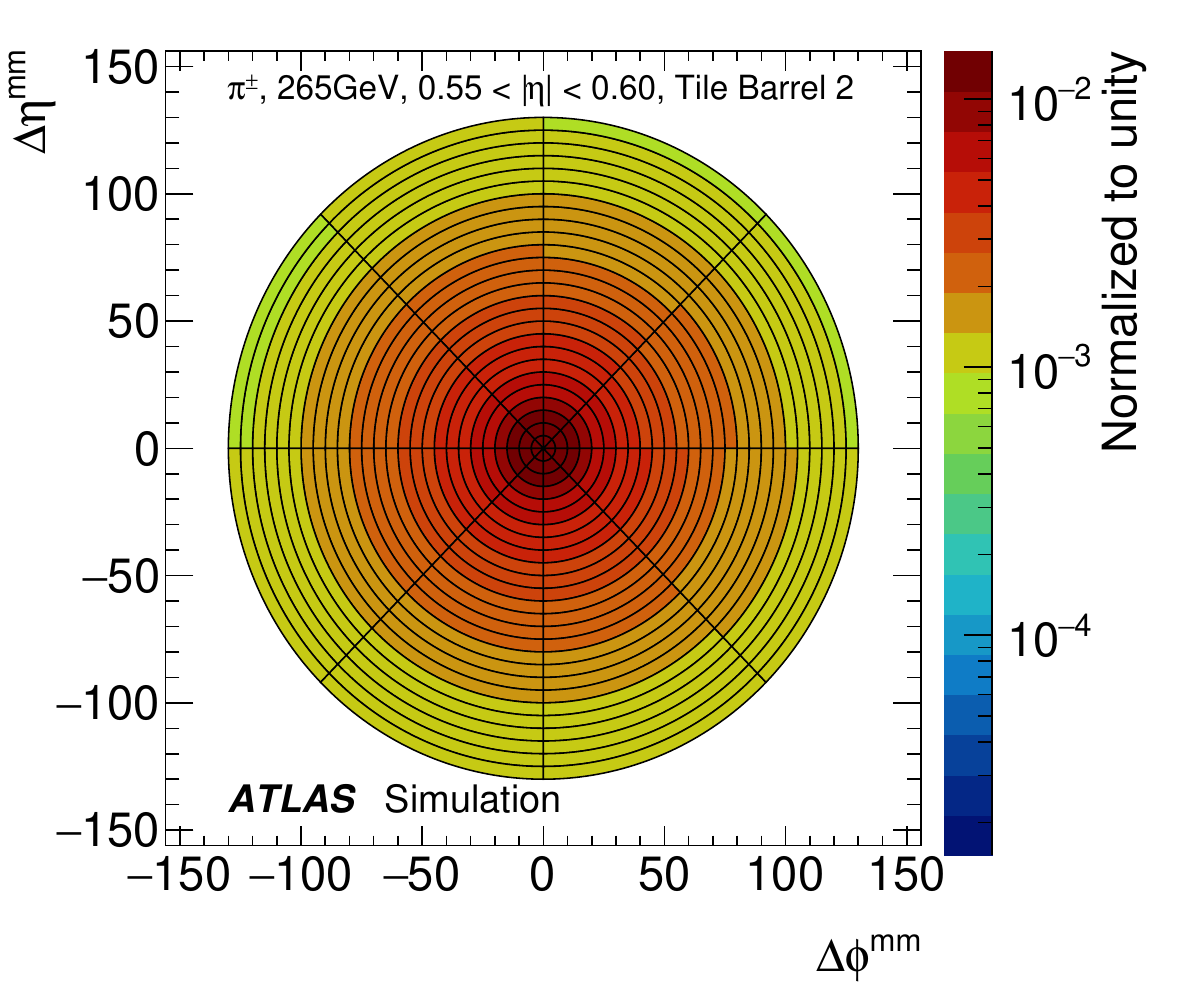}}
\label{fig:shape:param:pion_full}
\caption{ The lateral shower shape parameterization for (a) photons and (b) pions with energies of 265~\GeV{} in the range $0.55 < |\eta| < 0.60$ and parameterized in the second layer of the EM barrel  and Tile  barrel respectively. To visualize the core of the shower, these plots have a cut-off at $\Delta R^\text{mm} \sim 100~\SI{}{\milli\meter}$.}
\label{fig:shape:param_full}
\end{figure}

The lateral shower shape describes the lateral energy distribution in each calorimeter layer. The parameterization is derived in each relevant layer and for each PCA bin. The shower development is parameterized in voxels using the coordinates defined in Eq.~(\ref{eqn:shape:r_alpha}). To exclude hits far away from the centre of the shower, only 99.5\% of the total energy of each PCA bin cumulatively along $\Delta R^{\SI{}{\milli\meter}}$ is considered. The shower shape distribution in each PCA bin (of each layer) is then normalized to the energy in that PCA bin to create the probability density function for the average shower shape. Figure~\ref{fig:shape:param_full} shows the average lateral shower profile corresponding to the electromagnetic and hadronic showers in the second layer of the EM barrel and the Tile barrel, respectively.  The memory footprint of these histograms is reduced by storing only the $|\Delta\phi^{\SI{}{\milli\meter}}|$ coordinates for $0 \leq \alpha \leq \pi$, because the shower is symmetric in $\Delta\phi^{\SI{}{\milli\meter}}$.
 
\subsection{Simulation of hits}
\label{sec:NumberOfHits}
A key limitation of \AFII{} is that the lateral shower shape simulation is based on the average shower shape. This model works well for electrons and photons, but cannot reproduce the complex structure of hadronic showers. ATLAS extracts the shower structure of electrons, photons, hadrons and jets by clustering calorimeter cells using the  TopoCluster algorithm~\cite{PERF-2014-07}.
The clustering proceeds by starting from a seed cell with an energy 4$\sigma$ above the calorimeter noise threshold and adding cells with an energy at least 2$\sigma$ above the noise and finally adding adjacent cells of any energy. However, using the average shower shape means that the energy distribution and position of the hadronic clusters differ compared to \GEANT.
Instead of directly using the average shower shape, \FCaloSII{} uses the average shower shape as a probability distribution function (pdf) to generate hits which are subsequently mapped onto the calorimeter cell structure. For particles entering the calorimeter with a non-zero angle with respect to the calorimeter cell boundaries, the position of each hit is modified to account for the longitudinal position within each layer to improve the simulation of the shower shapes. For each PCA bin, the average value of the longitudinal position distribution from the reference sample is used to correct the lateral position at which the hits are produced.
The models used to assign energy to each hit are discussed in the following sections. The energies of hits are normalized so that their sum exactly matches the simulated energy in a layer as discussed in Section~\ref{sec:epara}.
 
\subsubsection{Electrons and photons}
\label{sec:NumberOfHits:egamma}
 
The number of generated hits for electrons and photons is calculated from the energy deposited in each calorimeter layer and the intrinsic resolution of the calorimeter technology in that layer.  For a given energy $E$ simulated within a calorimeter layer, the resolution is defined as:
\begin{eqnarray}
\label{eqn:shape:det_res}
\sigma_E / E &=& a / \sqrt{E/\GeV} \oplus c
\end{eqnarray}
where $a$ is the stochastic term and $c$ the constant term. The values used for $a$ and $c$ for the different detector technologies are listed in Table~\ref{tab:NumberOfHitsCalc} and are used to calculate the resolution, $\sigma_E$.
 
\begin{table}[hbtp!]
\centering
\caption{Stochastic and constant terms for the intrinsic calorimeter energy resolution for the different detectors and used to simulate the hits for electrons and photons~\cite{CERN-LHCC-96-040}.}
\vspace{0.2cm}
\begin{tabular}{lll}
\toprule
Calorimeter technology & Stochastic term $a$ & Constant term $c$ \\
\midrule
LAr EM barrel and endcap & 10.1\% & 0.2\% \\
Tile                    & 56.4\% & 5.5\% \\
LAr hadronic endcap    & 76.2\%  & 0     \\
FCal                   & 28.5\%   & 3.5\% \\
\hline
\end{tabular}
\label{tab:NumberOfHitsCalc}
\end{table}
 
The expected number of hits, $\lambda$, which would produce this resolution from a Poisson statistical process is calculated as:
 
\begin{eqnarray}
\label{eqn:shape:lambda}
\lambda &=& 1/ (\sigma_E /E)^2
\end{eqnarray}

A random number following a Poisson distribution $N = \text{Poisson}(\lambda)$ is used to simulate $N$ hits of equal energy $E_\text{hit}=E/N$. The positions of these hits are randomly distributed according to the average shower shape introduced in Section~\ref{sec:shapepara}.
 
For electrons and photons, which deposit most of their energy in the LAr EM calorimeters, the expected number of hits (see Eq.~(\ref{eqn:shape:lambda})) is dominated by the stochastic term, $a$, in the energy resolution. Assuming a stochastic term of 10.1\% and equal energy for each hit, electron and photon showers have hits with $E_\text{hit} \approx 10~\MeV$.
 
\subsubsection{Hadrons}
\label{sec:NumberOfHits:hadrons}
 
The number of generated hits for hadrons is calculated following the same procedure as described in Section~\ref{sec:NumberOfHits:egamma}. However, for hadrons, the stochastic and constant terms in each layer are $\eta$-dependent and much larger due to intrinsic fluctuations in hadronic showers. To derive these terms, a special simulation of charged pions is used, where in addition to the measurable energy deposited in the active material of the calorimeter, the total energies lost in both the active and inactive parts of the calorimeter are recorded. The ratio of these two energies in each layer is the sampling fraction per shower and varies with the total energy deposited, denoted by $f_{\text{sample}}(E)$. The relative resolution, $\sigma_E/E$, of $f_{\text{sample}}(E)$ is fitted with Eq.~(\ref{eqn:shape:det_res}) to extract the stochastic and constant terms for each $\eta$-slice. Only showers that deposit more than 1~\GeV{} of energy in a calorimeter layer are considered in the fit.
 
The stochastic terms obtained from simulation for pions are in the range of approximately 30\%--40\% for the EM calorimeters, 50\%--60\% for the Tile calorimeter, 60\%--80\% for the HEC calorimeter and 80\%--100\% for the FCal. The constant terms $c$ are in the range of 1\%--10\%.
 
Using $\eta$-dependent stochastic and constant terms significantly  improves the modelling of hadronic showers for most layers. A notable exception is observed in the highly granular calorimeter layers EMB1 and EME1. In these cases, the stochastic and constant terms shown in Table~\ref{tab:NumberOfHitsCalc} are used.
 
\begin{figure}[h]
\centering
\subfloat[\GEANT]{\label{fig:WeightedHits:G4}
\includegraphics[width=0.48\textwidth]{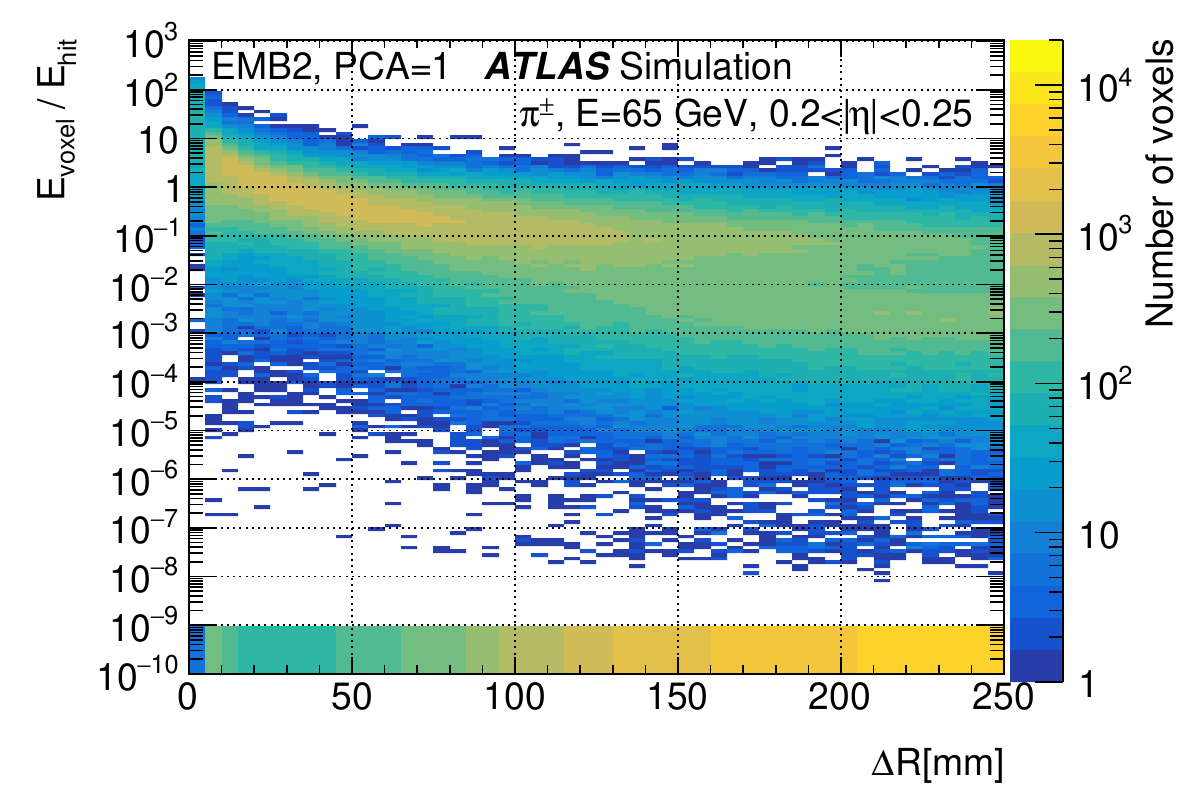} }
\subfloat[Model: equal hit energy]{\label{fig:WeightedHits:equal}
\includegraphics[width=0.48\textwidth]{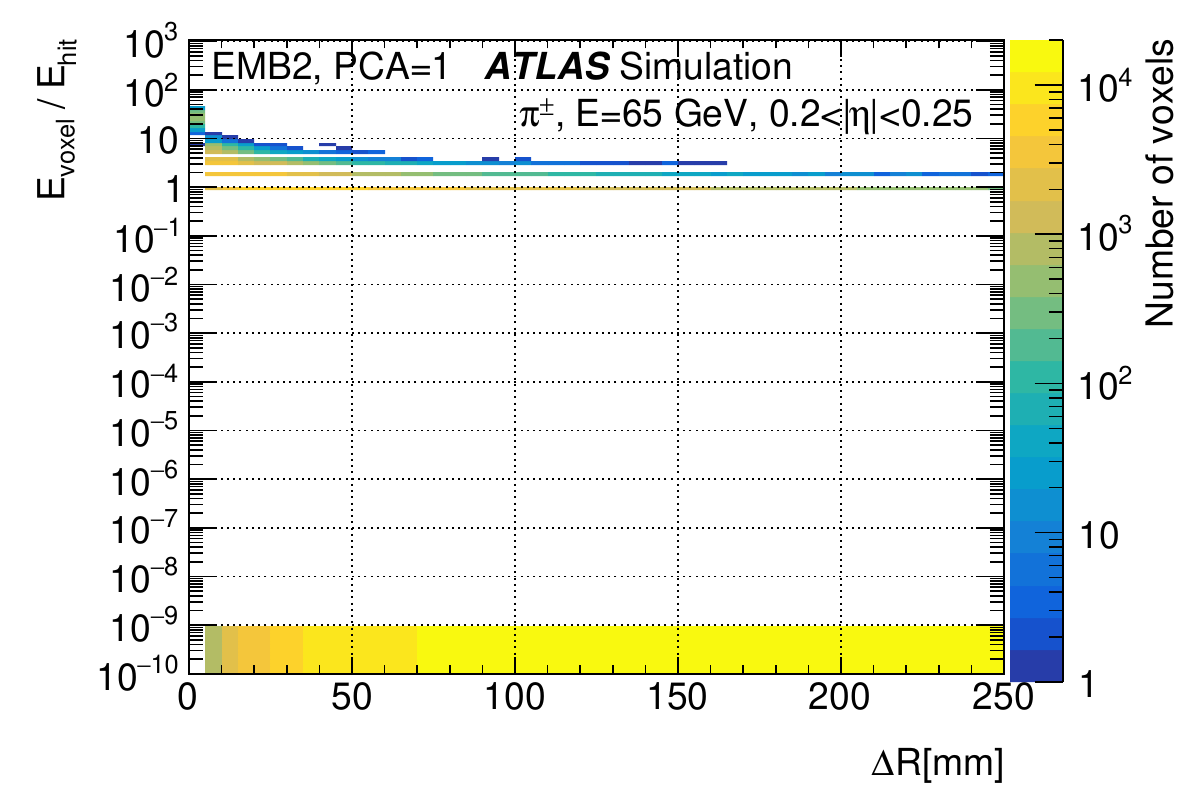} }\\
\subfloat[Model: weighted hit energy]{ \label{fig:WeightedHits:weight}
\includegraphics[width=0.48\textwidth]{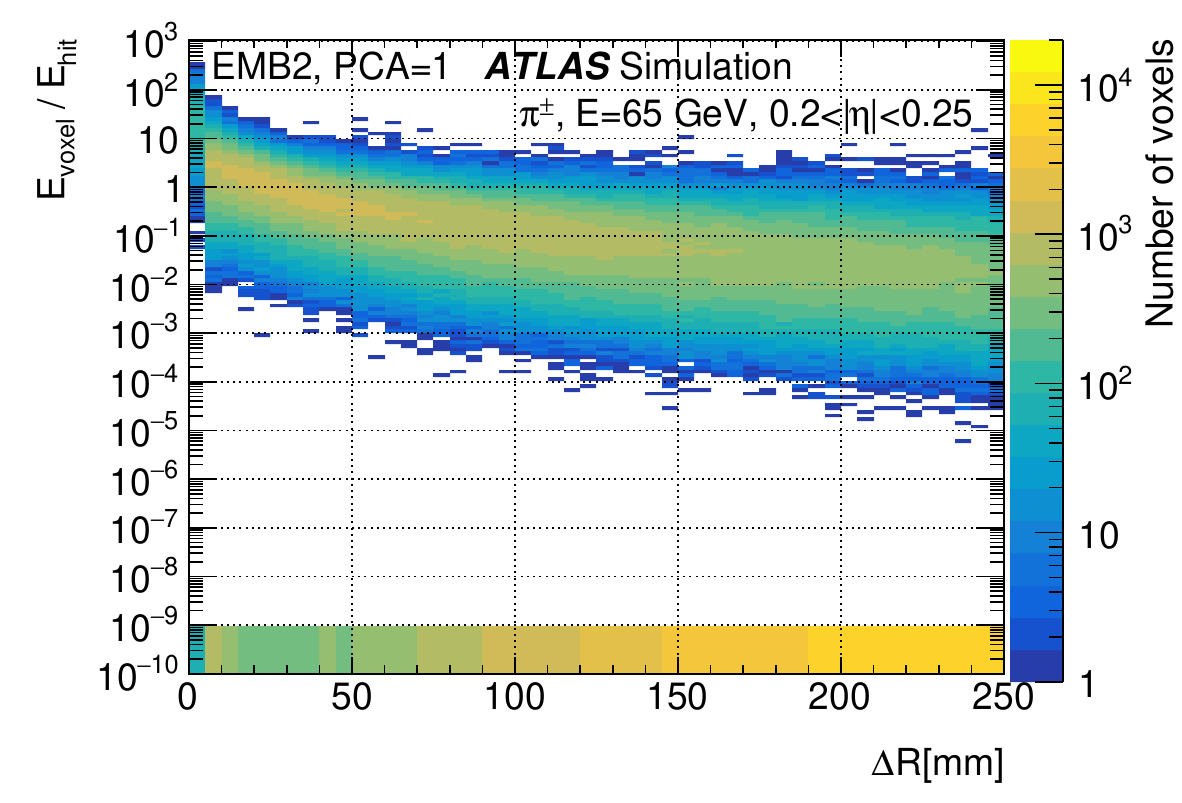} }
\subfloat[Comparison of mean and RMS of the models]{\label{fig:WeightedHits:comparison}
\includegraphics[width=0.48\textwidth]{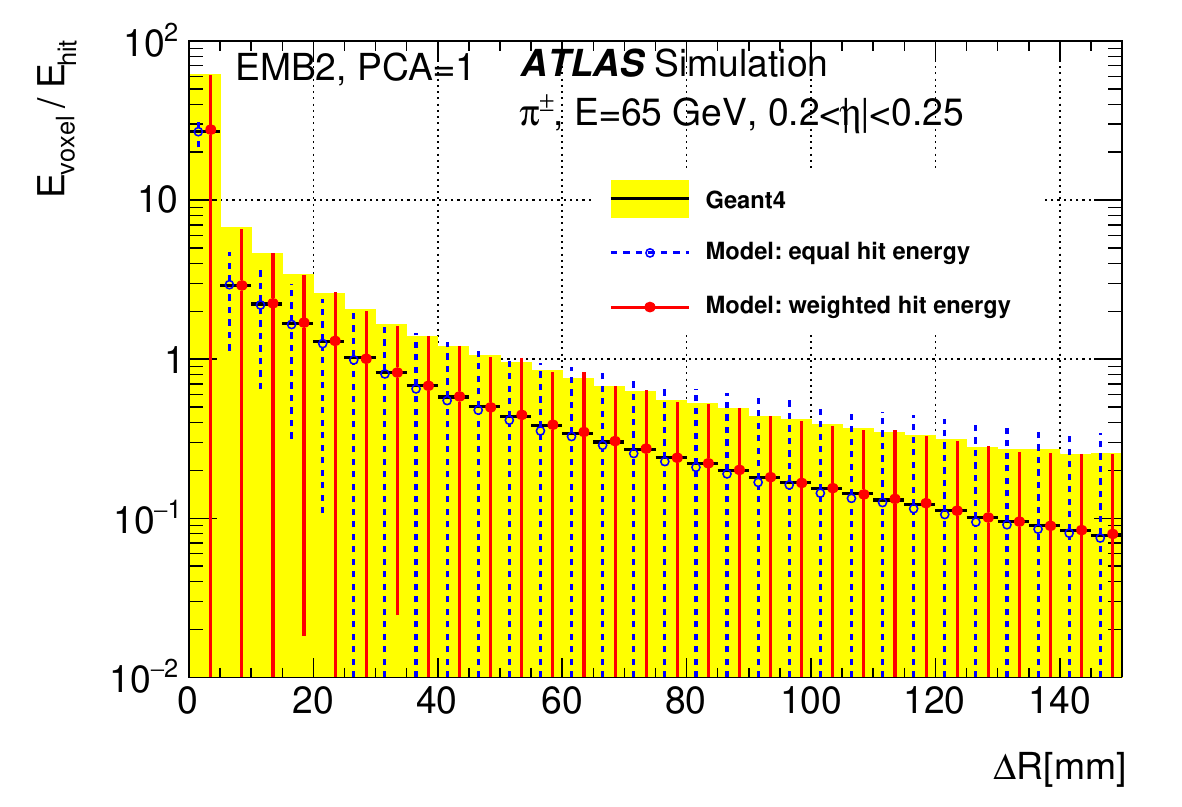}}
 
\caption{Ratio $E_\text{voxel}/E_\text{hit}$ as function of $\Delta R^{\SI{}{\milli\meter}}$ for deposited energy from a 65~\GeV\ charged pion in EMB2 in the range $0.20<|\eta|<0.25$ in the first bin of the leading PCA (PCA=1). Entries with $E_\text{voxel}=0$ are shown in the underflow bin below $10^{-9}$. Lateral shower shape model (a) in \GEANT, (b) in a model using equal deposited energies, (c) in a model using weighted hit deposited energies. (d) Comparison of the mean (central value) and the RMS (error bars) for the equal hit, weighted hit and \GEANT models. The yellow band indicates the $1\sigma$ uncertainty for \GEANT.}
\label{fig:PionEMB2HitEnergies}
\end{figure}

In Figure~\ref{fig:WeightedHits:G4} the energy fractions inside voxels along $\Delta R^{\SI{}{\milli\meter}}$ in the EMB2 layer of the calorimeter are shown for a 65\,\GeV\ charged pion in the range $0.20<|\eta|<0.25$ for the first bin of the leading PCA using \GEANT. The number of voxels with a particular energy fraction is represented on the $z$-axis. In the \GEANT distribution, away from the centre of the shower, only a small number of voxels have an energy of $O(E_\text{hit})$, while most voxels have a substantially lower energy. Due to the large stochastic terms for hadrons ($> 30\%$), $E_\text{hit}$ is approximately 100--300~\MeV{} for hadronic showers, which is similar to the energy needed to seed a cluster.
 
In Figure~\ref{fig:WeightedHits:equal} the same distribution is shown simulated using \FCaloSII where each hit is assigned equal energy. The mean and the RMS calculated by including the voxels with an energy of $O(E_\text{hit})$, for each distribution, are compared in Figure~\ref{fig:WeightedHits:comparison}.  Although the mean of the energy fraction is correctly reproduced by \FCaloSII, the number of voxels with an energy of $O(E_\text{hit})$ is substantially larger away from the centre of the shower. In many cases just one of these hits together with some noise is sufficient to seed the formation of a calorimeter cluster, which then leads to substantial differences in the cluster energy and position distribution compared to \GEANT.
 
To correct for this mismodelling, a second model is developed where instead of assigning equal energy, a hit weight is introduced. The weight is calculated such that the number of hits simulated in a certain $\Delta R^{\text{mm}}$ bin is changed to better reproduce the RMS of the distribution from \GEANT, denoted by $\text{RMS}_\text{G4}$. The steps involved to calculate this weight are discussed below.
 
As a first step, the voxels with sufficiently low energy are assigned $E_{\text{voxel}} = 0$. Only voxels that do not change the mean by more than 0.01\% are considered in this step. Then two Poisson distributions are calculated, one reproducing the fraction of voxels with $E_{\text{voxel}} = 0$ and a second one reproducing the RMS of the total \GEANT distribution. The smaller of the two RMS values is used and denoted by $\text{RMS}_\text{Poisson}$.
 
The $\text{RMS}_\text{Poisson}$ value is used to determine the number of simulated hits by calculating:
\begin{eqnarray*}
N_\text{Poisson} &=& 1/(\text{RMS}_\text{Poisson}/\lambda)^2.
\end{eqnarray*}
The weight is then calculated using $N_\text{Poisson}$ as follows:
\begin{equation*}
w=\langle E_\text{voxel}/E_\text{hit} \rangle/N_\text{Poisson}
\end{equation*}
and the energy is recalculated as:
\begin{equation*}
E^\prime_\text{hit}=E_\text{hit} \cdot w.
\end{equation*}
These weights are calculated for each average shower shape discussed in Section~\ref{sec:shapepara}. To ensure that the average shower shape is unchanged, a correction of $1/w$ is applied to the probability of all voxels at a distance of $\Delta R^{\text{mm}}$ from the shower center.

In addition, if the RMS of the Poisson distribution is smaller than that of \GEANT, i.e.\
$\text{RMS}_\text{Poisson} < \text{RMS}_\text{G4}$, additional fluctuations are added by applying a smearing function to the $N_\text{Poisson}$ value. The smearing function has the form $\text{e}^s$, where $s$ is a random number generated from a Gaussian distribution such that:
\begin{equation*}
\text{RMS}_{\text{e}^s} = \text{RMS}_\text{smearing}/\sqrt{N_\text{Poisson}}\,,
\end{equation*}
and $\text{RMS}_\text{smearing}$ is calculated as:
\begin{equation*}
\text{RMS}^2_\text{smearing}=\text{RMS}^2_\text{G4} - \text{RMS}^2_\text{Poisson}\,.
\end{equation*}
 
Combining these corrections, the hit energy is derived as:
\begin{equation*}
E^{\prime\prime}_\text{hit}=E_\text{hit} \cdot w \cdot \text{e}^s\,.
\end{equation*}
 
Figure~\ref{fig:WeightedHits:weight} shows the $E_\text{voxel}/E_\text{hit}$ distribution simulated using \FCaloSII with the weighted hit model. The number of voxels with an energy of $O(E_\text{hit})$ is seen to be substantially better modelled when compared with \GEANT. Additionally, the mean and the RMS of the equal hit model and the weighted hit model are overlaid with those of \GEANT in Figure~\ref{fig:WeightedHits:comparison}, demonstrating the improved modelling of the RMS for weighted hits.  The dependence of the weight $w$ and the $\text{RMS}_{\text{e}^s}$ parameter is stored as function of $\Delta R^{\SI{}{\milli\meter}}$ together with the average shower shape scaled by $1/w$ for all charged pions with energy above 1~\GeV. For hadrons, the weighted hit model is used instead of the equal hit energy model.
 
\FloatBarrier
\section{Calorimeter simulation with \FGAN}
\label{sec:fcsgan}
\FGAN is a fast calorimeter simulation tool that parameterizes the interactions of particles in the ATLAS calorimeter system using 300 GANs, one for each particle type and $\eta$ slice in which the reference samples are produced. \FGAN takes a different approach than \FCaloSII, which as seen in Section~\ref{sec:fcsv2} factorizes the shower parameterization into several components, i.e.\ longitudinal and lateral energy distributions for different energy points that requires interpolation between them. A GAN, instead, provides a comprehensive solution to the simulation of any particle of any energy. This results in a simpler model that has a lower memory requirement at the price of a significantly higher time needed for producing the parametrisation used in the simulation.
A detailed description of \FGAN is provided in Ref.~\cite{ATL-SOFT-PUB-2018-001}, and other studies of deep generative models for fast calorimeter simulation can be found in Refs.~\cite{de_Oliveira_2017,PhysRevLett.120.042003,CALOGAN,erdmann2018generating,Erdmann_2019,3dimGAN}. GANs were chosen because they have proven successful in generating realistic showers in calorimeters. A GAN~\cite{GAN} is a combination of two deep networks, a generator producing artificial showers and a discriminator trying to distinguish the generated images from real ones. The two networks compete against each other in a game resulting in a type of training that is unlike those for other machine-learning problems. For example, the loss functions cannot easily be used to assess the quality of the training or to select the best training point.
The architecture of the GANs is described in Section~\ref{sec:WGAN}. Section~\ref{sec:gan:result} discusses the training strategy, the selection of the best epoch and its performance. Finally, the strategy to map the energy from the voxels to the calorimeter cells is explained in Section~\ref{sec:gan:athena}.
 
\subsection{Architecture}
\label{sec:WGAN}
\FGAN uses the Wasserstein GAN~\cite{arjovsky2017wasserstein} with a
gradient penalty (WGAN-GP) term~\cite{WGAN-GP} in the loss function of the
discriminator. This configuration provides good performance
and training stability.
 
\begin{figure}[htb]
\begin{center}
\includegraphics[width=.7\linewidth]{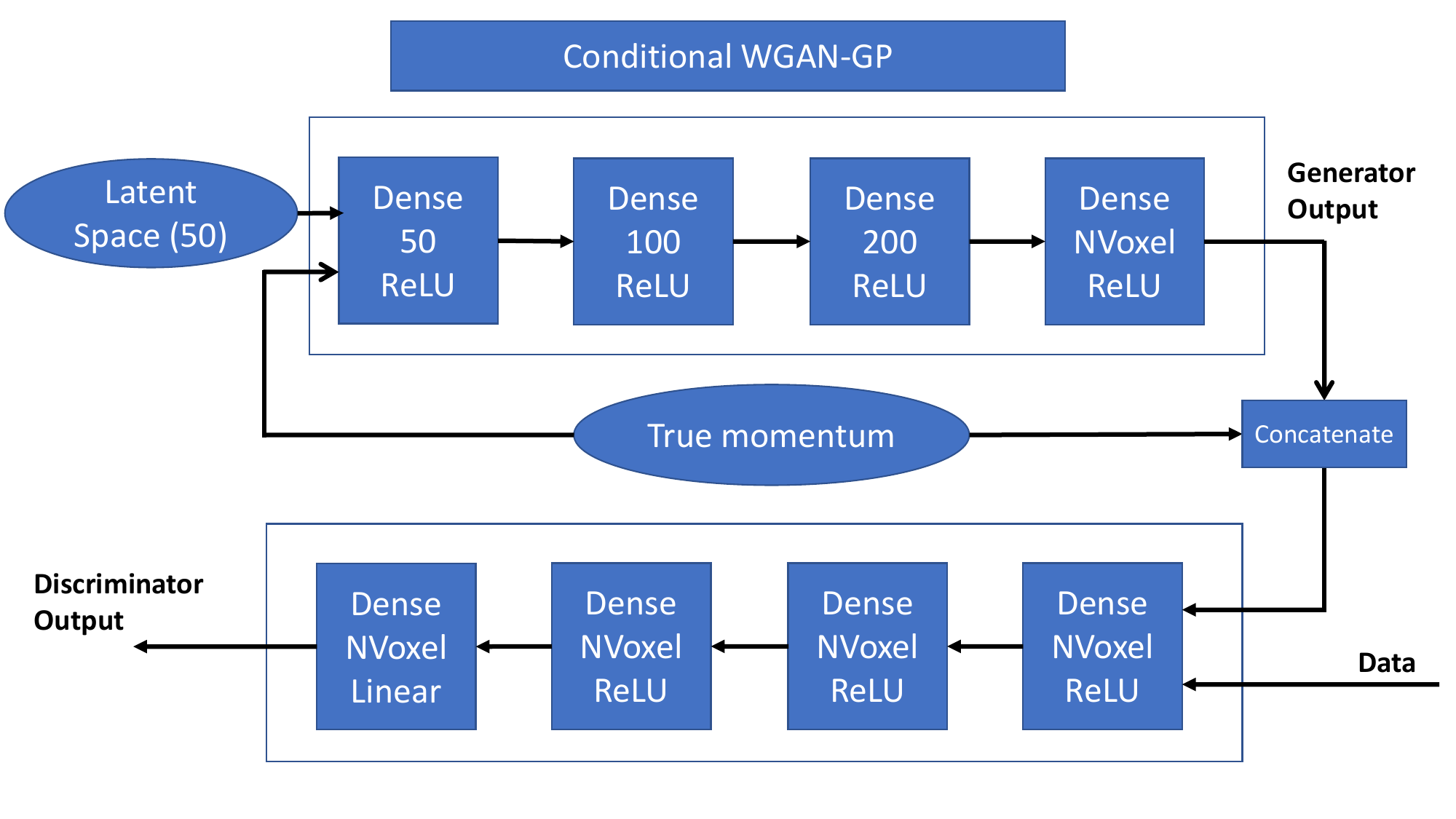}
\end{center}
\caption{Schematic representation of the architecture of the GANs used by \FGAN. The input to the generator is at the top left and the output from the discriminator is at the bottom left. The Rectified Linear Unit (ReLU) activation function is used in all layers of the discriminator with the exception of the last.}
\label{fig:gan:architecture}
\end{figure}

The WGAN-GP is implemented in TensorFlow
2.0~\cite{TF} such that the training can be performed on either CPUs
or GPUs. The architecture of the WGAN-GP is presented in
Figure~\ref{fig:gan:architecture}. The generator uses a latent space
of 50 values and has three hidden layers of increasing size. The
output layer of the generator and the input layer of the discriminator
have a number of nodes equal to the number of voxels (NVoxel) corresponding to the specific
particle type and $\eta$ slice. The discriminator maintains the same number of nodes until the last layer, which has a single output node. The GANs are conditioned on a single parameter, the true momentum of the particle.
 
Each node uses the Rectified Linear Unit (ReLU) activation function.
Both the generator and discriminator use the Adam~\cite{kingma2014adam} optimizer with a learning rate of $10^{-4}$. The exponential decay rate for the first moment
($\beta1$) is set to 0.5, while the second moment ($\beta2$) is set to the default value (0.999), as are all other parameters that are explicitly given here.
The training is performed using a batch size of 128 events, and the discriminator is trained five times for each training of the generator.
Finally, the gradient penalty, $\lambda$, is set to 10. These parameters are summarized in Table~\ref{tab:GANArchitecture}. This set of hyperparameters as well as the overall architecture were chosen as a compromise between the modelling performance and the time required to train the 300 GANs.
 
\begin{table}[]
\caption{Overview of the parameters of the WGAN-GP.   }
\vspace{0.2cm}
\begin{center}
\begin{tabular}{l | l}
\hline
NVoxel & Number of voxels \\
Generator nodes & 50, 50, 100, 200, NVoxel\\
Discriminator nodes & NVoxel, NVoxel, NVoxel, NVoxel, 1\\
Activation function & ReLU \\
Optimizer & Adam~\cite{kingma2014adam}\\
Learning rate & $10^{-4}$ \\
$\beta1$ & 0.5 \\
$\beta2$ & 0.999 \\
Batch size & 128 \\
Training ratio (D/G) & 5 \\
Gradient penalty ($\lambda$) & 10 \\
\hline
\end{tabular}
\end{center}
\label{tab:GANArchitecture}
\end{table}
 
\subsection{Training}
\label{sec:gan:result}
Each GAN is trained first on a single energy point and then the other energy points are added progressively to the training mixture starting from the energy points closest in energy to the initial sample. The training procedure can be summarized as follows:
\begin{enumerate}
\item Train the first 50\,000 epochs using the 32~\GeV{} sample.
\item Every 20\,000 epochs add a new sample, alternating between higher and lower energies.
\item Once all energy points have been added, continue training with all samples for the remaining epochs.
\end{enumerate}
 
The energy in each voxel is normalized to the true momentum of the primary particle entering the calorimeter, which means that the GAN only needs to learn the relative shape of the showers. The true momenta, which are used as labels for the conditioning, are also normalized to the highest value (4.2~\TeV), which results in a range of values (0,1] which is optimal for the training of the GANs.
 
The training is performed for 1~million epochs with a TensorFlow checkpoint saved every 1000 epochs to monitor the improvements in the training. The training time for each GAN is approximately 8~hours on the NVIDIA V100~\cite{NVIDIA_V100} GPUs available on the CERN HTCondor system~\cite{HTCondor_CERN}. The limited number of GPUs available to train the 300 GANs sets the limit of 1~million epochs, while the frequency at which the checkpoints are stored is limited by both speed and disk space.
 
\subsubsection{Best epoch selection}
\label{sec:bestEpochSelection}
Due to the interplay between the generator and discriminator, the final epoch is not necessarily the best one. The figure of merit used to select the best epoch is a $\chi^2$ between the reference samples and the GANs. The variable chosen is the sum of the energy in all voxels that corresponds to the total energy deposited in the calorimeter by the particle.
 
For each energy point, the range used for the distribution is defined to be a $\pm$3~RMS interval around the peak for electromagnetic showers in the \GEANT reference samples. As the energy distributions of the pions have longer tails, the range for hadrons is defined to be between $-4$~RMS and 3.5~RMS. A total of 30 bins are used for all energy points. The $\chi^2$ is then evaluated between the binned distributions produced from all events in the reference samples and 10\,000 events generated from the GAN and weighted by the statistical uncertainty. The overflow and underflow bins are not used in the $\chi^2$ evaluation. The total $\chi^2$ for a checkpoint is the sum of the $\chi^2$ for each of the 15 energy points. The checkpoint with the lowest $\chi^2$ sum is chosen for each GAN. This selection criterion, as opposed to selecting the last trained epoch, avoids the problem of selecting an epoch with an unfavourable fluctuation in the training.
 
\begin{figure}[htb]
\begin{center}
\includegraphics[width=.7\linewidth]{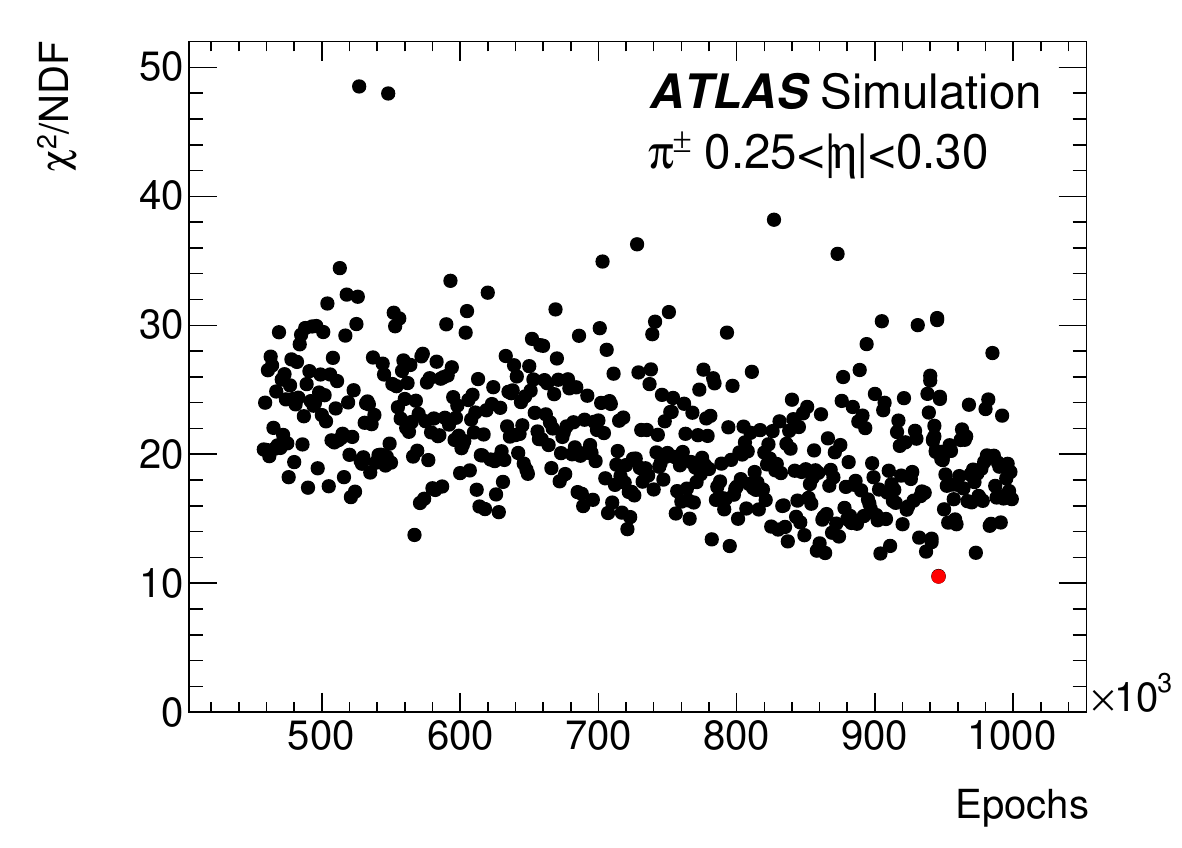}
\end{center}
\caption{The $\chi^2$ sum divided by the number of degree of freedom (NDF) calculated between the reference samples and the GAN as a function of the number of epochs. The lowest point (in red) represents the selected epoch.}
\label{fig:chi2}
\end{figure}

The evolution of this $\chi^2$ as a function of the epoch is shown in Figure~\ref{fig:chi2} for pions with $0.25 < |\eta| < 0.3$. The average $\chi^2$ decreases with increasing epoch and the fluctuations around the average are typical of GAN training. The point with the lowest $\chi^2$ sum, which in the example presented in Figure~\ref{fig:chi2} occurs at epoch number 946\,000, is the checkpoint used for the simulation of pions in that $\eta$ range. This procedure is repeated for all 300 GANs.
 
\subsubsection{Performance}
 
\begin{figure}[!htb]
\begin{center}
\includegraphics[width=\linewidth]{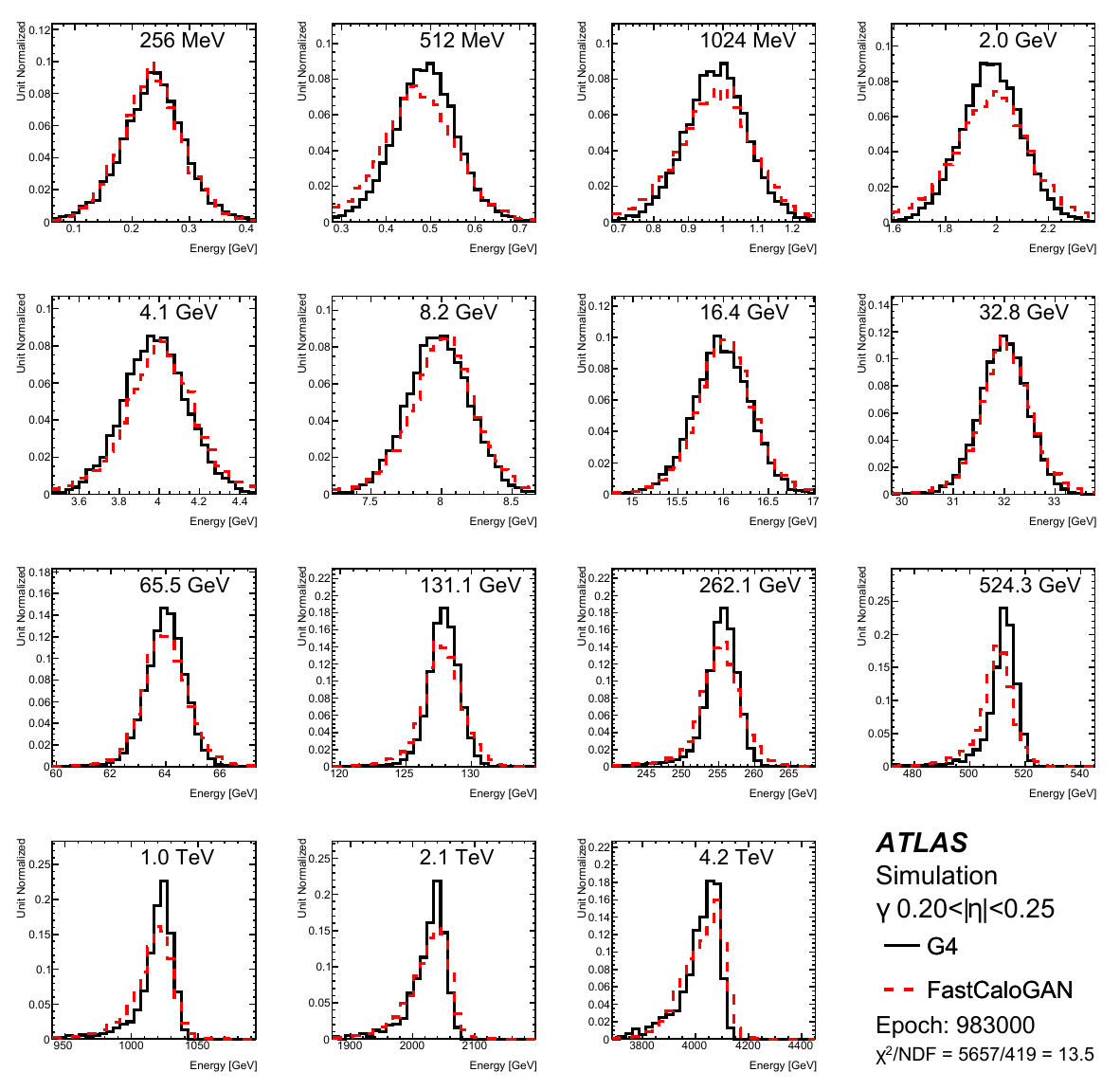}
\end{center}
\caption{Sum of the energy in all voxels for photons with $0.2 < |\eta| < 0.25$. The calorimeter response for \GEANT (solid black line) compared with \FGAN (dashed red line).}
\label{fig:gan:gamma_20}
\end{figure}

The performance of the best epoch for photons with $0.2 < |\eta| < 0.25$ is shown in Figure~\ref{fig:gan:gamma_20}. For each of the 15 energy points, the distribution of the total energy, defined as the sum of the energy in all voxels, is shown for the \GEANT input samples and the events generated with the GAN. In most cases, the means of the two distributions are comparable and so are their widths.
 
\begin{figure}[!htb]
\begin{center}
\includegraphics[width=\linewidth]{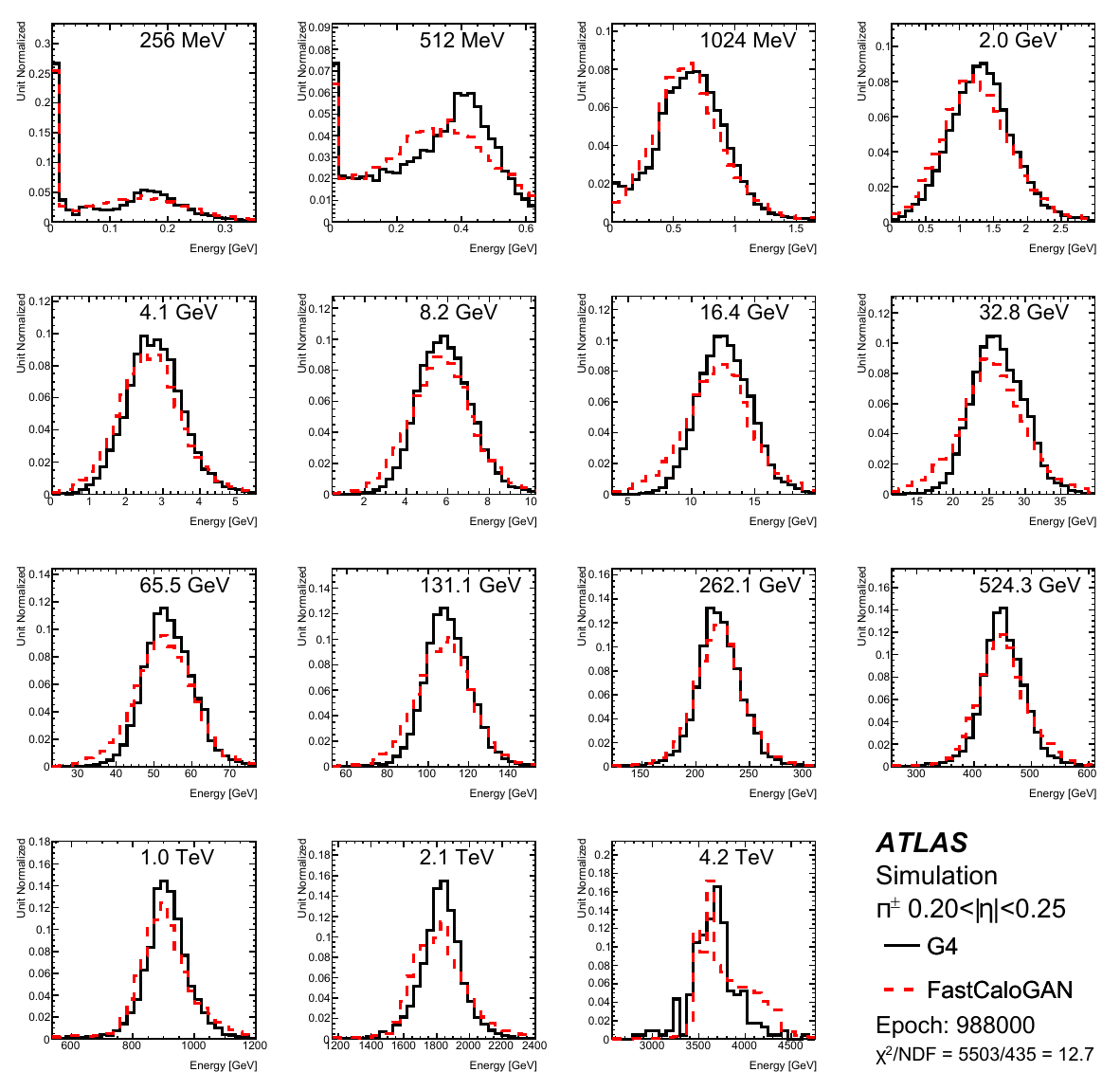}
\end{center}
\caption{Sum of the energy in all voxels for pions with $0.2 < |\eta| < 0.25$. The calorimeter response for \GEANT (solid black line) is compared with \FGAN (dashed red line). }
\label{fig:gan:pion_20}
\end{figure}

Similarly, the performance of the GAN for pions with $0.2 < |\eta| < 0.25$ is shown in Figure~\ref{fig:gan:pion_20}. The first two energy points show a  different shape than the other energy points and are not well described. The description of the highest energy point is poor due to the difficulties in reproducing the irregular shape and low number of events in the reference sample; given the extreme rarity of such high-energy hadrons in physics samples, the poor modelling is not of significant concern. Furthermore, in its final configuration described in Section~\ref{sec:af3_comb}, \AFIII does not use GANs in this energy range.
 
\begin{figure}[hbtp!]
\centering
\subfloat[]{\includegraphics[width=.47\textwidth]{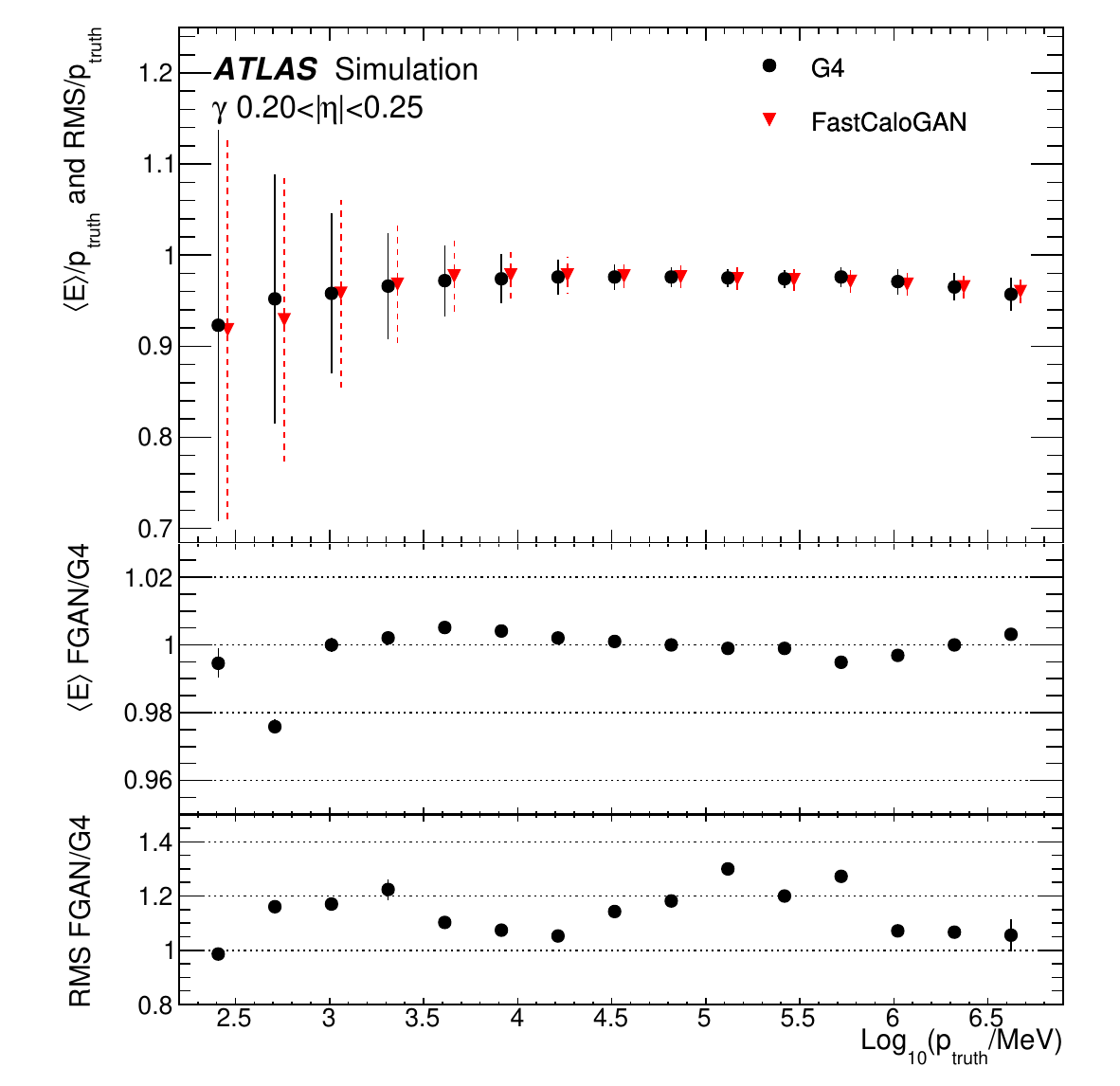}}
\subfloat[]{\includegraphics[width=.47\textwidth]{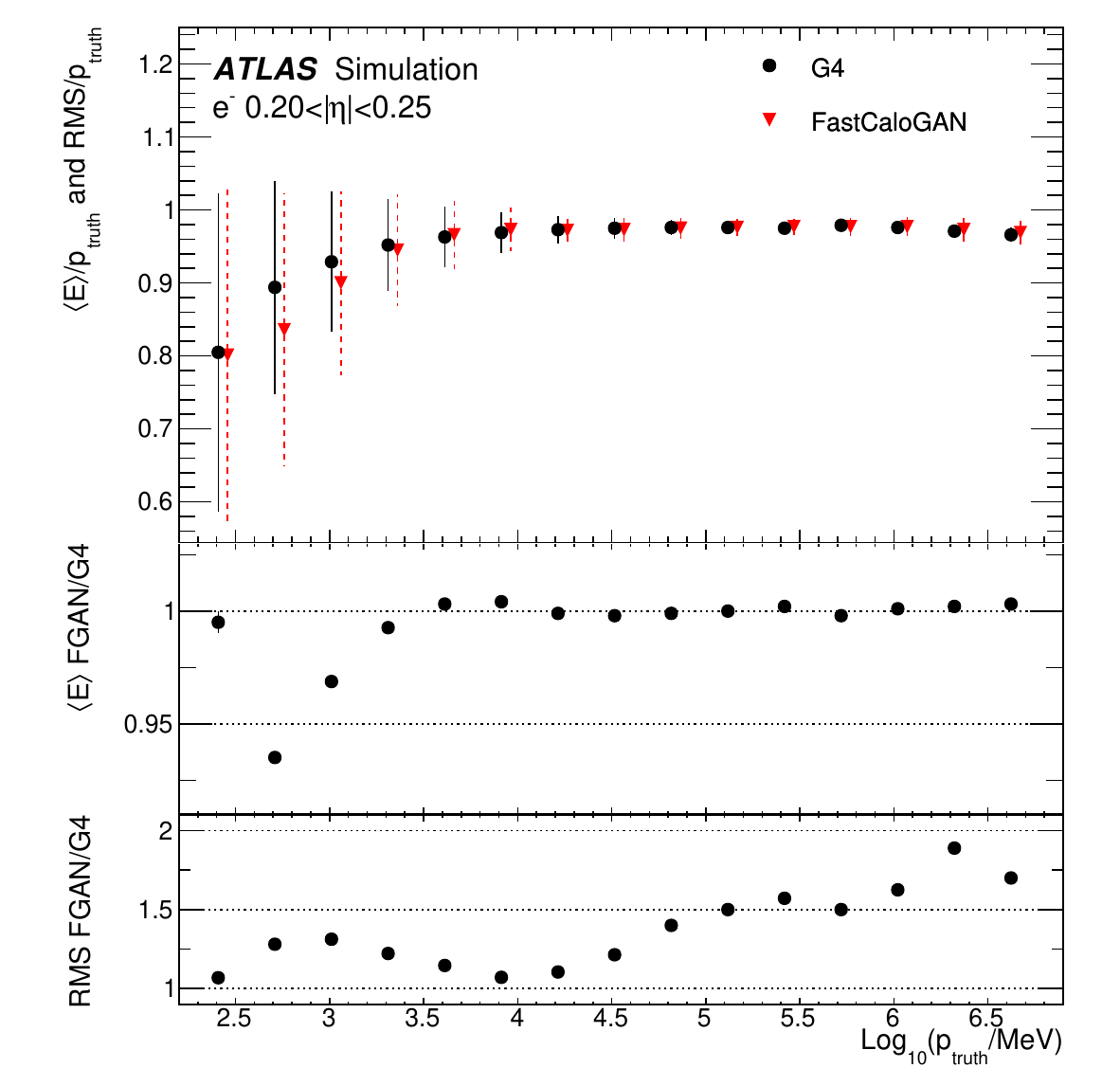}} \\
\subfloat[]{\includegraphics[width=.47\textwidth]{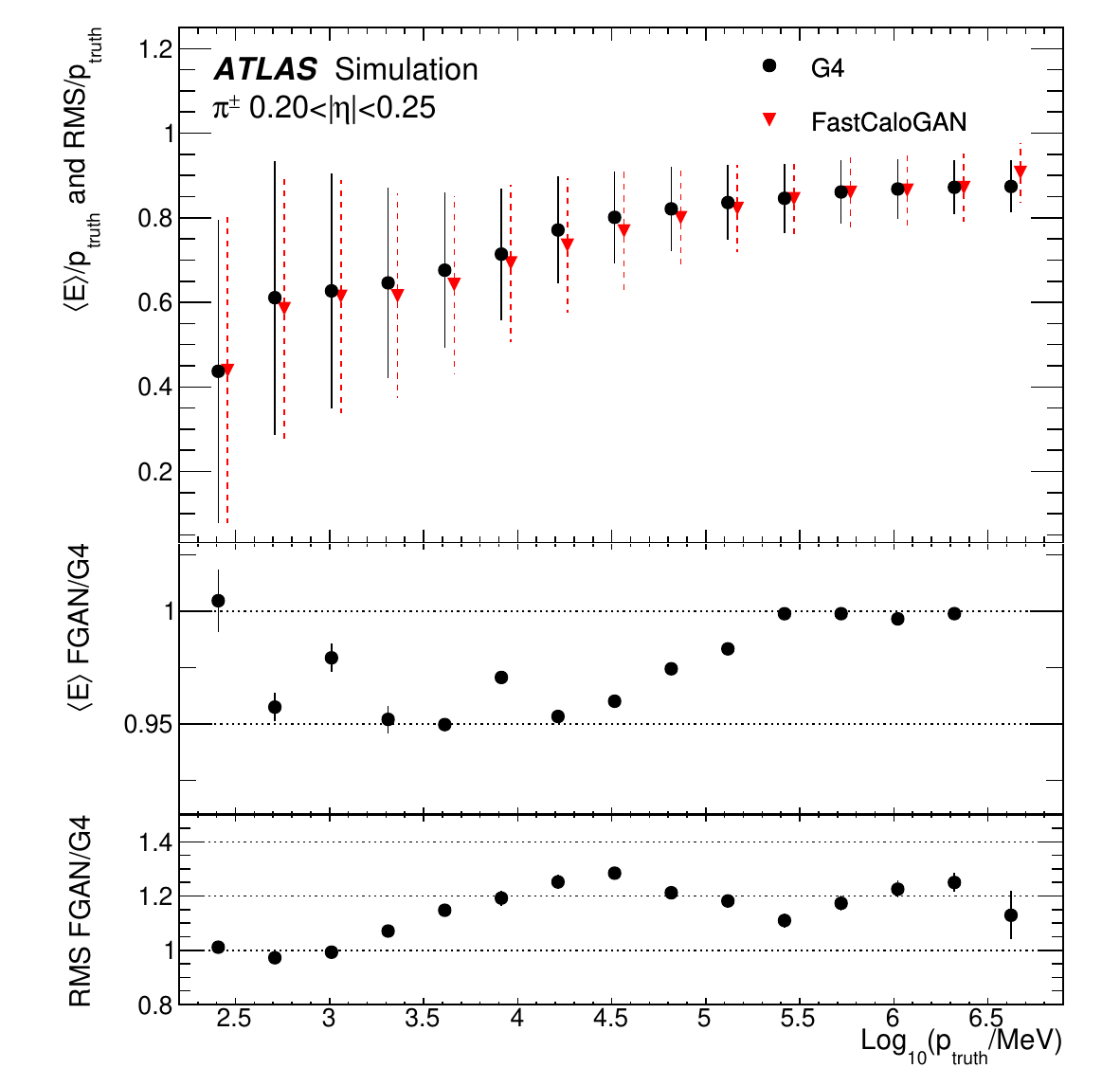}}
\caption{Sum and RMS of the energy in all voxels normalized to the true momentum for (a) photons, (b) electrons and (c) pions with $0.2 < |\eta| < 0.25$  as a function of the true momentum. The calorimeter response for \GEANT (solid black line) is compared with \FGAN (dashed red line), which is also abbreviated to FGAN. The uncertainty bars in the top panel indicate the RMS of the total energy distribution. The ratio of the means of the two energy distributions is shown in the middle panel, and the ratio of the RMS values is shown in the bottom panel. The error bars in the ratio indicate its statistical uncertainty. For most points, this uncertainty is smaller than the size of the markers.}
\label{fig:gan:summary:photons}
\end{figure}

The mean and RMS, indicated by the size of the uncertainty bars, of the total energy as a function of the true particle momentum ($p_\mathrm{truth}$) is shown in Figures~\ref{fig:gan:summary:photons}(a), (b) and (c) for photons, electrons and pions, respectively. For photons and electrons, the GANs reproduce the mean energies of the reference samples except at the low momentum points. The RMS from the GANs is larger than that of the reference sample for all energies. For pions, the GANs generate distributions with a lower mean and a larger RMS for a wider energy range.
 
\begin{figure}[hbtp!]
\centering
\subfloat[]{\includegraphics[width=.47\textwidth]{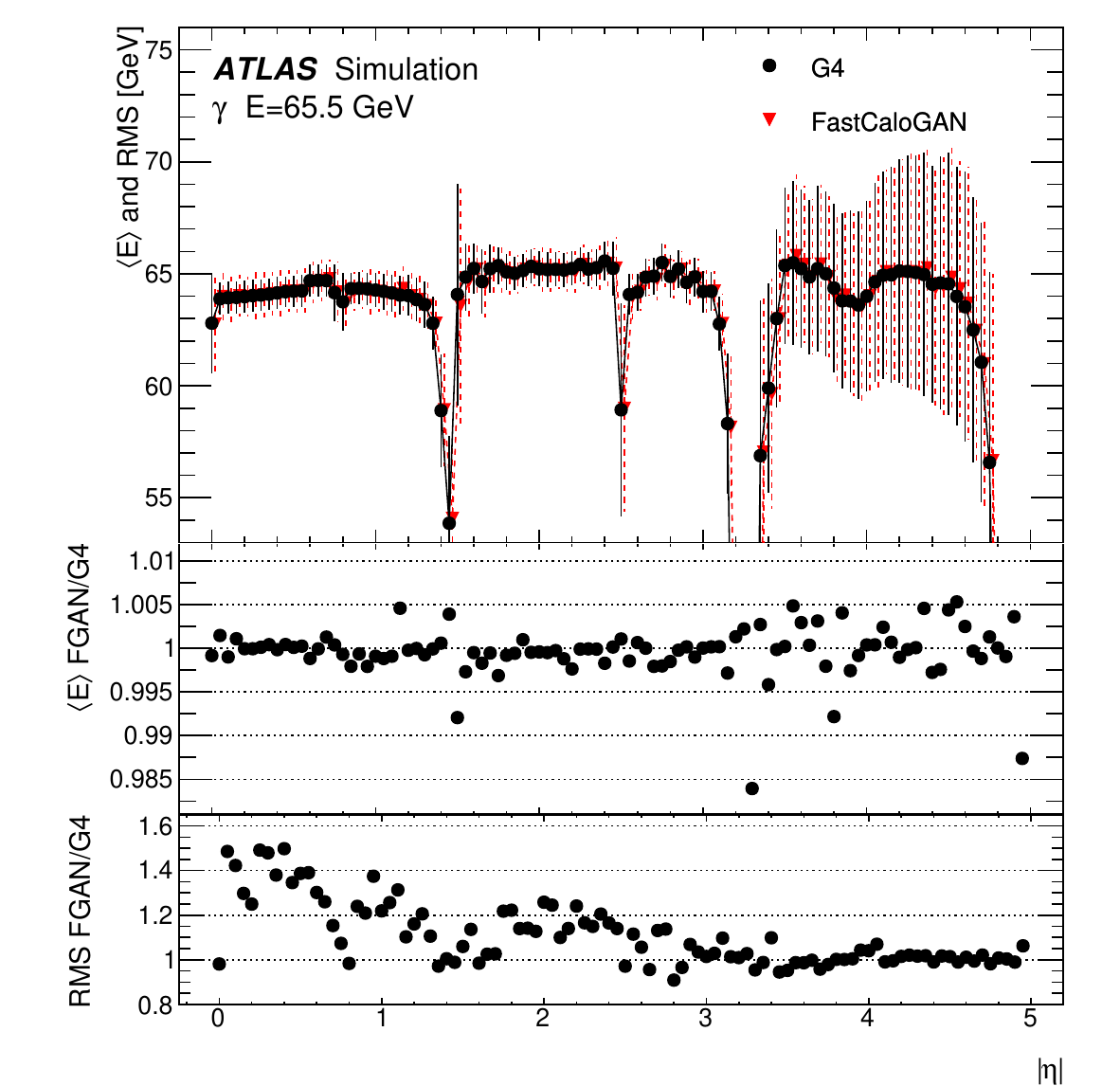}}
\subfloat[]{\includegraphics[width=.47\textwidth]{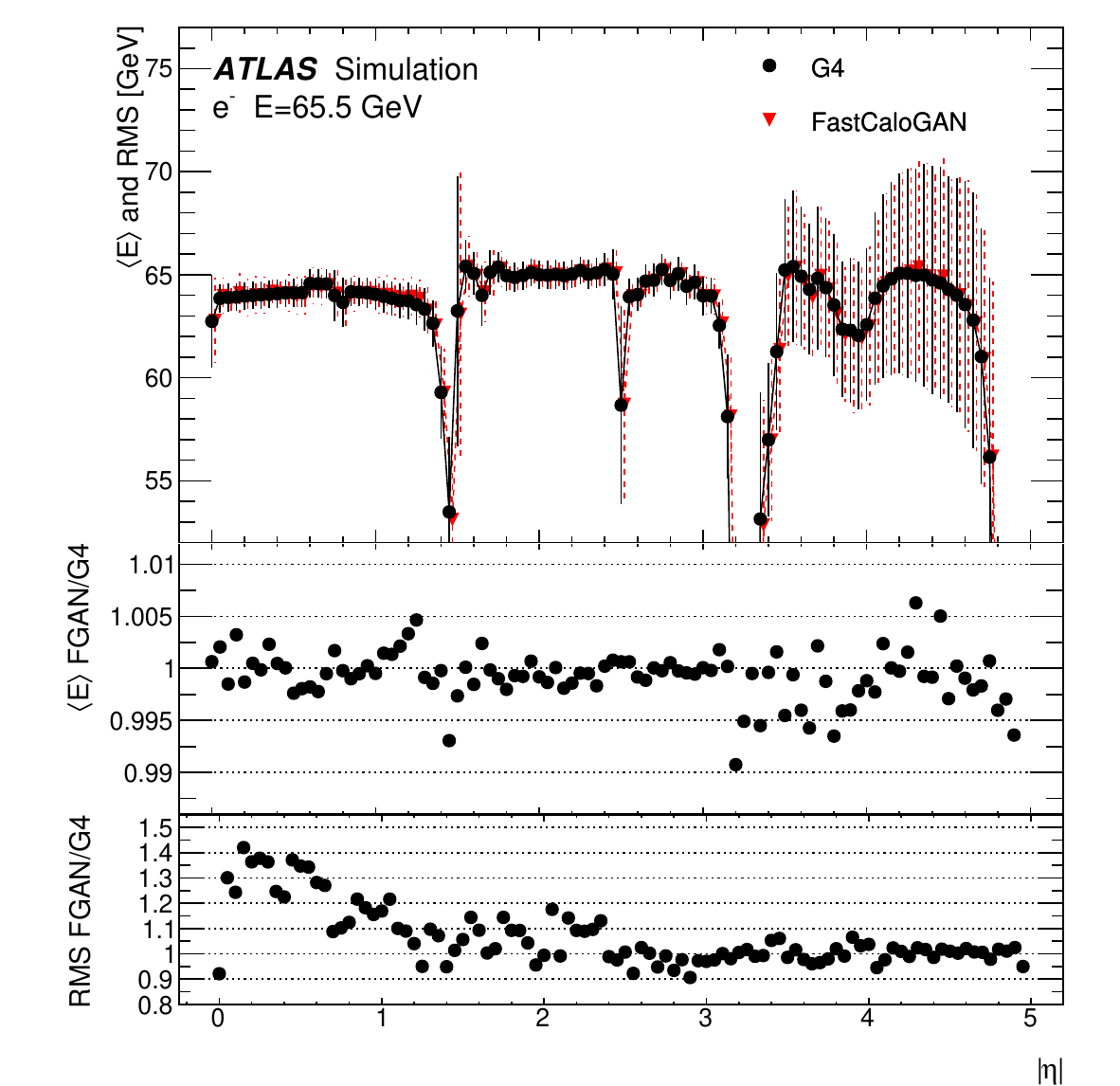}} \\
\subfloat[]{\includegraphics[width=.47\textwidth]{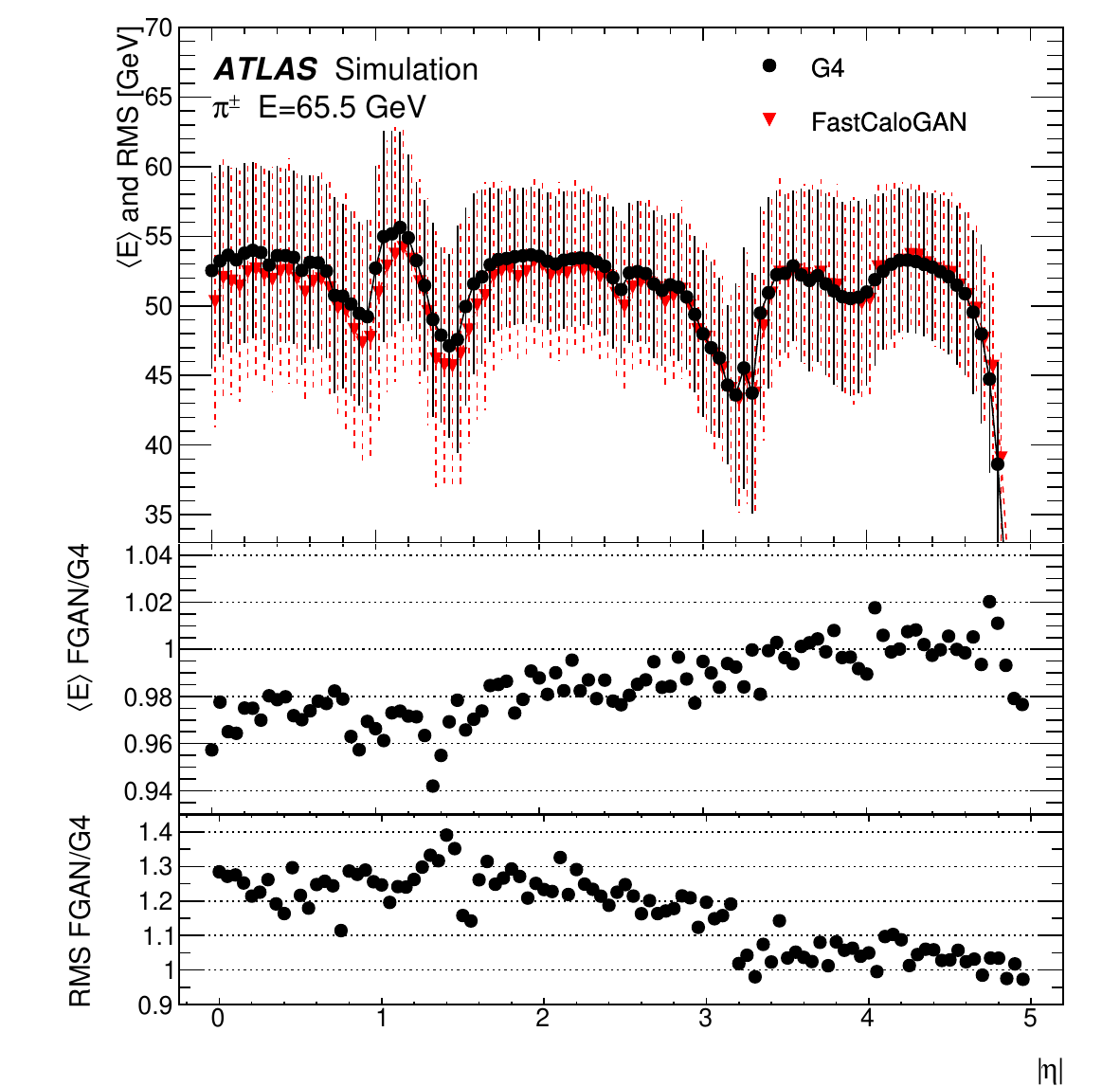}}
\caption{Sum and RMS of the energy in all voxels as a function of $|\eta|$ for (a) photons, (b) electrons and (c) pions of momentum 65~\GeV{}. The calorimeter response for \GEANT (solid black line) is compared with \FGAN (dashed red line), which is also abbreviated to FGAN, while their ratio is shown in the ratio plots. The uncertainty bars in the top panel indicate the RMS of the total energy distribution. The ratio of the means of the two energy distributions is shown in the middle panel, and the ratio of the RMS is shown in the bottom panel. The error bars in the ratio indicate its statistical uncertainty. For most points, this uncertainty is smaller than the size of the markers.}
\label{fig:gan:eta}
\end{figure}

The total energy, defined as the sum of the energy in all voxels, for particles with momentum 65~\GeV{} as a function of $\eta$ for photons, electrons and pions is shown in Figures~\ref{fig:gan:eta}(a)--(c). For photons and electrons, the GAN and reference sample means agree to better than 1\% in almost all the regions, while the distributions generated by the GANs are wider than the reference samples in the barrel region. Small discrepancies are observed in the transition regions between detectors, where the energy response is non-Gaussian. For pions, the means agree to within 4\%, with larger discrepancies observed in the barrel region, where the energy is slightly lower. The \FGAN RMS is larger in both the barrel and endcap regions.
 
\subsection{Simulation of hits}
\label{sec:gan:athena}
The GAN models trained to generate showers in \FGAN are implemented in the ATLAS Athena software framework using the Lightweight Trained Neural Network (LWTNN)~\cite{daniel_hay_guest_2019_3249317}.
 
The kinetic energy \Ekin of the particle is used as the conditional parameter of the GAN. The output of the GAN is the energy assigned to each voxel. Each one of these energies must be assigned to a variable number of cells because the voxels in \FGAN can be larger than the ATLAS cells. To assign the correct amount of energy to each cell, the voxel surface defined in Eq.~(\ref{eqn:shape:r_alpha}) is sampled uniformly,  generating a grid of hits. Layers that are not binned along the angular direction have their energy uniformly distributed across the whole annulus surface. The granularity used to sample the voxel is \SI{1}{\milli\meter}  in the high-granularity EMB1 and EME1 layers, while \SI{5}{\milli\meter} is used in the other layers. A maximum of 10 hits are created in either direction to limit the number of hits that are generated; this is required to have a small simulation time. The energy generated by the GAN in the voxel is divided uniformly between the hits. The hits are then assigned to the calorimeter cells using the simplified geometry. The longitudinal mid-position in each layer is used for the calculation of the hit position.
 
\FloatBarrier
\section{Simulation of muon punch-through}
\label{sec:muonPunchThrough}
\begin{figure}[hbtp!]
\begin{center}
\includegraphics[width=0.7\textwidth]{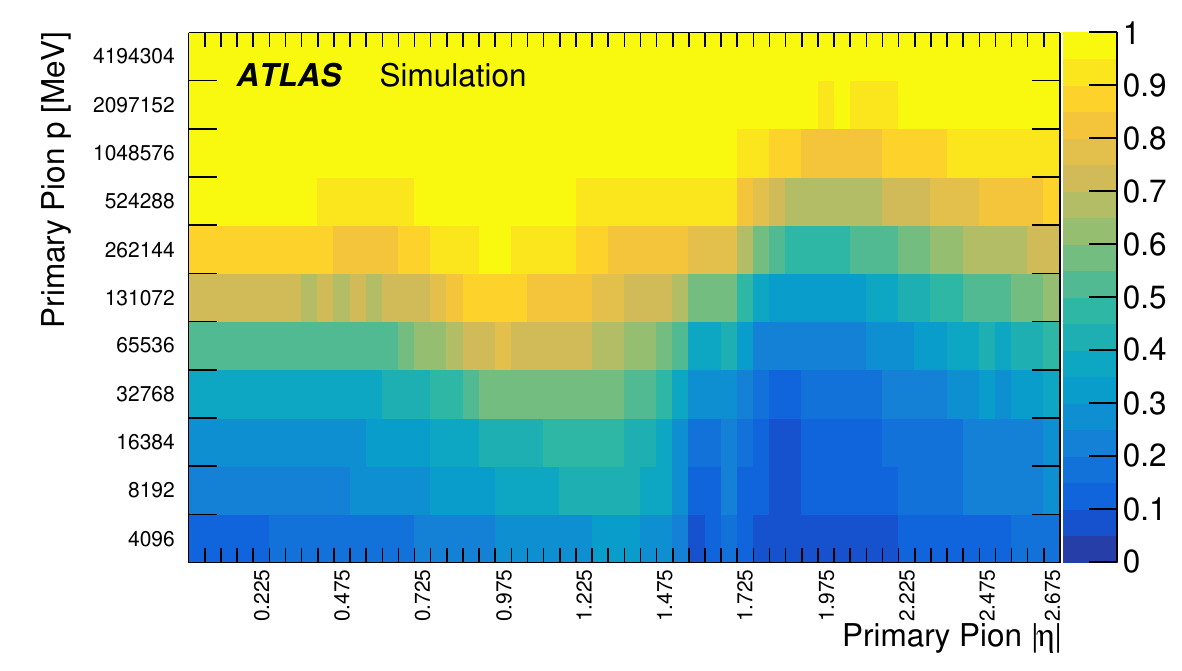}
\end{center}
\caption{The probability of a single-pion event to produce at least one punch-through particle with an energy of at least 50~\MeV\ as a function of the $\eta$ and $p$ of the incoming pion determined from \GEANT.}
\label{fig:mpt_pt_v_eta}
\end{figure}

Secondary particles created in hadronic showers inside the calorimeter can escape through the back of the calorimeter and generate hits in the muon spectrometer. This effect is referred to as \textit{muon punch through}. These particles are reconstructed in the muon spectrometer and need to be well modelled to accurately describe the backgrounds of reconstructed muons. A dedicated treatment of these particles is required because the information about the path of the particles is lost due to the parameterization of the calorimeter response in \AFIII. Figure~\ref{fig:mpt_pt_v_eta} shows the probability of a single pion entering the calorimeter to create at least one secondary particle which escapes the calorimeter volume with an energy of at least 50~\MeV{}  determined using the \GEANT simulation. The probability increases with increasing momentum $p$ and varies as a function of $\eta$. Particles with energies below 50~\MeV{}  are not simulated in the muon spectrometer because they would have negligible impact.
 
The \AFIII punch-through parameterization is derived separately for the five types of secondary particles that can emerge from the back of the calorimeter: photons, electrons, pions, muons, and protons. These account for 92\% of the total punch through. The parameterizations of their multiplicity and kinematics are determined from single-pion samples simulated using \GEANT. As the properties of the secondary particles depend significantly on the $\eta$ direction and energy of the incoming pion, the reference samples within the acceptance of the muon spectrometer $|\eta| \leq 2.7$ and with momenta between 65~\GeV\ and 4.2~\TeV\ are used to determine the parameterization. The small number of secondary particles in lower-energy samples did not allow a parametrization of primary particles with an energy lower than 65~\GeV.
 
\begin{figure}[htbp]
\subfloat[]{\includegraphics[width=.5\textwidth]{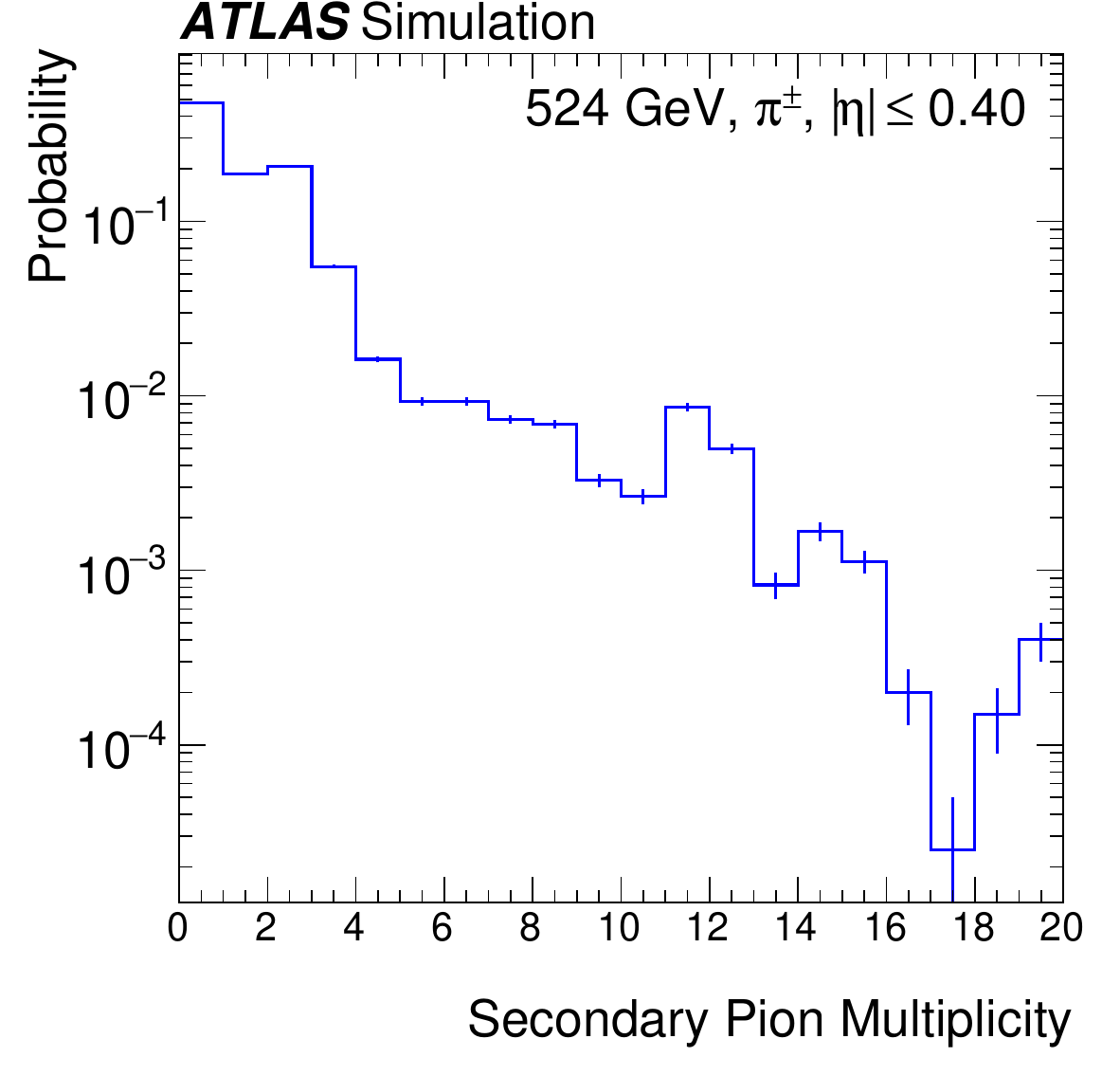}}
\subfloat[]{\includegraphics[width=.5\textwidth]{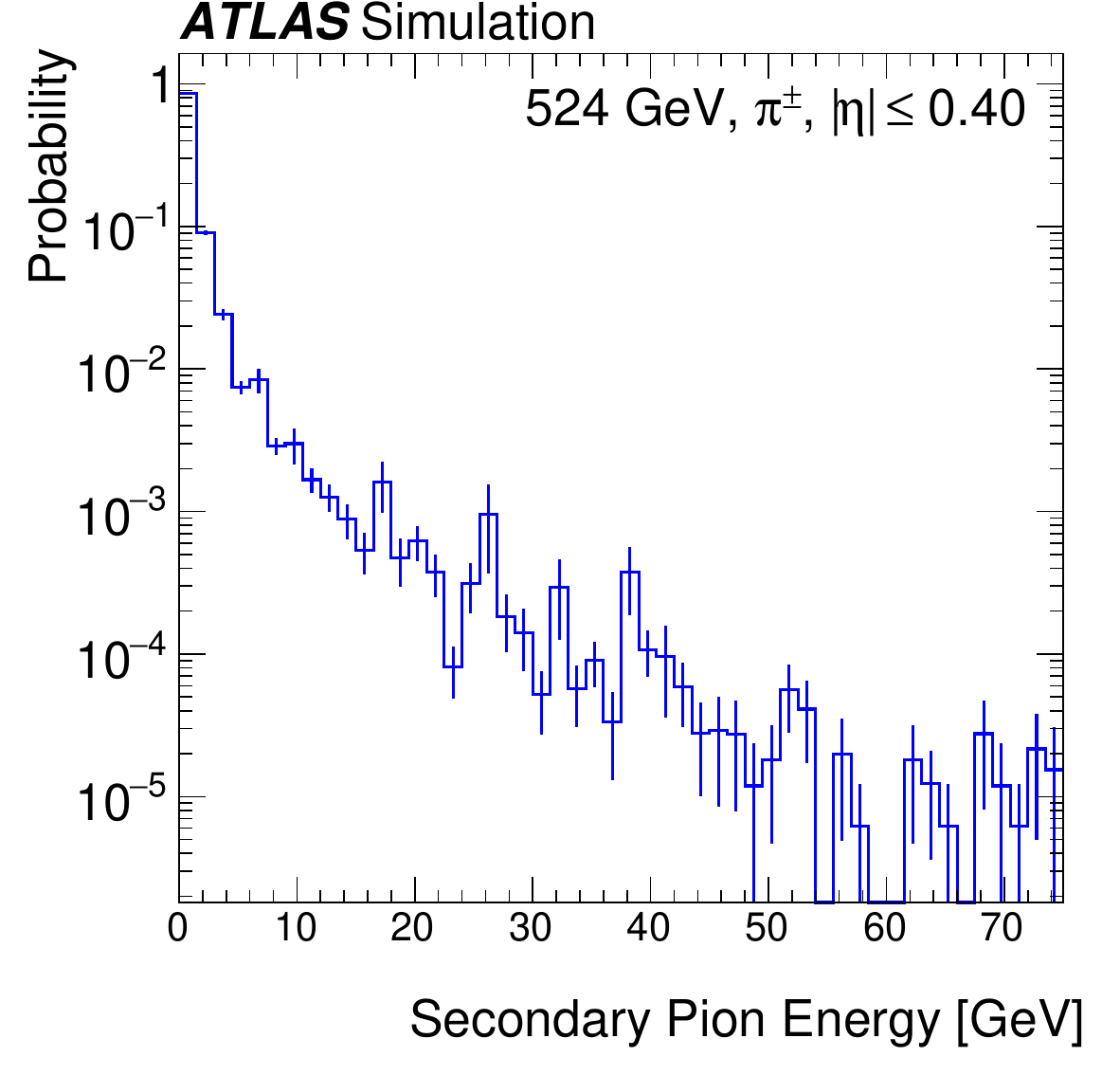}}
\caption{The punch-through probability as a function of the punch-through pion (a) multiplicity and (b) energy. The error bars indicate the statistical uncertainty and the overflow is not included in the final bins.}
\label{fig:param_punchthrough}
\end{figure}

\begin{figure}[htbp]
\subfloat[]{\includegraphics[width=.5\textwidth]{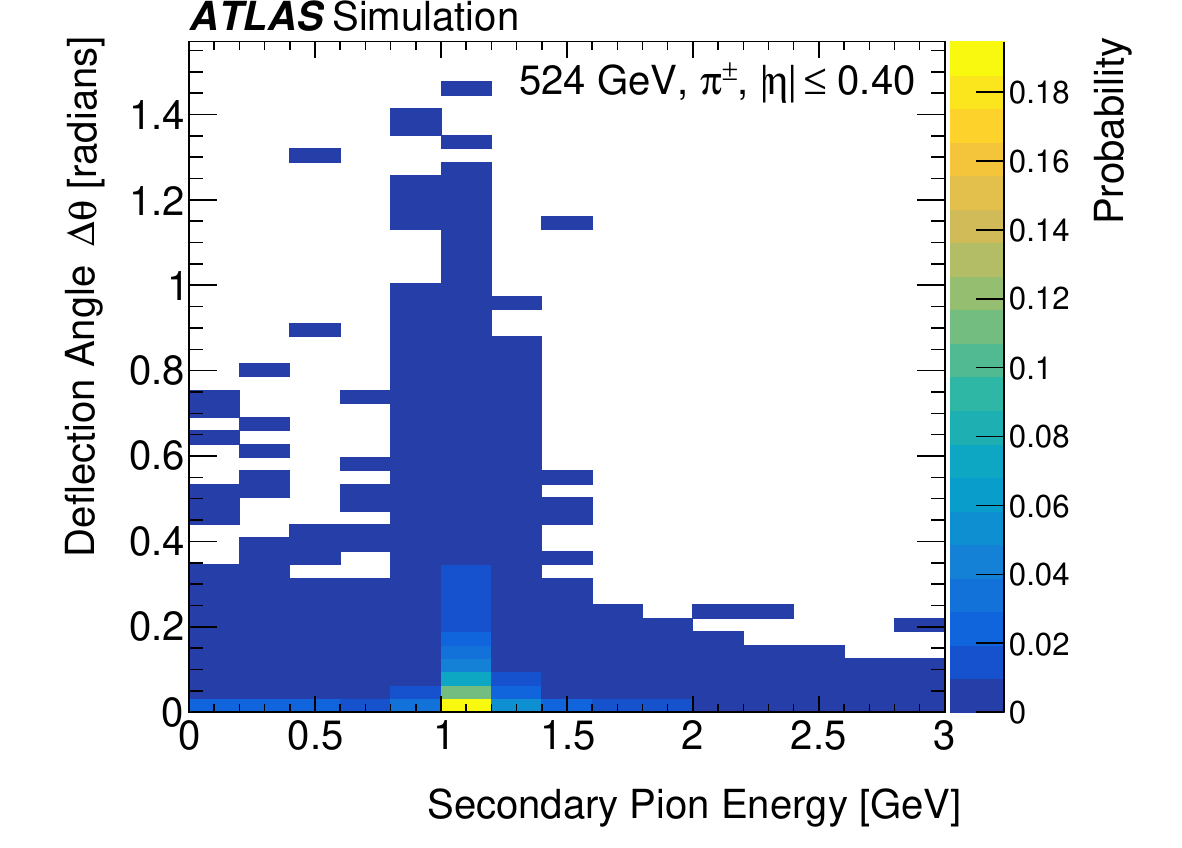}}
\subfloat[]{\includegraphics[width=.5\textwidth]{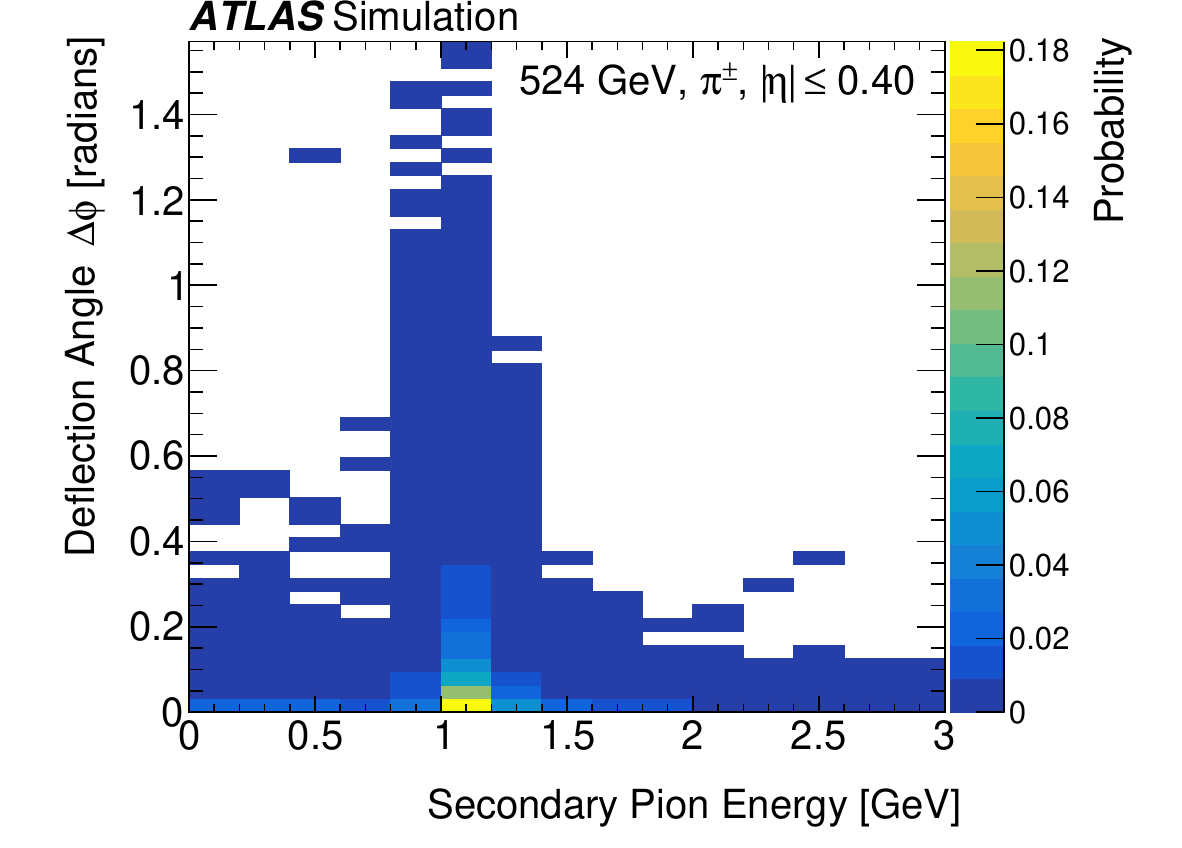}} \\
\subfloat[]{\includegraphics[width=.5\textwidth]{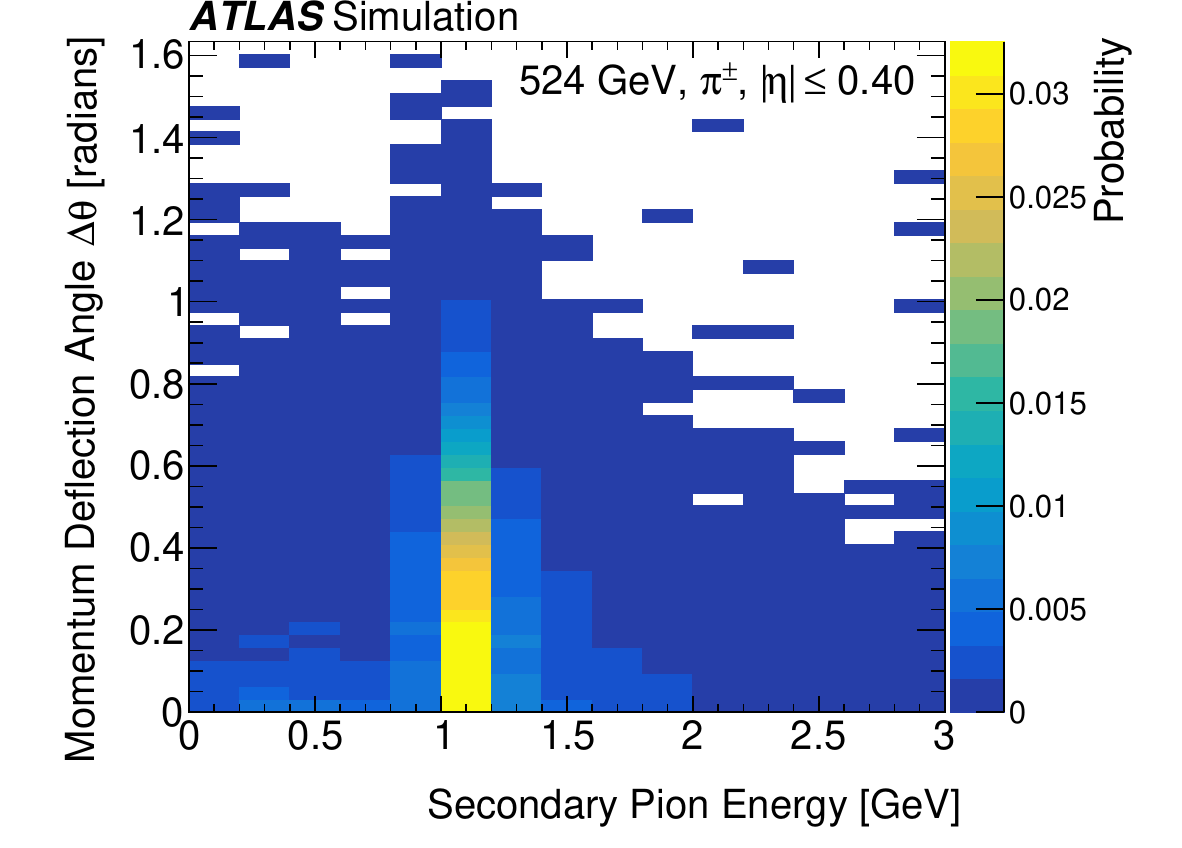}}
\subfloat[]{\includegraphics[width=.5\textwidth]{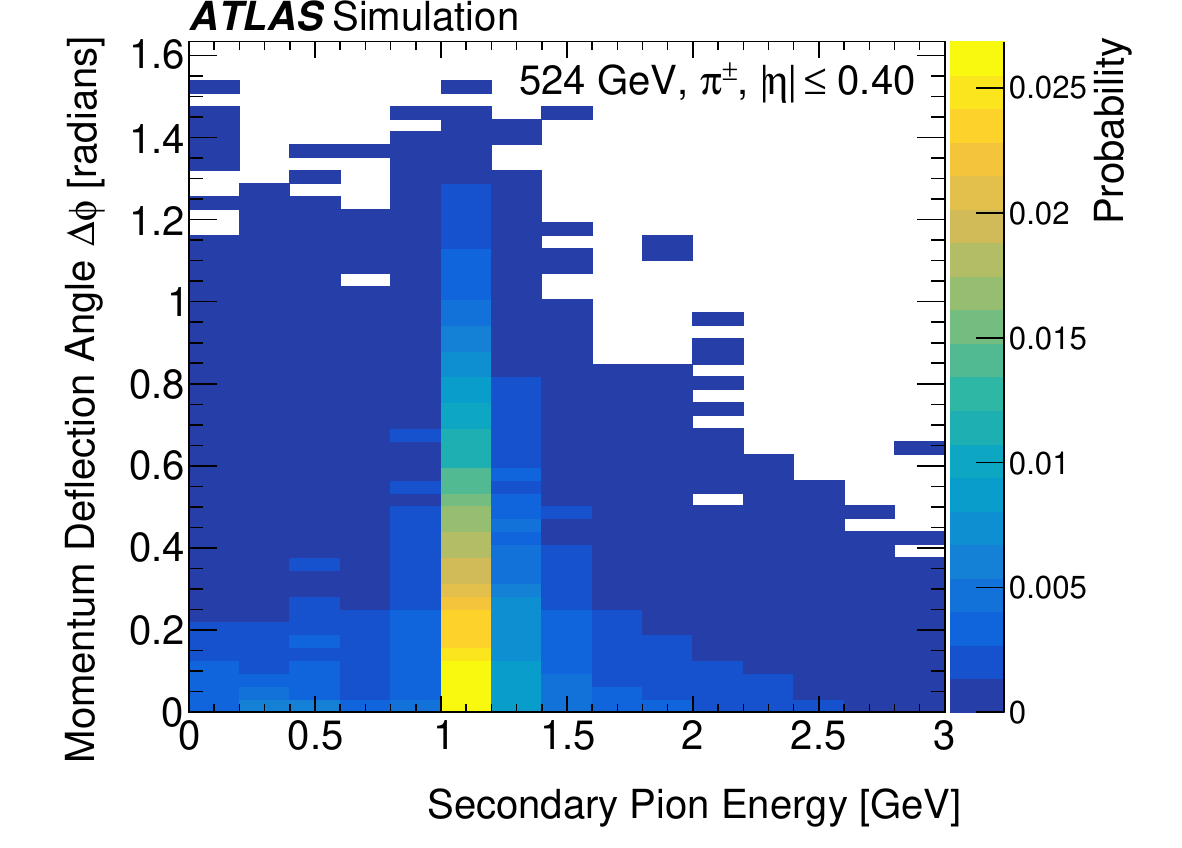}}
\caption{The punch-through probability as a function of (a) deflection angle in $\theta$ and energy, (b) deflection angle in $\phi$ and energy, (c) relative momentum deflection in $\theta$ and energy, and (d) relative momentum deflection in $\phi$ and energy. Secondary pions with an energy of 524~\GeV\ in the region $|\eta| \leq 0.4$ from the \GEANT reference samples were used.}
\label{fig:param_punchthrough2}
\end{figure}

The properties of the secondaries described by the parameterization include their energy, and their position and momentum relative to that of the incoming pion. The position and momentum of the secondaries are determined via deflection angles, $\Delta \theta$ and $\Delta \phi$, relative to the direction of propagation of the incoming pion. As an example, Figures~\ref{fig:param_punchthrough} and~\ref{fig:param_punchthrough2} show the histograms extracted from the \GEANT simulation and used to parameterize the secondaries produced by primary pions with an energy of 524~\GeV\ and $|\eta| \leq 0.4$. The peak at 1~\GeV{} is the most probable value of the energy of the secondary pions emerging form the calorimeter.
 
During the simulation of \AFIII, the muon punch-through parameterization is invoked whenever particles that have some probability of punching through enter the calorimeter. For each incoming particle, the number of secondaries and their energy, position and momentum are selected randomly from the punch-through parameterization histograms (see Figures~\ref{fig:param_punchthrough} and~\ref{fig:param_punchthrough2}), using them as probability density functions. The parameterization is interpolated linearly for $\eta$ and logarithmically for \pT to values between the discrete points used to determine the parameterization. Two sets of correlations are accounted for in the modelling of the secondaries: the correlations between the relative position and energy and correlations between the relative momentum and energy. After the multiplicity and properties of  the punch-through secondaries have been determined, their propagation through the muon spectrometer is simulated using \GEANT.
 
\section{The combination of \FCaloSII and \FGAN: \AFIII}
\label{sec:af3}
 
\subsection{Configuration of \AFIII}
\label{sec:af3_comb}
\begin{figure}[h]
\begin{center}
\includegraphics[width=0.8\linewidth]{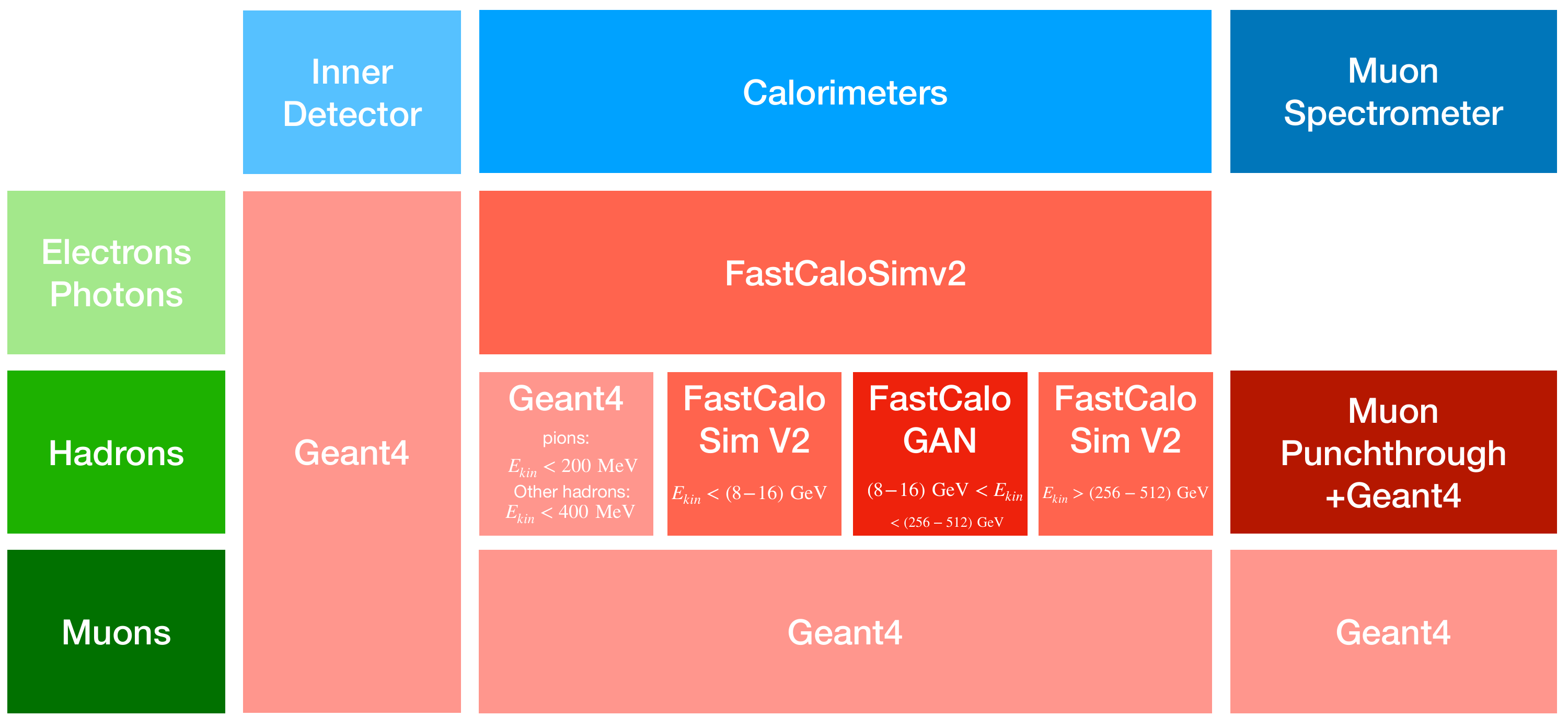}
\end{center}
\caption{The configuration of the different tools used for \AFIII, which depends on the particle type, the detector and the particle energy.}
\label{fig:af3:config}
\end{figure}

The new fast simulation tool, \AFIII, is defined by combining the fast simulation tools described above in a way that balances modelling performance needs with CPU requirements. \AFIII uses the Integrated Simulation Framework (ISF), which allows different simulation tools to be combined in a flexible way~\cite{Lukas:1458503}. \AFIII uses the following configuration as illustrated in Figure~\ref{fig:af3:config}:
 
\begin{itemize}
\item \GEANT is used to simulate all particles in the inner detector and muons in all detectors. Hadrons with kinetic energies below 400~\MeV\ (200~\MeV\ for pions) in the calorimeter are also simulated in \GEANT.
\item \FCaloSII is used to simulate electrons and photons of all energies and hadrons with kinetic energies $\Ekin < \text{8--16}$~\GeV\ or $\Ekin > \text{256--512}$~\GeV\ in the calorimeter. A transition range of energies is given because the response is interpolated linearly between the two models as discussed later.
\item \FGAN is used to simulate hadrons with $\text{8--16~\GeV} \leq \Ekin \leq \text{256--512}$~\GeV\ in the calorimeter.
\item The muon punch-through tool parameterizes the properties of particles exiting the calorimeter, which are then simulated with \GEANT in the muon spectrometer.
\end{itemize}
 
\subsection{Configuration of the Fast Calorimeter Simulation}
The configuration of \AFIII is determined by comparing the performance of \FCaloSII and \FGAN.
 
\subsubsection{Electrons and Photons}
\begin{figure}[h]
\begin{center}
\includegraphics[width=0.6\linewidth]{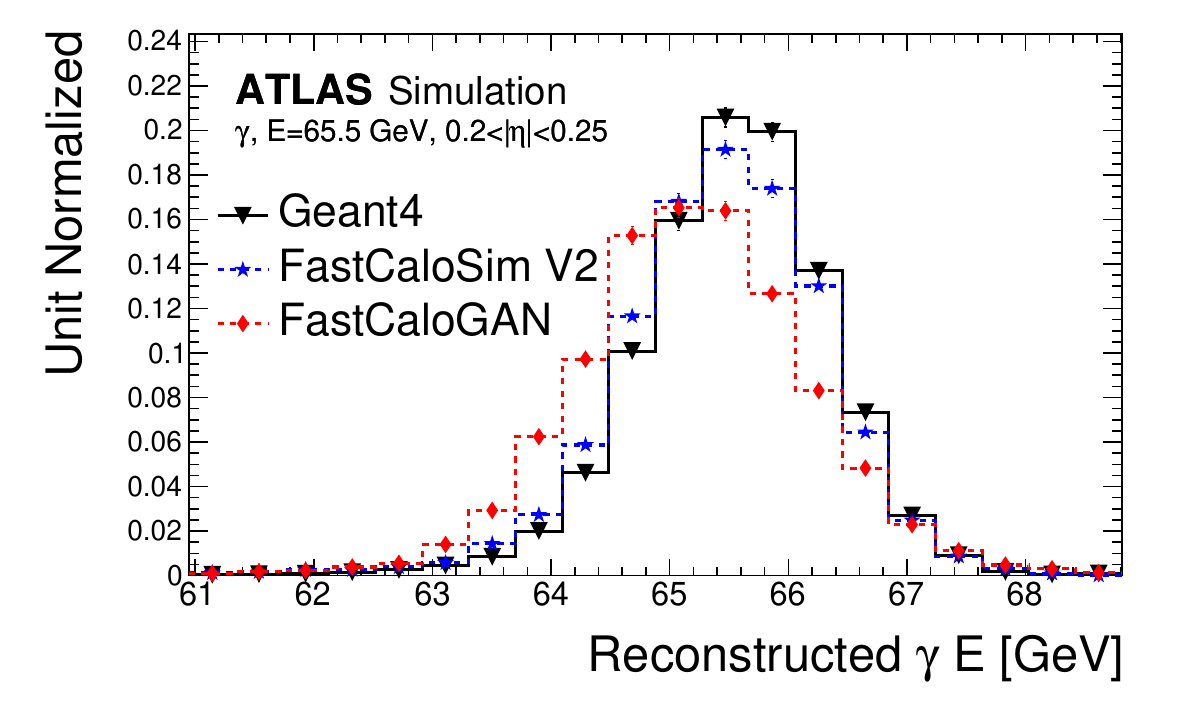}
\end{center}
\caption{Reconstructed photon energy for photons generated at the calorimeter surface with an energy of 65~\GeV{} and $0.2 < |\eta| < 0.25$ by \GEANT (solid black line), \FCaloSII{} (dashed blue line), and \FGAN (dashed red line). The statistical uncertainties are shown but are similar in size to the points or smaller.}
\label{fig:af3:photon_reco}
\end{figure}

The simulation of electrons and photons relies on the accurate simulation of electromagnetic showers in the electromagnetic calorimeter. The total reconstructed energy for 65~\GeV\ photons is shown in Figure~\ref{fig:af3:photon_reco} for \GEANT, \FCaloSII and \FGAN. \FGAN does not model the photon energy correctly and a similar poor performance is observed for electrons; therefore, \FCaloSII is selected to simulate all electromagnetic showers. The poor modelling of electromagnetic showers in \FGAN can be explained as follows: the GANs are trained without the energy resolution correction for the accordion structure of the calorimeter (see Section~\ref{sec:energy_correction}), and the energy scale of the detailed \GEANT hits, used in the training of \FGAN, is slightly lower than that of the full \GEANT hits (see Section~\ref{sec:refsamples}).
Both these effects can be corrected for in future versions of \FGAN but could not be included here due to time constraints.
 
\subsubsection{Low-energy hadrons}
\label{af3:energy:low}
\begin{figure}[h]
\subfloat[]{\includegraphics[width=0.48\textwidth]{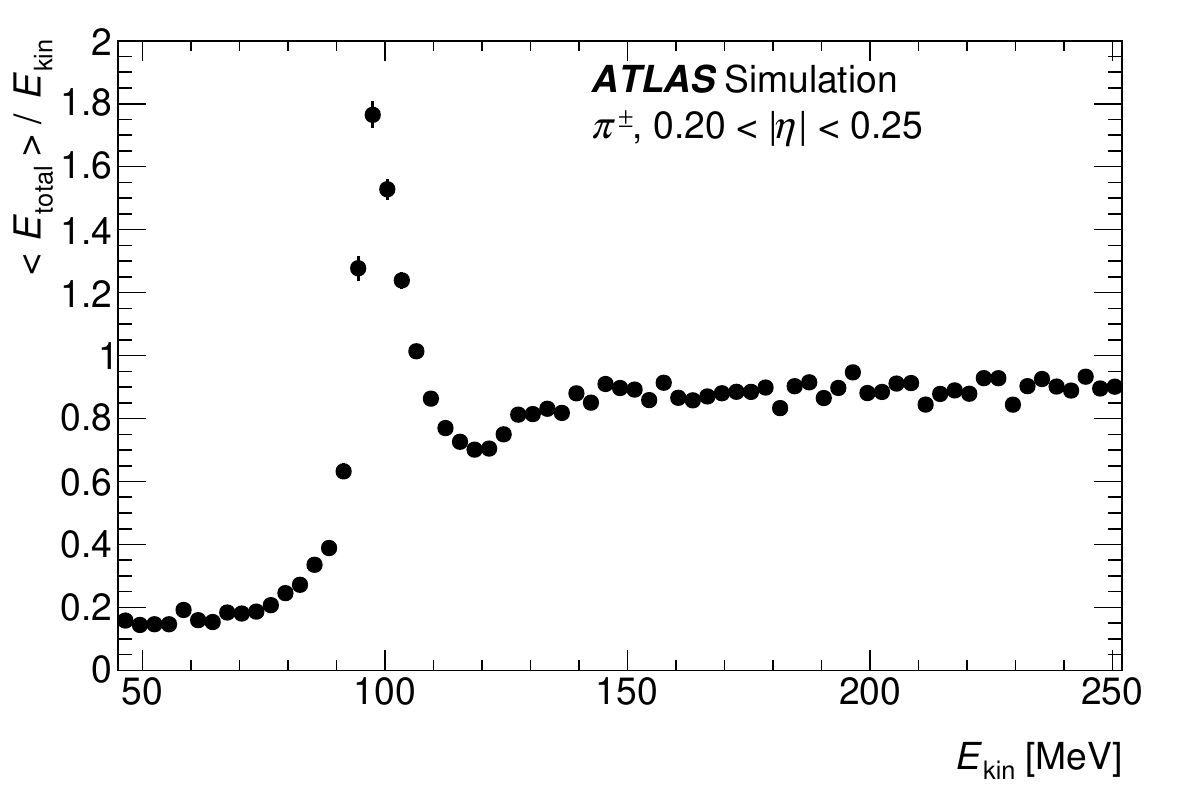}\label{fig:low:pion_slice}}
\subfloat[]{\includegraphics[width=0.48\textwidth]{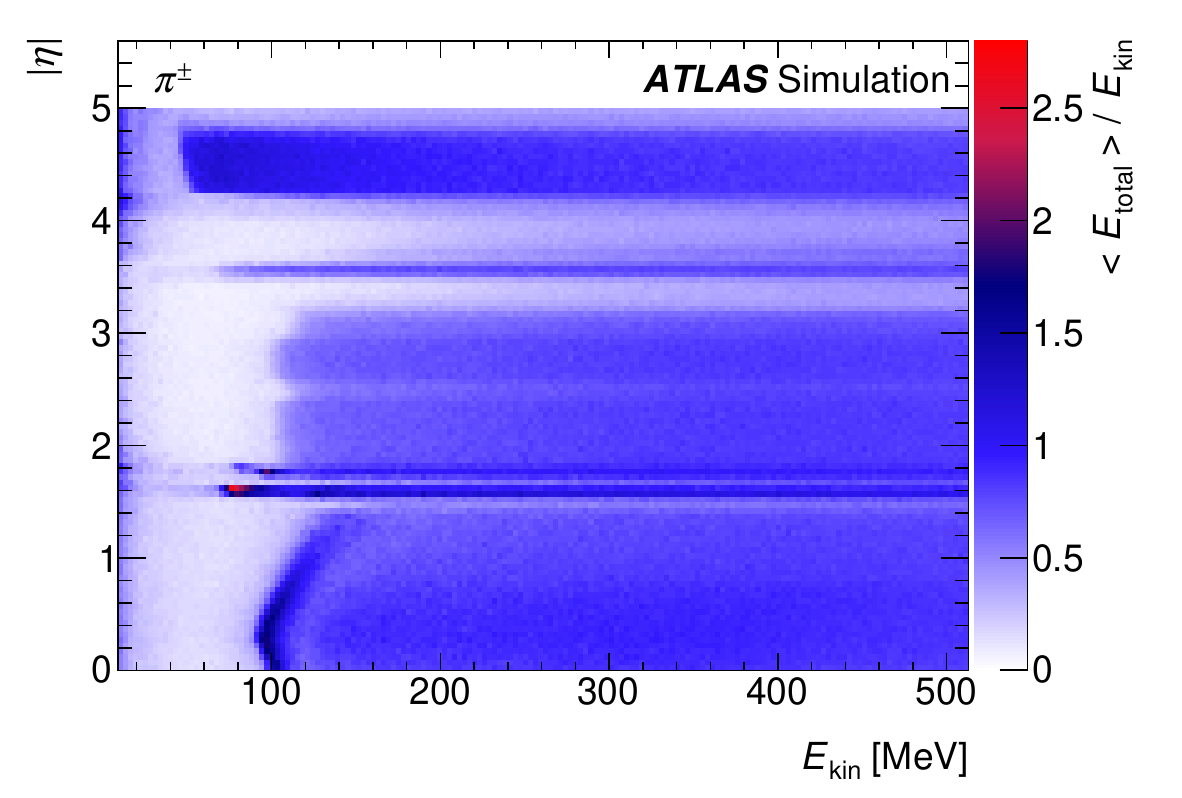} \label{fig:low:pion_2D}}
\caption{
Ratio of the average energy response to the generated energy for $\pi^{\pm}$ for (a) $0.20 < |\eta| < 0.25$ and (b) as a function of $|\eta|$ and \Ekin. The error bars indicate the statistical uncertainty of the mean. For most points, this uncertainty is smaller than the size of the markers.
}
\label{fig:eresp_lowE_pionMap}
\end{figure}

At low energies, the distribution of the average hadron energy response becomes complex and has a significant dependence on both \Ekin and $|\eta|$  as shown in Figure~\ref{fig:low:pion_2D}. This is because the measured energy depends strongly on the extent to which these shorter showers develop within the active liquid argon of the electromagnetic calorimeter or within the inactive lead absorbers. As an example, Figure~\ref{fig:low:pion_slice} shows the ratio of the average energy response to \Ekin as a function of \Ekin for charged pions in the range $0.20 < |\eta| < 0.25$. For pions with a kinetic energy of 100~\MeV\, the largest amount of deposited energy is typically within the liquid argon of the Presampler, which leads to a spike in the energy response. On the other hand, pions with a kinetic energy of 10~\MeV\ deposit far less energy in the active liquid-argon regions and more in the inactive regions. In addition, the energy calibration of the Presampler is derived using high-energy particles, which deposit much less energy in the Presampler, which means that the measured fraction of shower energy in the Presampler increases further for $\Ekin \approx 100$~\MeV.
 
The dependence of the energy response to low-energy charged pions on $\eta$ is due to the different amount of material that the charged pion passes through, which shifts the values of the kinetic energy at which the spike in the response occurs. Deriving a parameterization for such low-energy hadrons would require a significantly more complex method for deriving parameterizations in order to achieve high accuracy. Therefore, in \AFIII pions below 200~\MeV\ and all other hadrons below 400~\MeV\ (as shown in Table~\ref{tab:eresp_lowE_passToG4}) are instead simulated by \GEANT. Above these energy thresholds their total energy response is modelled using \AFIII. This choice does not significantly affect the speed of \AFIII because the simulation of low-energy hadrons requires only a comparatively small amount of CPU time.
 
\begin{table}[h]
\begin{center}
\caption{Hadron energies below which \AFIII relies on \GEANT for their simulation}
\vspace{0.2cm}
\begin{tabular}{cc}
\hline
Particle     & \Ekin[\MeV]  \\
\hline
$\pi^{\pm}$  & 200   \\
$K^\pm$, $K_\text{L}$, $p/\bar{p}$, $n/\bar{n}$       & 400    \\
\hline
\end{tabular}
\label{tab:eresp_lowE_passToG4}
\end{center}
\end{table}
 
\subsubsection{Medium-energy hadrons}
\begin{figure}[hbtp!]
\begin{center}
\includegraphics[width=0.6\textwidth]{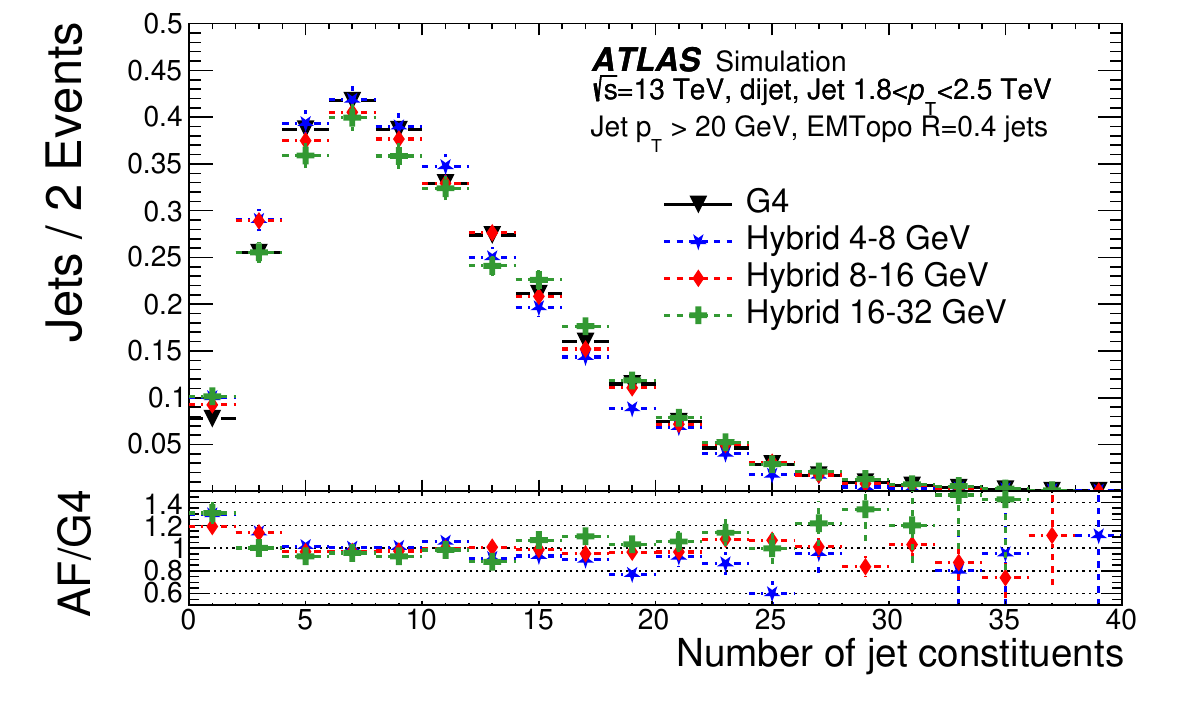}
\end{center}
\caption{Distribution of the number of constituents in the jets in a $1.8 < \pT < 2.5$ TeV dijet sample in \GEANT (black triangles) and the combination of \FCaloSII and \FGAN with transitions in the range 4--8~\GeV\ (blue stars), 8--16~\GeV\ (red diamonds) and 16--32~\GeV\ (green crosses). Here `hybrid' refers to the combination of \FCaloSII and \FGAN. The statistical uncertainties are shown but may be smaller than the markers.}
\label{fig:JetOptimisation:nclusters_mix}
\end{figure}

For hadronic showers, the number of clusters in a jet plays an important role in modeling the jet substructure and is therefore used as a metric to compare the performance of \FCaloSII and \FGAN. Differences in the modelling of the number of clusters between \FCaloSII and \FGAN are expected because \FGAN can model the correlations within a single event, while \FCaloSII cannot. Figure~\ref{fig:JetOptimisation:nclusters_mix} compares the modelling of the number of clusters in a jet for three different combinations of \FCaloSII and \FGAN. The hybrid models differ in the energy range over which the transition between \FCaloSII and \FGAN occurs; for example in the Hybrid 4--8~\GeV\ model, \FCaloSII is used up to 4~\GeV\ and \FGAN is used above 8~\GeV. Between 4 and 8~\GeV, the response is interpolated linearly between the two models as described in Section~\ref{sec:energy_interpol}. The Hybrid 4--8~\GeV\ model underestimates the number of constituents, while the Hybrid 16--32~\GeV\ model overestimates the number of constituents. Therefore, the Hybrid 8-16\,GeV model is chosen as the configuration for \AFIII. Other key jet variables, including the number of jets,  the \pT and $\eta$ distributions and variables used for substructure, are also checked for these different configurations, which provides additional support for choosing the Hybrid 8--16~\GeV\ model. Section~\ref{sec:recoJets} discusses the performance of \AFIII in modelling jet variables.
 
\subsubsection{High-energy hadrons}
\begin{figure}[hbtp!]
\centering
\subfloat[]{\includegraphics[width=0.47\textwidth]{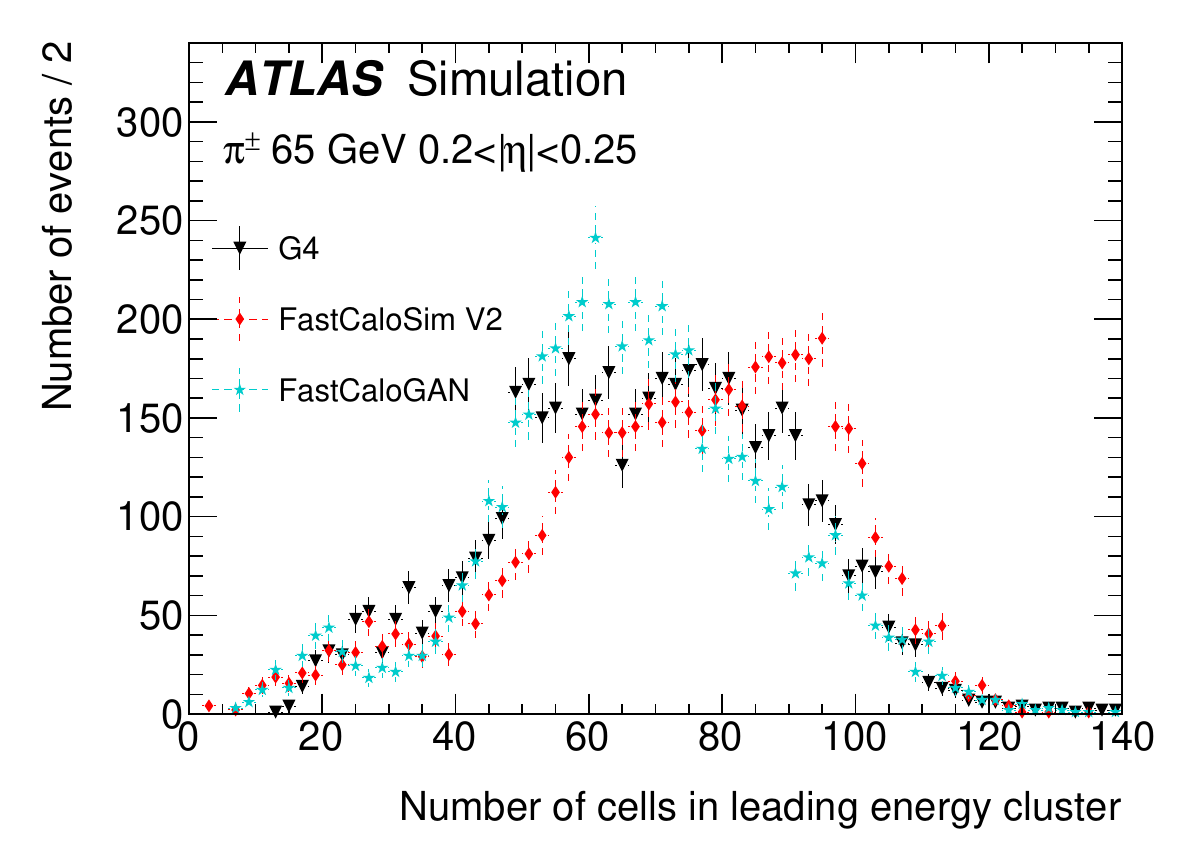}}
\subfloat[]{\includegraphics[width=0.47\textwidth]{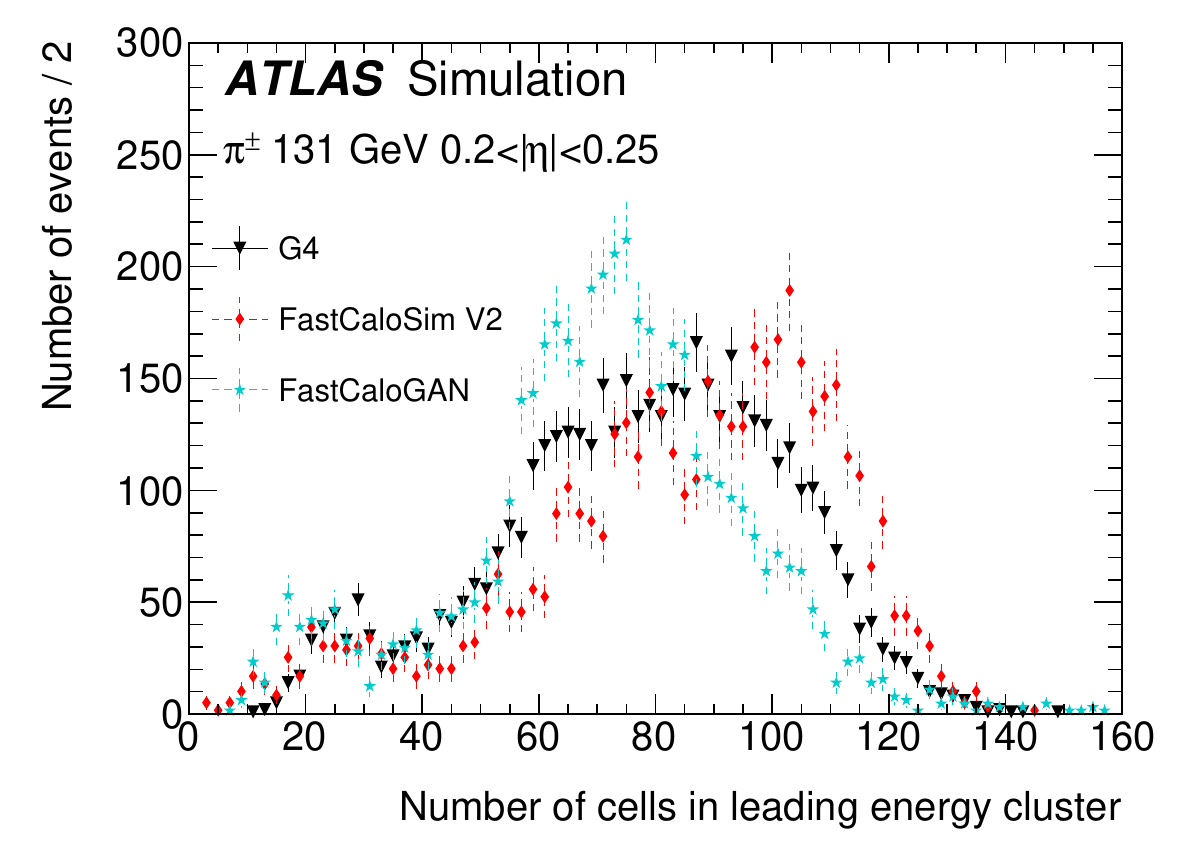}}\\
\subfloat[]{\includegraphics[width=0.47\textwidth]{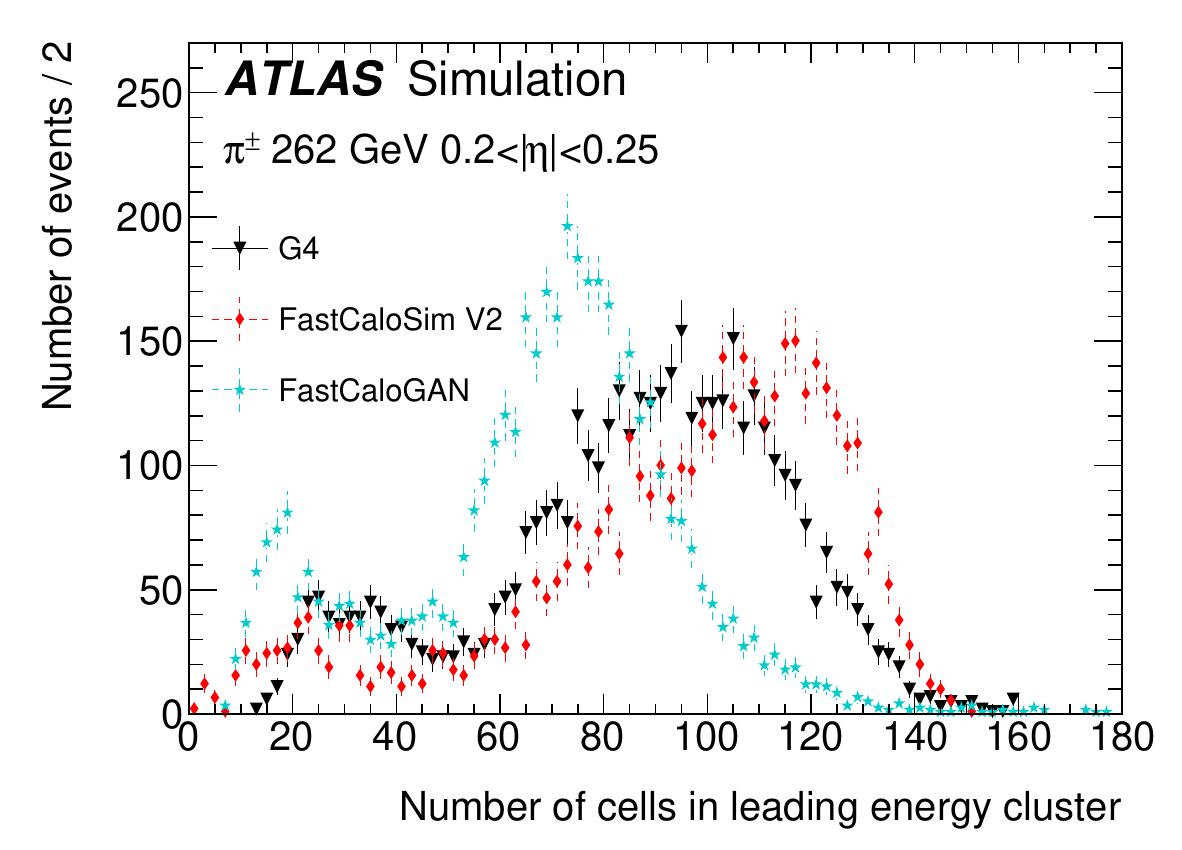}}
\subfloat[]{\includegraphics[width=0.47\textwidth]{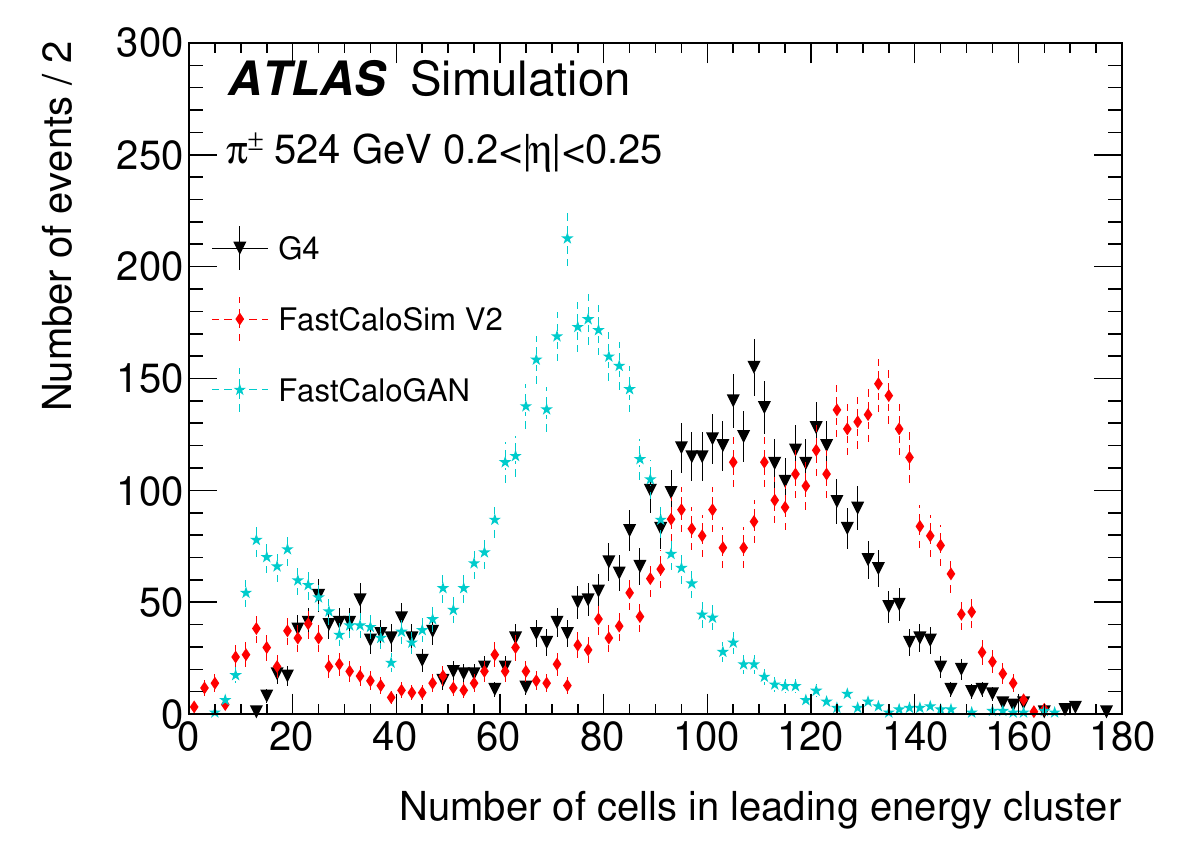}}
\caption{Number of cells in the leading cluster for pions in the barrel at different energies in \GEANT (black triangles), \FCaloSII (red diamonds) and \FGAN (blue stars). The statistical uncertainties are shown but may be smaller than the markers.}
\label{fig:AF3_optimisation:pions}
\end{figure}

At higher energies the modelling of the properties of individual clusters becomes important. Figure~\ref{fig:AF3_optimisation:pions} compares the number of cells in the calorimeter clusters in \GEANT with \FCaloSII and \FGAN for pion energies ranging from 65~\GeV\ to 524~\GeV.
Although \FCaloSII slightly overestimates the number of cells for all energies, \FGAN significantly underestimates the number of cells and this becomes more pronounced at higher energy. Studies of additional jet variables, many of which are shown in Section~\ref{sec:recoJets}, confirmed that \FCaloSII has better modelling for higher-energy hadrons. Therefore, \FCaloSII is used to simulate hadrons with $\Ekin > \text{256--512}$~\GeV. As shown  in Section~\ref{sec:recoJets}, despite these discrepancies, the modelling of higher-level objects such as jets is sufficient for physics analysis.
 
\subsubsection{Muon punch-through}
 
The muon punch-through parameterization described in Section~\ref{sec:muonPunchThrough} is used to simulate particles punching through the calorimeter. After the multiplicity and properties of the secondaries are determined using the punch-through parameterization, their path through the muon spectrometer is simulated using \GEANT.

\subsection{Energy interpolation}
\label{sec:energy_interpol}
\begin{figure}[h]
\subfloat[Photons]{\includegraphics[width=0.45\textwidth]{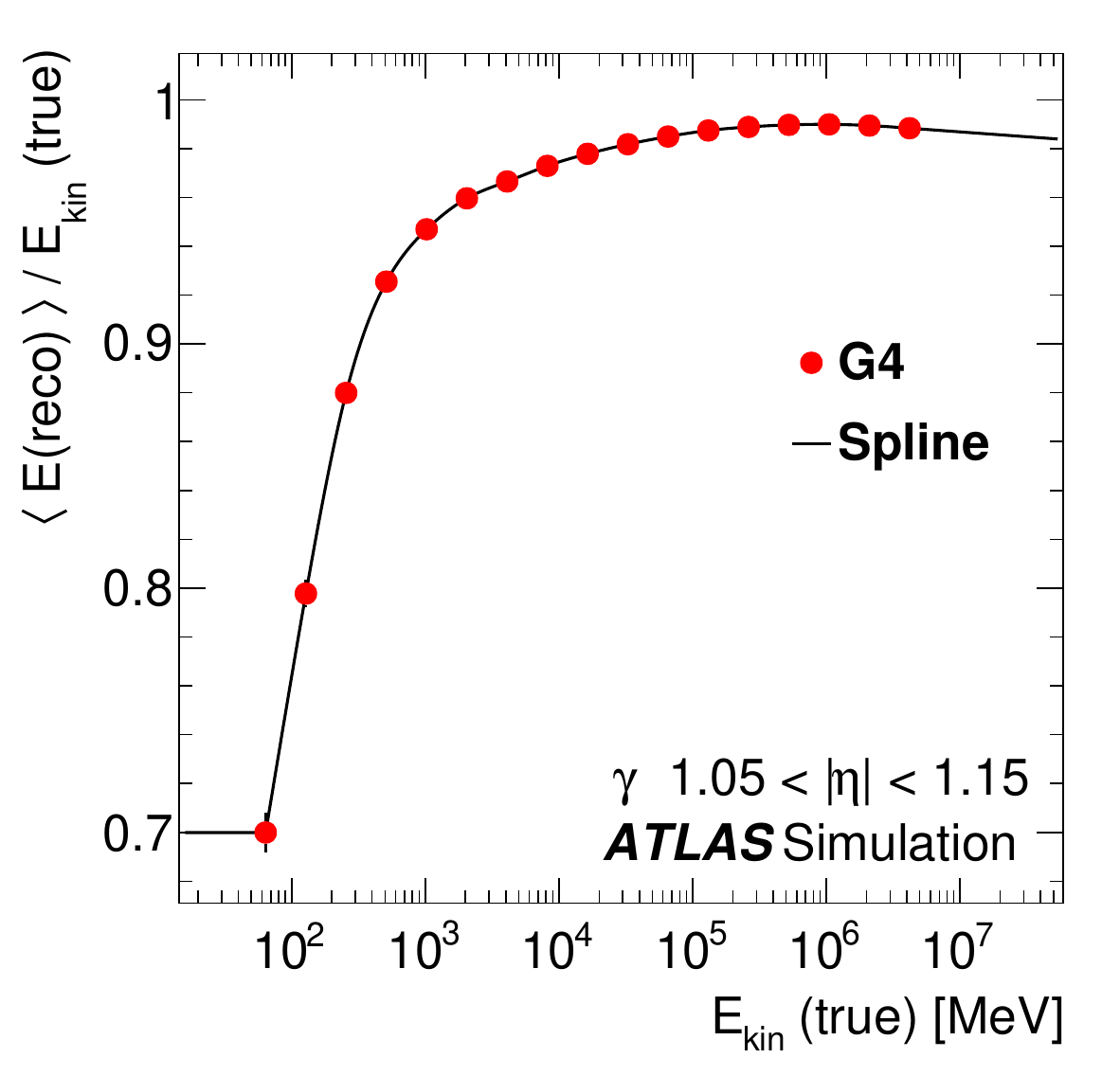}}
\subfloat[Pions]{\includegraphics[width=0.45\textwidth]{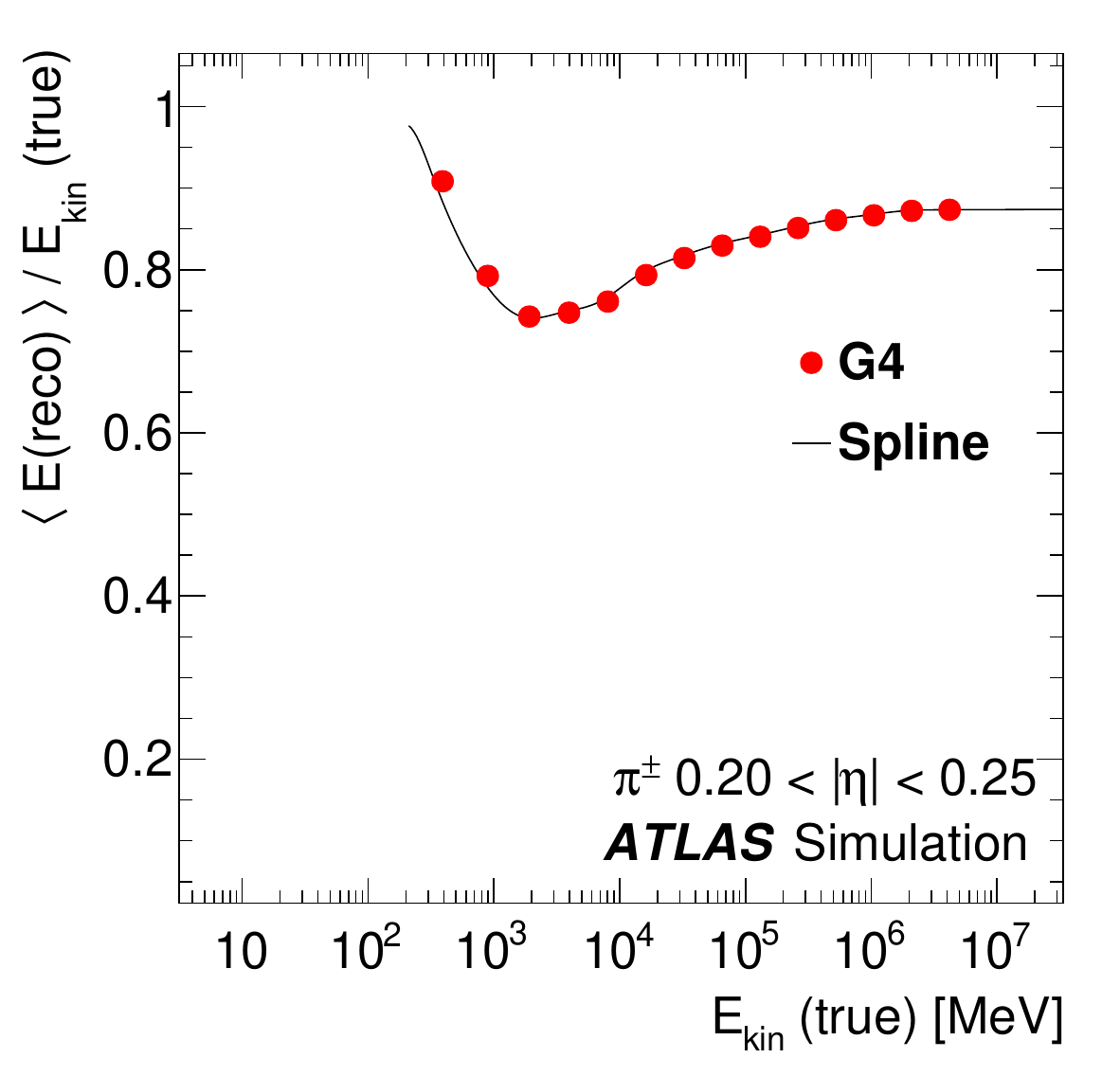}}
\caption{ Energy response, defined as the ratio of the reconstructed energy in the calorimeter cells to the kinetic energy of the particle, for (a) photons in $1.05 < |\eta| < 1.10 $ and (b) pions in $0.20 < |\eta| < 0.25$. The red dotted points represent the response derived at discrete energies, using \GEANT simulated single particles. The black line is a spline fit used to interpolate between discrete energy points. The statistical uncertainties are shown but are similar in size to the points or smaller.}
\label{fig:energy:interpolation}
\end{figure}

The \FCaloSII and \FGAN parameterizations are derived using samples with logarithmically spaced discrete energies, which need to be extrapolated to particles of all energies. In \FCaloSII, a piece-wise third order polynomial spline function is fitted to the total energy response in order to interpolate to intermediate energies. Furthermore, linear extrapolation is used to reach energies beyond those of the simulated input samples. The spline interpolations are generated for each particle and each $\eta$ slice and are used to rescale the total energy response from the parameterization points. An example of the energy response and fitted splines for photons and pions in the barrel region is shown in Figure~\ref{fig:energy:interpolation}. The energy response for high-energy photons is slightly reduced due to leakage into the Tile calorimeter. In \FGAN, the conditioning on the particle momentum creates a model that can produce particles of any energy.
 
In addition to the interpolation of the total energy response, the other longitudinal and lateral shower shape properties also need to be interpolated. In \FGAN the shape properties are interpolated automatically by the GANs, while in \FCaloSII the shape interpolation is done by randomly selecting the parameterization from the nearest energy point with a probability linear in $\log(\Ekin)$ and fitted such that unit probability is reached for the grid energy points.
 
In the two transition regions between \FCaloSII and \FGAN (for hadrons in the ranges 8--16~\GeV{} and 256--512~\GeV), a spline is used to interpolate between the two models. A smooth energy-response transition between the two models is obtained since the simulated energies are always scaled to the energy from \GEANT.
For electrons and photons the spline for the energy response is fitted down to 16~\MeV, below which a linear extrapolation is used. For hadrons the energy response is fitted down to a kinetic energy of 200~\MeV, below which \GEANT is used for the simulation.
 
\subsection{Corrections}
Four different corrections are applied to the calorimeter parameterization in \AFIII. However, the energy resolution correction discussed in Section~\ref{sec:energy:res_corr} and the energy  $\phi$-modulation correction discussed in Section~\ref{sec:energy:phi_mod} are only applied to \FCaloSII.
 
\subsubsection{Energy resolution correction}
\label{sec:energy:res_corr}
\begin{figure}[h]
\centering
\includegraphics[width=0.48\textwidth]{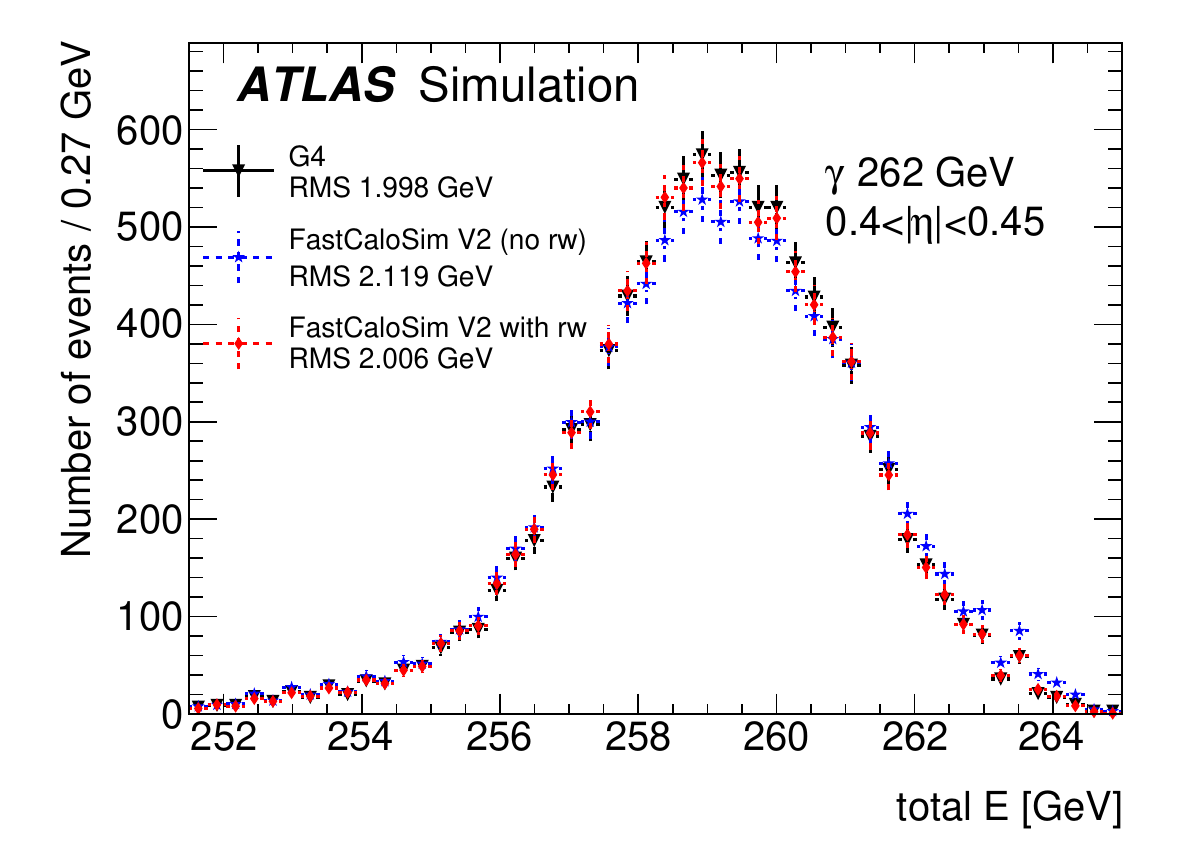}
\caption{
The simulated total energy before (blue stars) and after (red diamonds) probabilistic reweighting for a photon of energy 262~\GeV{} in the range $0.4 < |\eta| < 0.45$ compared with \GEANT (black triangles). The RMS of each distribution is indicated in the legend. The statistical uncertainties are shown but may be smaller than the markers.\label{fig:epara:res_corr}
}
\end{figure}

The simulation of the resolution of the total energy in \FCaloSII is improved by reweighting the distribution of simulated energies produced by \FCaloSII to the distribution from \GEANT. The ratio of the \GEANT simulated energy to the \FCaloSII simulated energy for each PCA bin is used to create a pdf. For each simulated total energy the pdf returns an associated probability. During simulation, for each simulated energy a uniform random number in [0,1] is drawn and if the number is smaller than the probability obtained from the pdf, the simulated energy is accepted. If the energy is rejected, then the energy simulation step discussed in Section~\ref{sec:epara} is repeated. The RMS is calculated using at least 99\% of all events and in a $\pm3\sigma$ range around the mean. This probabilistic reweighting (rw) obtains good agreement with the \GEANT distribution. Figure~\ref{fig:epara:res_corr} shows the resolution for photons, as an example, before and after the correction, and the RMS of the distribution is indicated.
 
\subsubsection{Energy $\phi$-modulation correction}
\label{sec:energy:phi_mod}
\begin{figure}[t]
\centering
\subfloat[]{ \label{fig:phimod}
\includegraphics[width=0.48\textwidth]{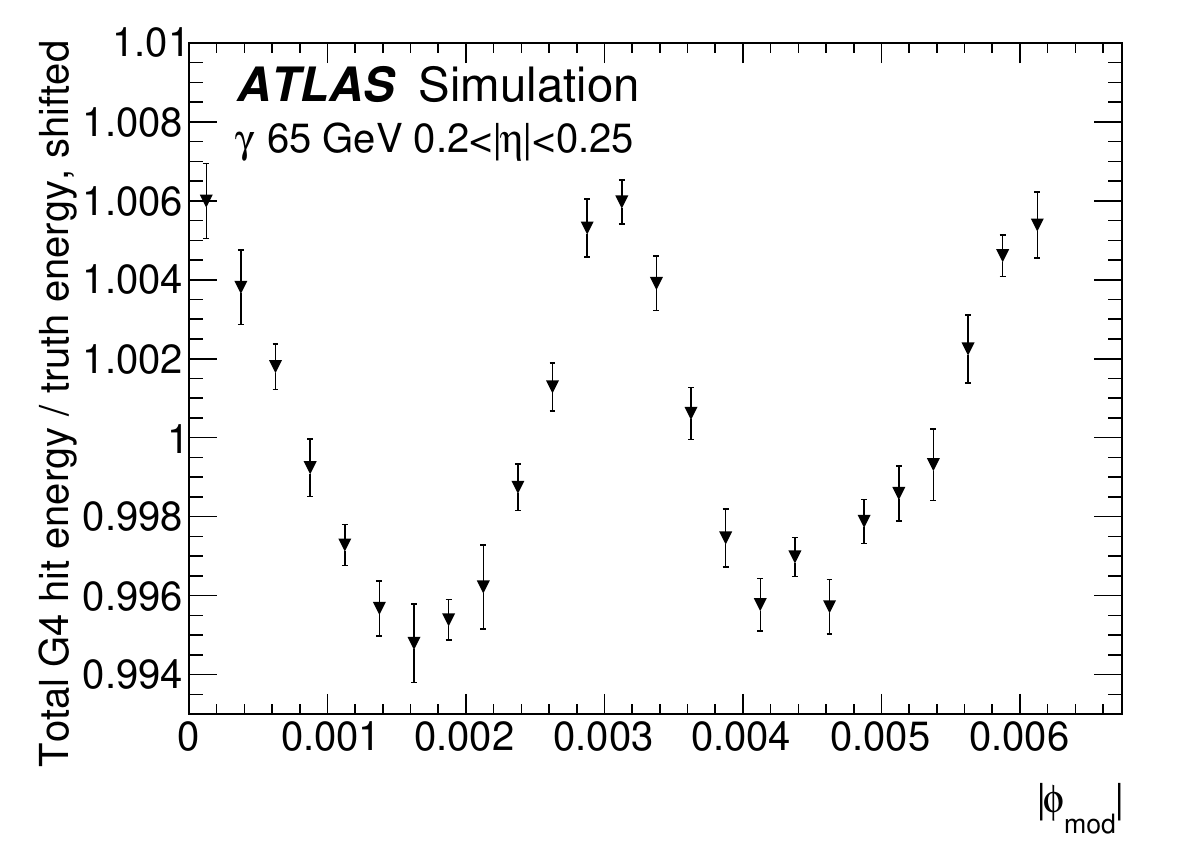}  }
\subfloat[]{\label{fig:phimod_eresponse}
\includegraphics[width=0.48\textwidth]{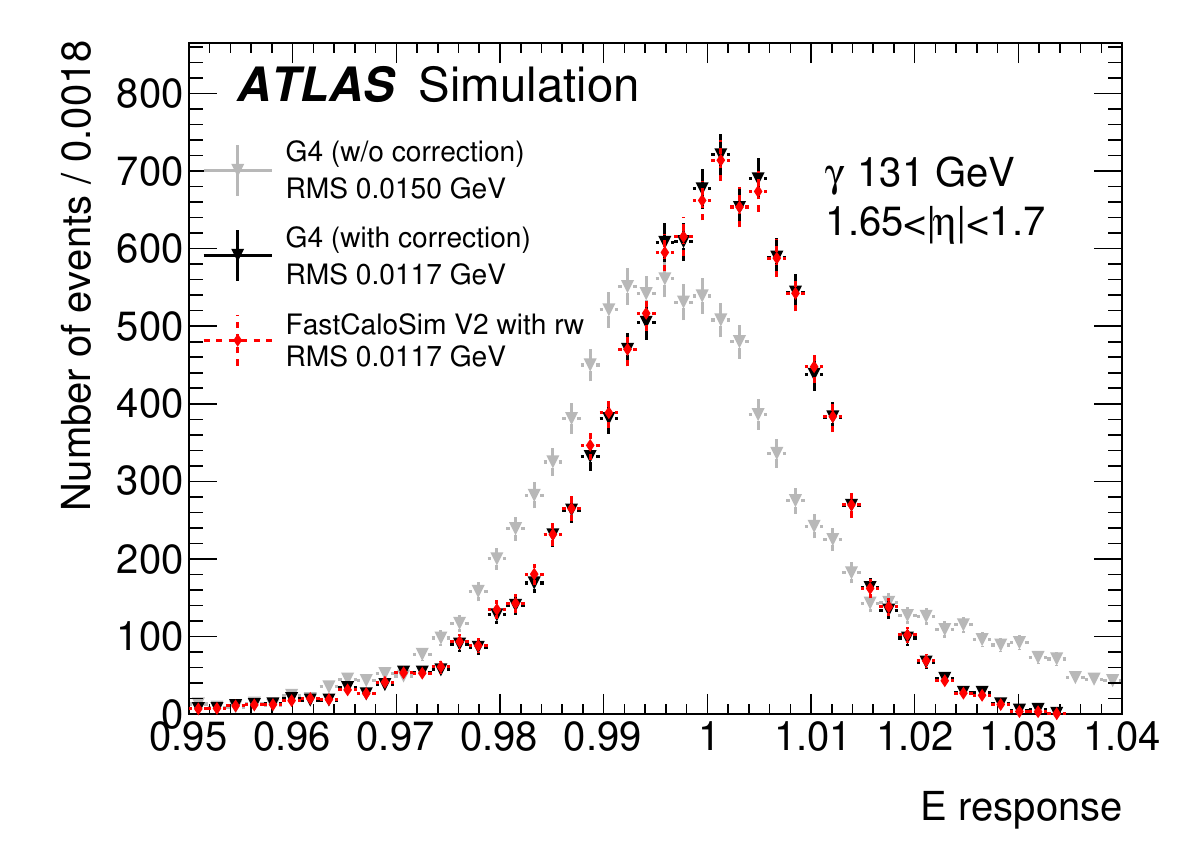} }
 
\caption{
(a) The total energy response exhibits a dependence on the impact position in $\phi$ of the particle in the calorimeter cell ($|\phi_\mathrm{mod}|$), shown for 65~\GeV\ photons with $0.2 < |\eta| < 0.25$ (\GEANT). The ratio has been shifted such that mean ratio of the energy from \GEANT to the true energy is unity. (b) The impact of the correction on \GEANT simulation (gray triangles are without correction, black are with corrections) and the result of the stand-alone simulation for 131~\GeV\ photons with $1.65 < |\eta| < 1.7$ to which the correction has been applied as well as the reweighting described in Section~\ref{sec:energy:res_corr}. The statistical uncertainties are shown in the error bars.
}
\end{figure}

Due to the accordion structure of the EM calorimeter, the total deposited energy is modulated in the $\phi$-direction as shown in Figure~\ref{fig:phimod}, where $|\phi_\mathrm{mod}| = |\mathrm{mod}(\phi_{\mathrm{calo}}, \pi/512)|$. The calibration applied during the ATLAS electron and photon reconstruction makes a correction for the $\phi$-modulation in the energy response observed in \GEANT; this calibration impacts the resolution of the reconstructed energy. The modulation is not reproduced in \FCaloSII, because it does not have a functional dependence on $\phi$. The resolution of showers in the electromagnetic calorimeter produced by \FCaloSII is corrected by deriving the energy parameterization of Section~\ref{sec:epara} after removing the modulation of the energy as a function of $\phi$ in the reference samples. This procedure is applied to particles with energies of at least 16~\GeV{}; below this threshold the effect is negligible and can be ignored. Figure~\ref{fig:phimod_eresponse} shows the energy response from \GEANT for photons with and without the removal of the $\phi$ modulation compared with the prediction from \FCaloSII. Good agreement in the modelling of the resolution between \FCaloSII and \GEANT is obtained for the \GEANT samples with the $\phi$-modulation removed. As a consequence of this strategy, during the reconstruction of electrons and photons simulated with \AFIII, a set of calibrations without a correction for the energy modulation in the $\phi$-direction must be applied, differing from the calibrations used for full simulation samples. This procedure particularly improves the modelling of the resolution of the calibrated energies for photons and electrons in \AFIII.
 
\subsubsection{Hadron total energy correction}
\label{sec:hadron_correction}
The hadron total energy correction accounts for the difference between the charged-pion response, which is used to derive the calorimeter parameterizations, and the response to other hadron species. It is particularly important at low energies, where the kinetic energy of a hadron is close to its mass.
 
The hadron total energy correction is derived using simulated samples of (anti)protons, (anti)neutrons and (anti)kaons as described in Section~\ref{sec:datasets}.
Using \GEANT, the parameterized energy is corrected by the ratio of the mean simulated hadron energy response, $\langle E_{\text{G4}}^{\text{h}} \rangle$, to the mean simulated pion energy response, $\langle E_{\text{G4}}^{\pi} \rangle$.
A further rescaling must be applied because the reference samples were generated using the momentum of the particle while the \Ekin is used for the parameterization. This is achieved by calculating the pion-to-hadron ratio of kinetic energies for each true momentum in the reference samples.
During \AFIII simulation, hadrons are then simulated using the charged-pion parameterization that provides the total energy $E_{\text{Total}}$ given the kinetic energy of the pion, $E_{\text{kin}}^{\pi,\text{true}}$, but with an additional correction based on the kinetic energy of the hadron, $E_{\text{kin}}^{\text{h,true}}$. The corrected energy response is then given by:
 
\begin{equation*}
E_{\text{Total}}^{\text{corr h}} = \frac{\langle E_{\text{G4}}^{\text{h}} \rangle}{ \langle E_{\text{G4}}^{\pi} \rangle} \times \frac{E_{\text{kin}}^{\pi,\text{true}}}{E_{\text{kin}}^{\text{h,true}}} \times  E_{\text{Total}}
\end{equation*}
 
\begin{figure}[t]
\centering
\includegraphics[width=0.45\textwidth]{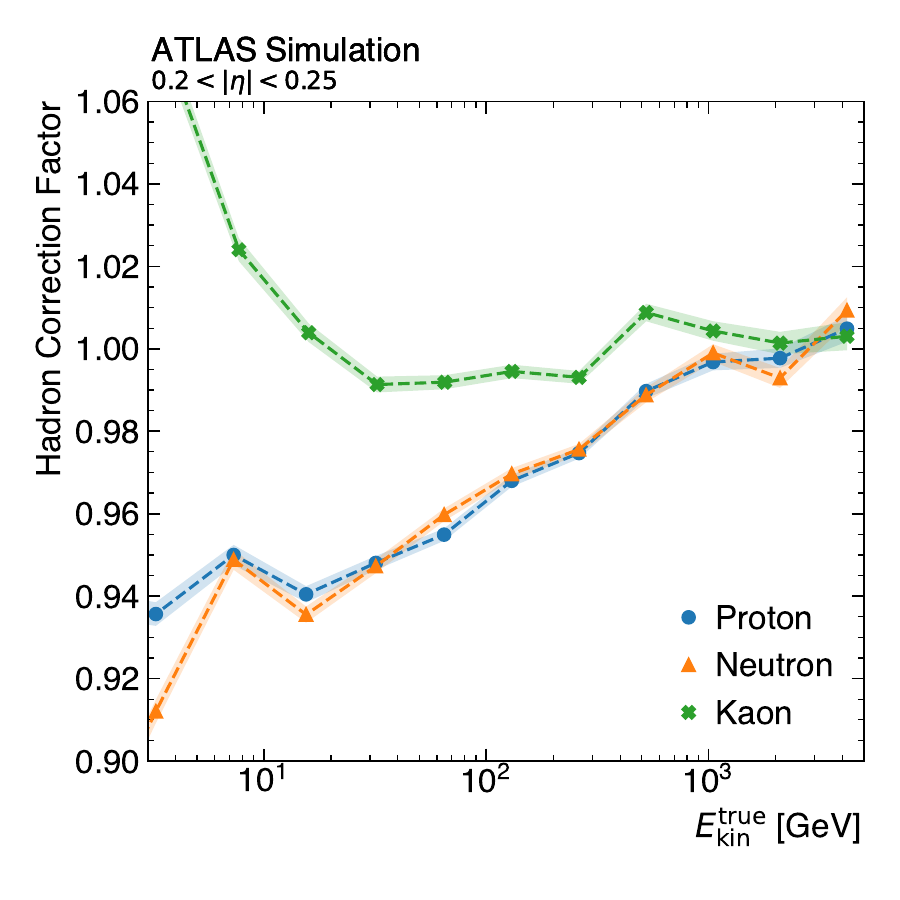}
\includegraphics[width=0.45\textwidth]{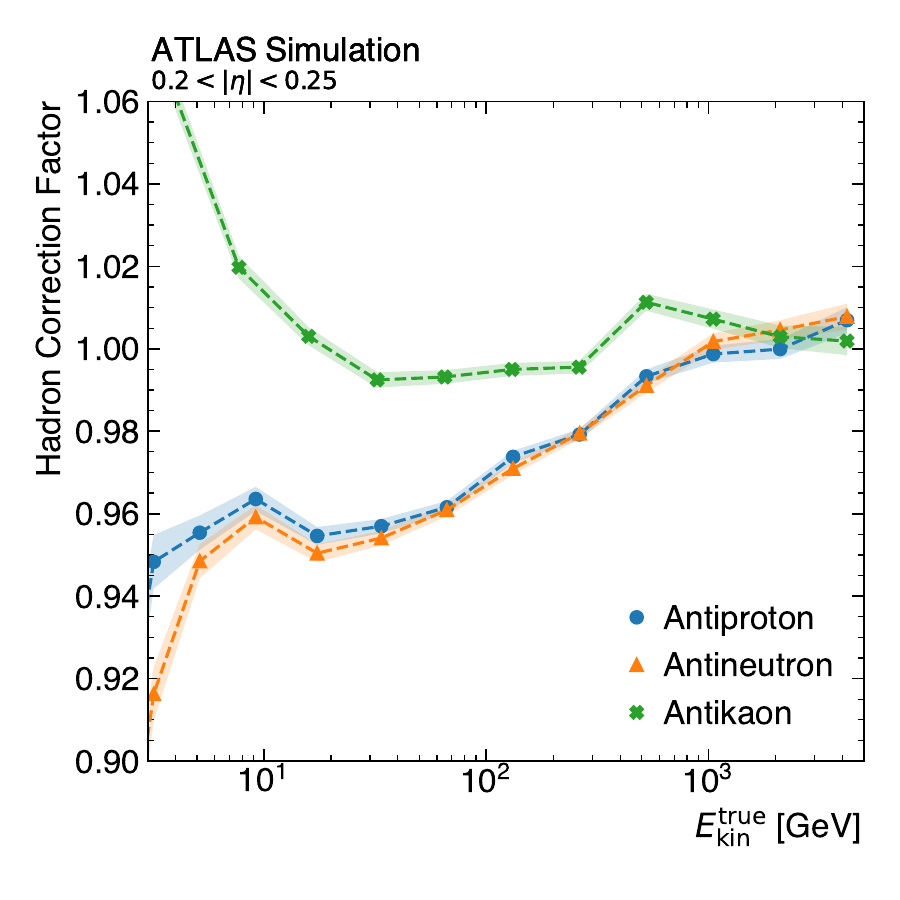}
\caption{
\label{fig:energy:Hadron}
Energy response correction factors as a function of the true kinetic energy for protons, neutrons and kaons (left) in the barrel and their antiparticles (right). The kinetic energy for antiparticles includes their mass.
The coloured bands indicate the size of the statistical uncertainty in the correction.
}
\end{figure}

The value $E_{\text{Total}}^{\text{corr h}}$ is the corrected energy. The hadron total energy correction is linearly interpolated between the logarithmically spaced energy grid points. Figure~\ref{fig:energy:Hadron} shows an example of the factor applied for the hadron total energy correction as a function of the true \Ekin for protons, neutrons and kaons. The hadron total energy correction is largest at small kinetic energies and decreases with increasing energy. It does not depend strongly on $\eta$ and is similar for protons and neutrons.
 
\subsubsection{Residual energy response correction}
\label{sec:energy_correction}
The residual energy response correction is applied to correct the total energy response for electrons, photons and pions from the parameterizations to match the average response of \GEANT after the full ATLAS simulation and reconstruction chain.  This correction can, therefore, correct for differences introduced during digitization and reconstruction. The residual energy response correction is the ratio of the average reconstructed energy when using \GEANT, $\langle E_{\text{G4}} \rangle$ to the average reconstructed energy from \AFIII, $\langle E_{\text{AF3}} \rangle$. The residual energy response correction is calculated and applied as follows:
 
\begin{equation*}
E_{\text{Total}}^{\text{corr res}} (\mathrm{p}) =  \langle E_{\text{G4}} (\mathrm{p}) \rangle / \langle E_{\text{AF3}} (\mathrm{p}) \rangle \times E_{\text{Total}} (\mathrm{p})
\end{equation*}
where $\mathrm{p} = [e, \gamma, \pi]$.
 
\begin{figure}[t]
\centering
\includegraphics[width=0.52\textwidth]{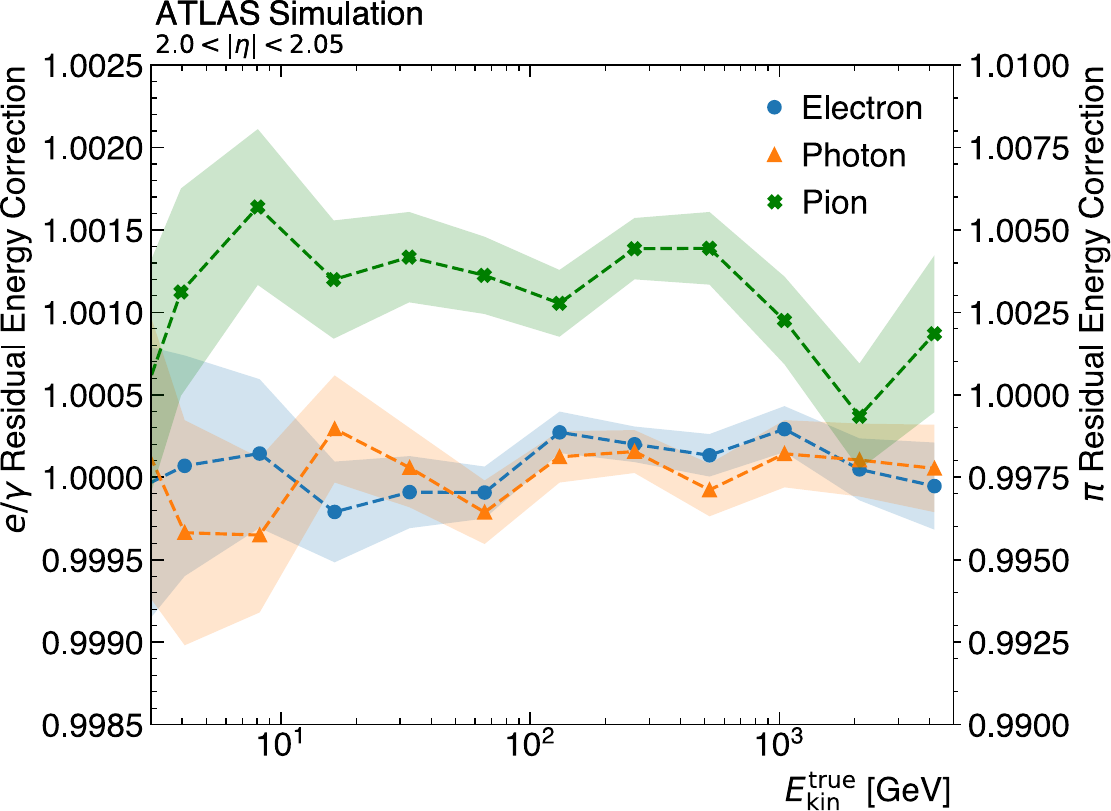}
\caption{ \label{fig:energy:corrections}
Residual energy response correction factors as a function of the true kinetic energy for photons, electrons and pions in the endcap.
The coloured bands indicate the size of the statistical uncertainty in the correction.
}
\end{figure}

It is derived for each parameterization grid point and linearly interpolated between the simulated energy points. Figure~\ref{fig:energy:corrections} shows an example of the derived residual energy response correction as a function of the true \Ekin for photons, electrons and pions. The residual energy response correction is at the per-mil level for electrons and photons and only slightly larger for pions and hence only applied when statistically significant.
 
\subsubsection{Simplified geometry shower shape correction}
\label{af3:wiggle_correction}
The hits generated by \FCaloSII or \FGAN are assigned to calorimeter cells using a simplified cuboid geometry. This introduces a bias in the energy distribution, which can result in a significant number of hits being assigned to neighbouring calorimeter cells.  To account for this effect while maintaining the reduced simulation CPU time afforded by the simplified geometry, a small displacement in $\phi$ is assigned to each hit before geometrically matching it to a cell with the simplified geometry. This procedure is substantially easier and faster than geometrically matching a hit to the cells in the complex ATLAS liquid-argon accordion structure.
 
\begin{figure}[h]
\subfloat[before correction]{\includegraphics[width=0.45\textwidth]{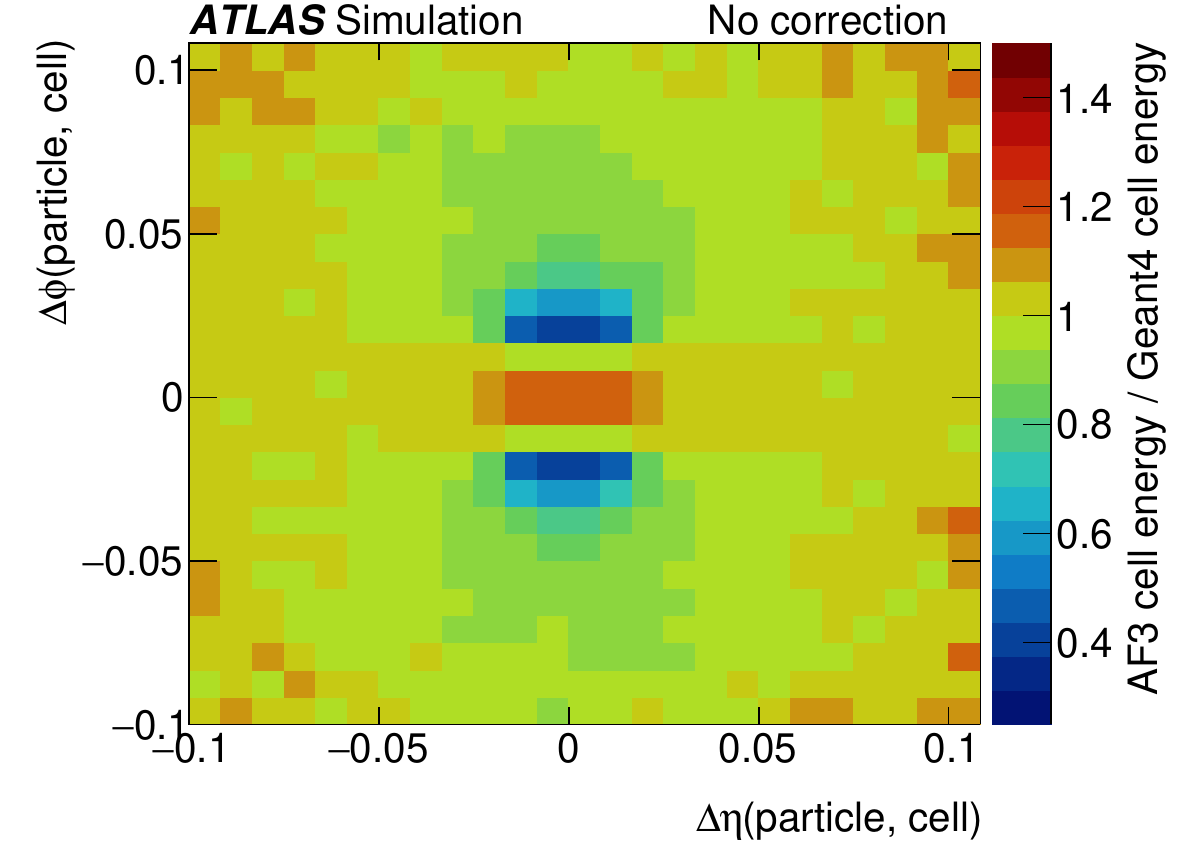}}\label{fig:wiggle:pre}
\subfloat[after correction]{\includegraphics[width=0.45\textwidth]{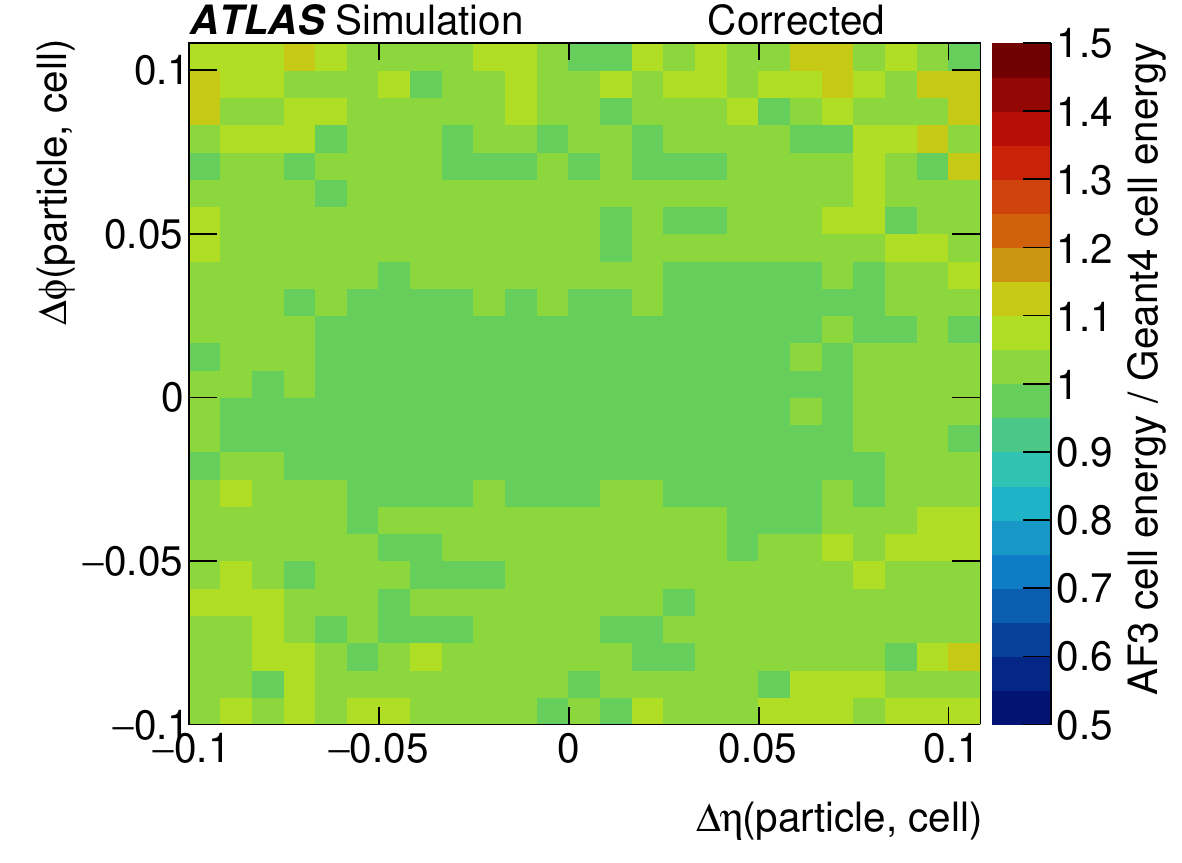}} \label{fig:wiggle:post}
\caption{The ratio of the energies assigned to each cuboid of the second electromagnetic barrel layer in \AFIII and \GEANT for a photon of 65~\GeV{}  in the range $0.20 < |\eta| < 0.25$ using a simplified cuboid geometry and after applying the correction for the simplified geometry.}
\label{fig:shape:wiggle}
\end{figure}

A pdf is derived from the difference between the cell assignment probabilities calculated in \GEANT and \FCaloSII. The correction is made using the pdf to randomly assign a displacement in $\phi$ to a hit. Figure~\ref{fig:shape:wiggle} shows the bias in deposited energy in each cuboid before and after this correction has been applied. Good agreement in the cell energy between \AFIII and \GEANT is observed once this correction has been applied.
 
\section{Performance of \AFIII}
\label{sec:af3perf}
The performance of \AFIII is studied by comparing the modeling of reconstructed quantities (Section~\ref{sec:recoObjects}) and important kinematic variables from physics analyses (Section~\ref{sec:physana}) in \AFIII, \GEANT, and \AFII. The reconstructed objects that depend on the performance of the calorimeter are electrons and photons, jets, and $\tau$-leptons.  For $b$-tagging and for other particles, such as muons, the performance of \AFIII depends primarily on the performance of the tracking detectors. This is studied as part of the validation of \AFIII{}, and we focus the specific case of muon punch-through in this paper because no significant differences from \GEANT are observed. Section~\ref{sec:techperf} discusses the CPU performance of \AFIII.
 
\subsection{Performance of \AFIII on objects for physics analysis}
\label{sec:recoObjects}
\subsubsection{Reconstructed photons and electrons}
\label{sec:recoEgamma}
 
\begin{figure}[t]
\centering
\subfloat[]{\label{fig:reco:egamma_elIDa}
\includegraphics[width=0.48\textwidth]{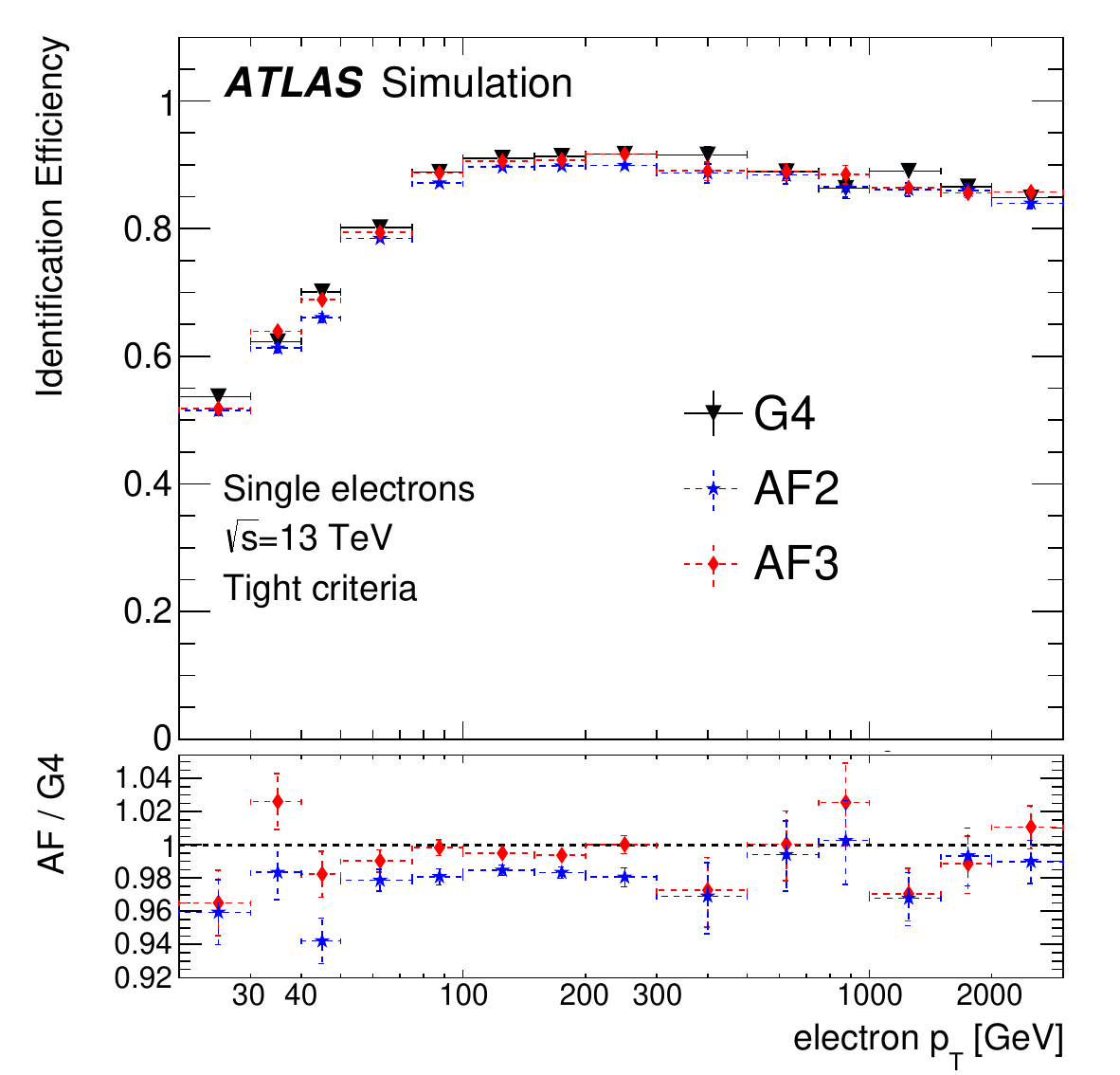}}
\subfloat[]{\label{fig:reco:egamma_elIDb}
\includegraphics[width=0.48\textwidth]{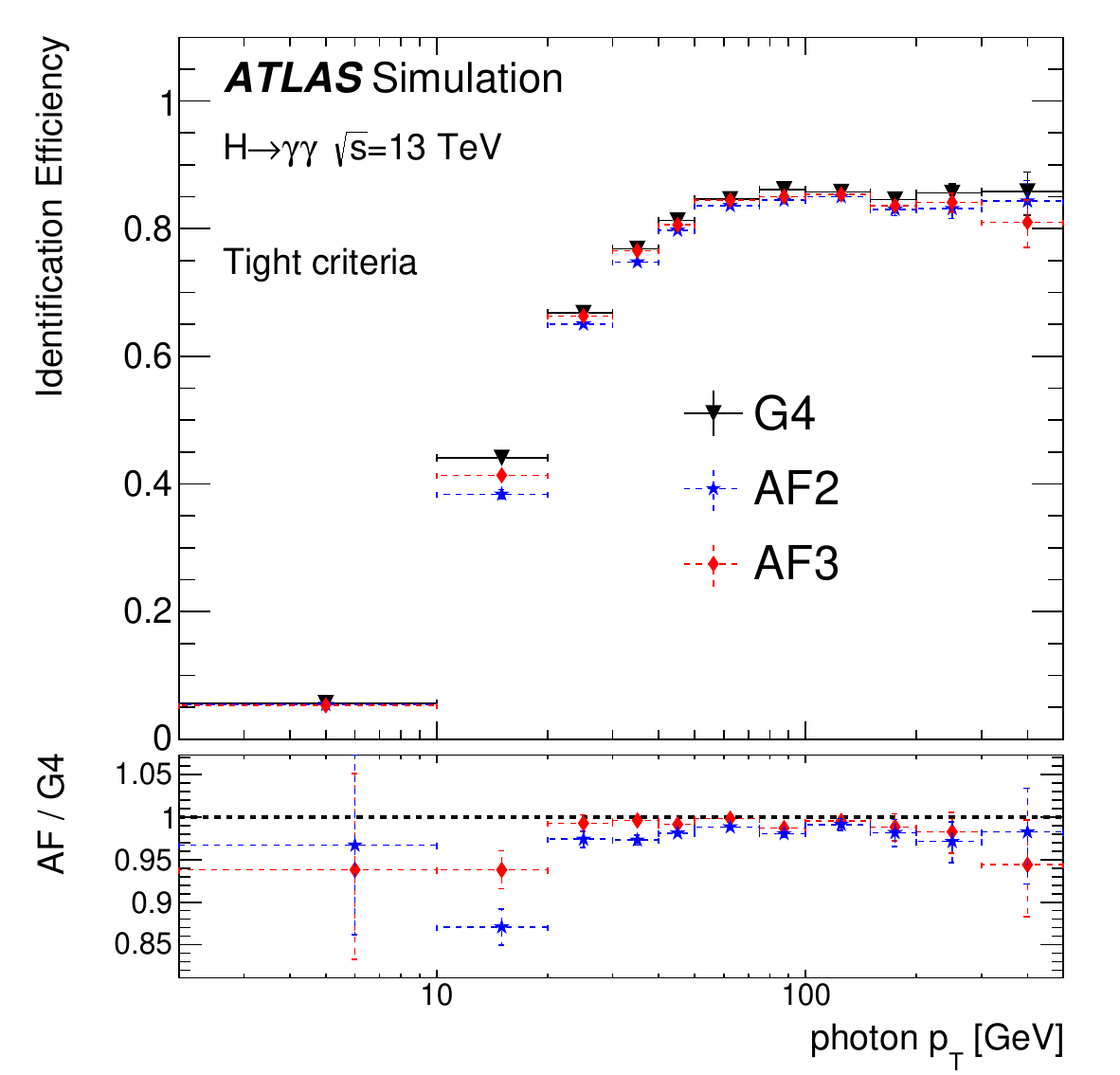}}
\caption{ \label{fig:reco:egamma_elID} `Tight' identification efficiencies for single electrons with true energy greater than 20\,\GeV (a) and photons from $H \rightarrow \gamma\gamma$ decays (b) inclusive in $|\eta| < 2.5$ as a function of their reconstructed \pT for \GEANT (black triangles), \AFII (blue stars), and \AFIII (red diamonds). The statistical uncertainties are shown but may be smaller than the size of the markers.
}
\end{figure}

Electron and photon candidates are reconstructed from topological clusters of deposited energy in the electromagnetic calorimeter, and in the case
of electrons, tracks in the inner detector are matched to the clusters~\cite{EGAM-2018-01}. For physics analysis, identification criteria are defined by requirements on shower shape and track quality. These identification criteria are labelled as `loose', `medium', and `tight'. The identification of electrons uses information from the inner detector, so a single electron sample with uniform \pT{}, $\eta$ and $\phi$-coverage are generated at the center of the ATLAS detector for validation in this section. These samples are then simulated with \pileup{} overlaid. The efficiencies for both electrons and photons are validated on an inclusive $\eta$ and \pT{} range that extends beyond what is typically considered for performance studies. Figure~\ref{fig:reco:egamma_elIDa} compares the electron identification efficiencies for the `tight' criterion as a function of the reconstructed \pT for \AFII, \AFIII and \GEANT. While \AFII agrees with \GEANT to better than 5\% in the electron \pT{} range from 30~\GeV{} to 300~\GeV{}, \AFIII agrees with \GEANT to within 2\% in most of the phase space. The `tight' photon identification efficiency is shown in Figure~\ref{fig:reco:egamma_elIDb} as a function of the photon \pT  for \AFII, \AFIII and \GEANT. Except at very low \pT, \AFIII agrees with \GEANT to within a few percent with better modelling than \AFII.
 
\subsubsection{Reconstructed jets}
\label{sec:recoJets}
 
Jets are reconstructed using a variety of reconstruction algorithms and inputs from the calorimeters and the inner detector. In addition, different \pileup mitigation and jet grooming algorithms are applied.
 
Particle-flow (EMPFlow) jets, which are constructed using EM-scale topological clusters~\cite{PERF-2015-09}, are reconstructed with the \akt algorithm~\cite{Cacciari:2008gp,Fastjet} with a radial distance parameter $R=0.4$, using charged constituents associated with the primary vertex~\cite{PERF-2015-01} and neutral particle-flow constituents as inputs~\cite{PERF-2015-09}. Large-radius jets ($R=1.0$) are reconstructed by applying the anti-$k_t$ algorithm to locally calibrated topological clusters (LCTopo)~\cite{PERF-2014-07} and the newer alternative of Unified Flow Objects (UFO)~\cite{atlascollaboration2020optimisation}.
 
\begin{figure}[h]
\centering
\subfloat[]{\label{fig:recoObjects:jetsa}
\includegraphics[width=0.49\textwidth]{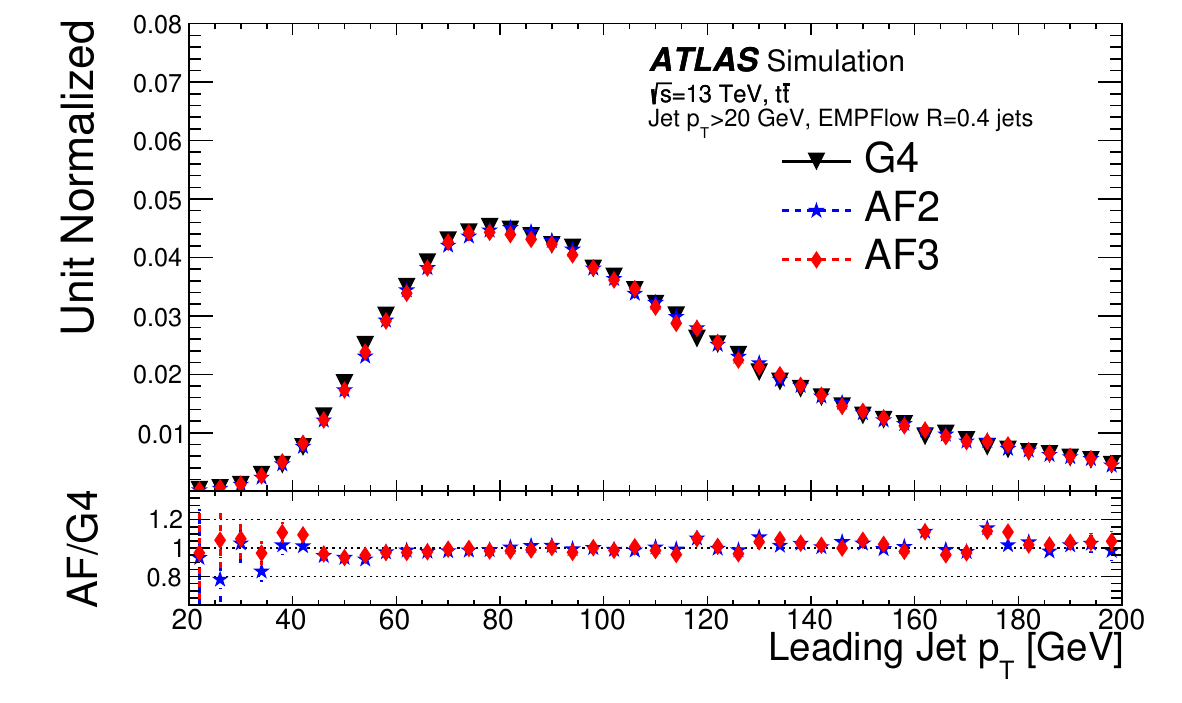}}
\subfloat[]{\label{fig:recoObjects:jetsb}\includegraphics[width=0.49\textwidth]{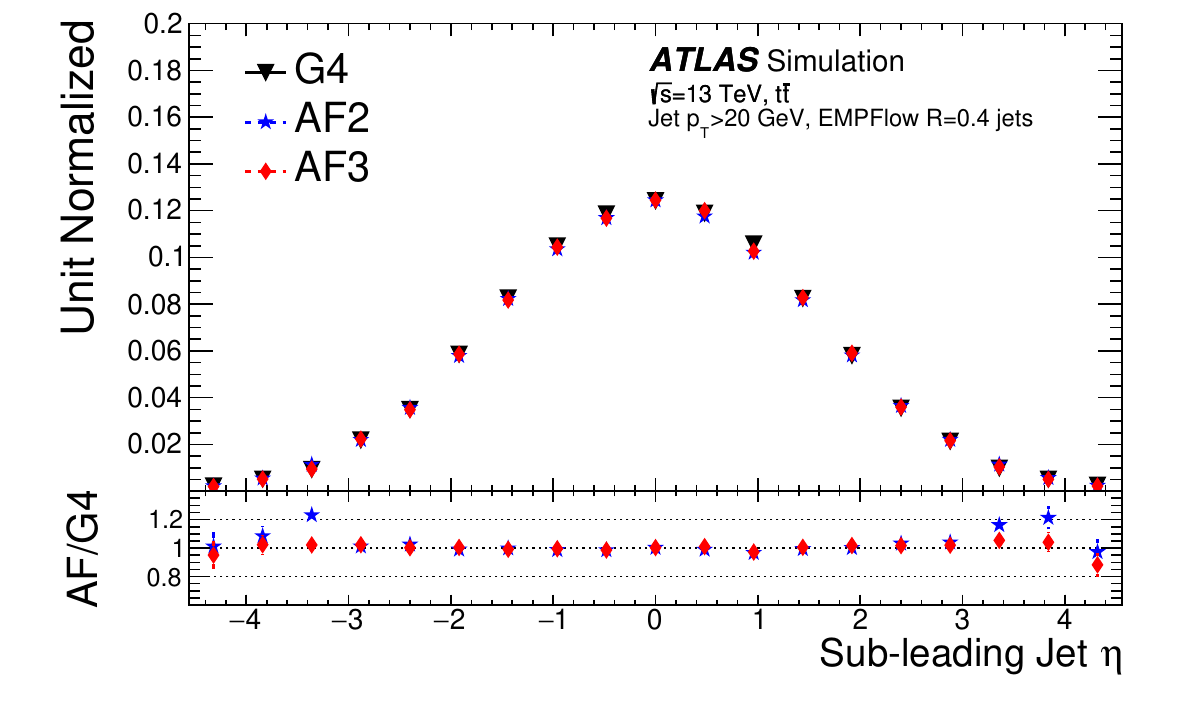}}
\caption{The transverse momentum distribution of the leading jets (a) and the pseudorapidity distribution of the sub-leading jets (b) in a \ttbar sample with \GEANT (black triangles), \AFII (blue stars), and \AFIII (red diamonds). The jets are EMPFlow jets with $R=0.4$. The statistical uncertainties are shown but may be smaller than the size of the markers.}
\label{fig:recoObjects:jets}
\end{figure}

The performance of \AFIII with EMPFlow jets is assessed using the \pT{} of the leading jet and the pseudorapidity distribution of the sub-leading jet in a \ttbar sample, which are shown in Figure~\ref{fig:recoObjects:jets}. For both distributions, \AFII and \AFIII are consistent with \GEANT at the percent level. In the forward $\eta$ regions of Figure~\ref{fig:recoObjects:jetsb}, \AFIII shows better agreement than \AFII with \GEANT thanks to the updated parameterization in the forward region of the detector.
 
\begin{figure}[h]
\centering
\subfloat[]{\label{fig:recoObjects:jetconsta}
\includegraphics[width=0.49\textwidth]{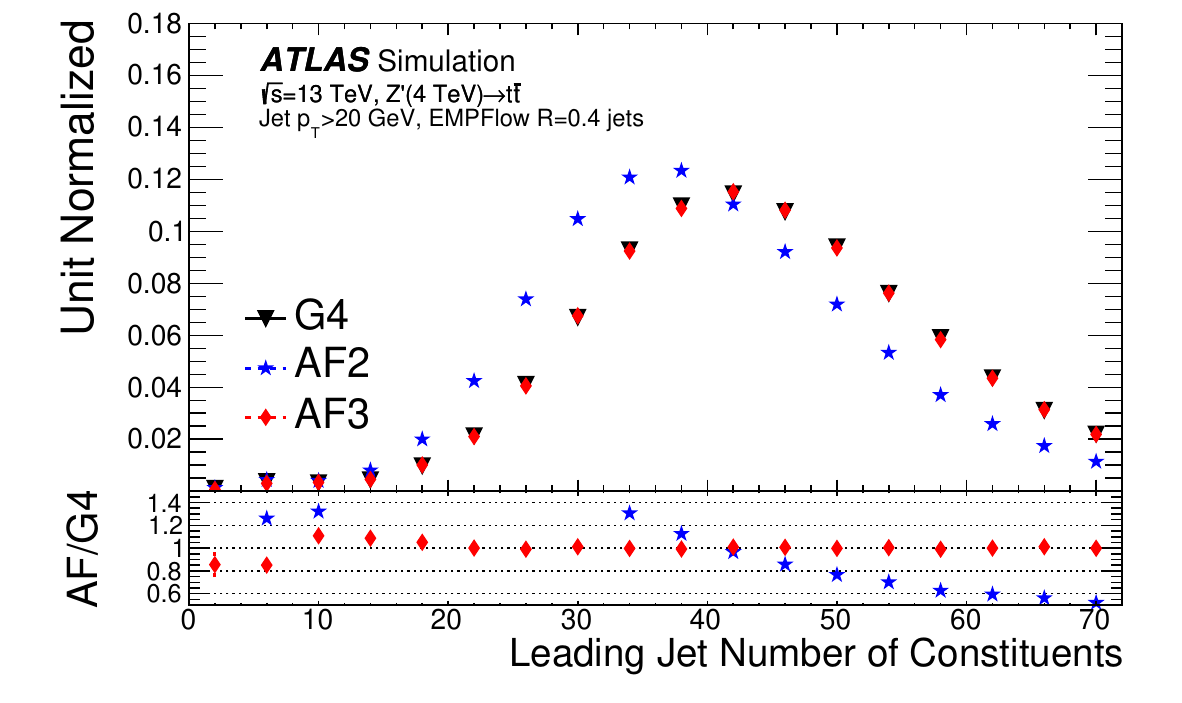}}
\subfloat[]{\label{fig:recoObjects:jetconstb}\includegraphics[width=0.49\textwidth]{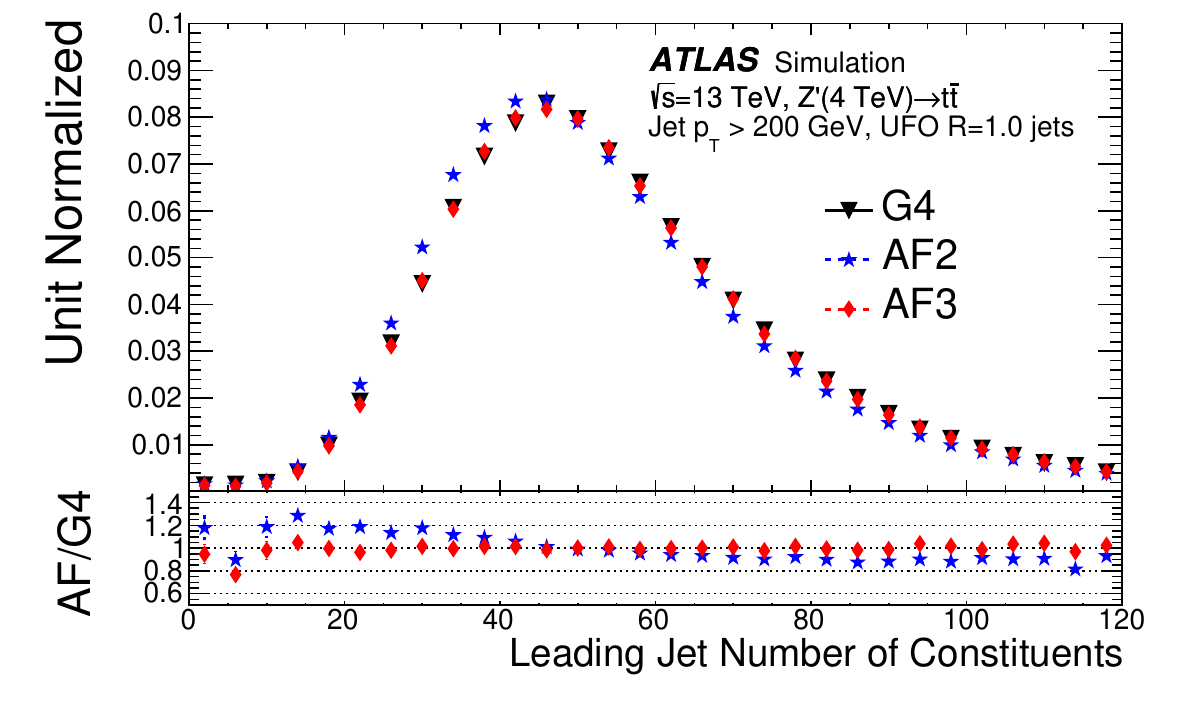}}
\caption{Distribution of the number of constituents in the leading jets for  EMPFlow jets with $R=0.4$ (a) and UFO jets with $R=1.0$ (b) in the $Z^\prime$ sample in \GEANT (black triangles), \AFII (blue stars), and \AFIII (red diamonds). The statistical uncertainties are shown but may be smaller than the size of the markers.}
\label{fig:recoObjects:jetconst}
\end{figure}

For higher-\pT{} jets, the simulation of the detailed structure within the jet plays an important role in the efficiency and classification. To provide better coverage for higher jet \pT{}, the $Z^\prime$ and $W^\prime$ boson samples were reweighted to have a flat leading-jet \pT{} spectrum as described in Section~\ref{sec:validation}. Figure~\ref{fig:recoObjects:jetconst} shows the number of charged constituents for leading jets with $\pT > 200$~\GeV\ from a sample containing $Z^\prime \rightarrow t\bar{t}$ events. Figure~\ref{fig:recoObjects:jetconsta} shows EMPFlow jets reconstructed with the \akt algorithm with a radius parameter $R=0.4$, while Figure~\ref{fig:recoObjects:jetconstb} shows UFO jets reconstructed with the \akt algorithm with $R=1.0$. The number of constituents in the EMPFlow jets is significantly underestimated by \AFII, while \AFIII reproduces the distribution from \GEANT within statistical uncertainties for jets with more than 14 constituents. For events with fewer constituents, \AFIII slightly underestimates the number of constituents. For the UFO jets, agreement with \GEANT improves significantly, going from a 20\% difference in \AFII to less than 10\% in \AFIII.
 
\begin{figure}[h]
\centering
\includegraphics[width=0.49\textwidth]{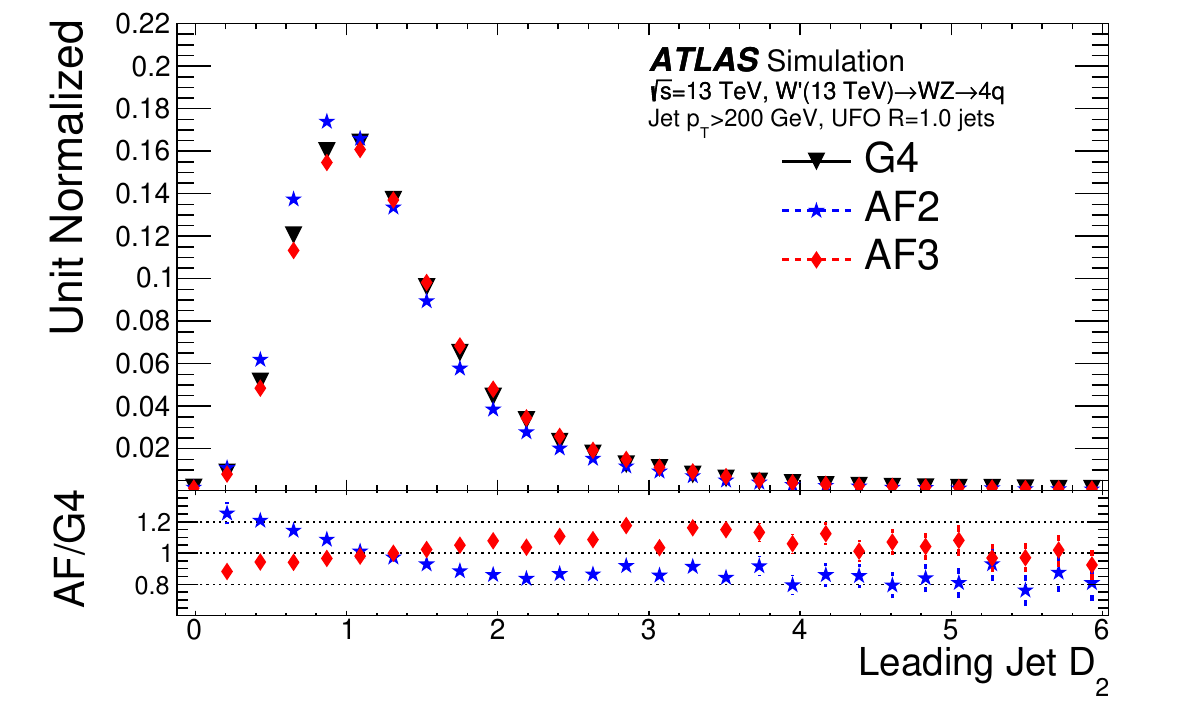}
\caption{The $D_2$ variable for the leading jets in a $W^\prime$ sample reconstructed using the UFO algorithm with radius parameter $R=1.0$ with \GEANT (black triangles), \AFII (blue stars), and \AFIII (red dimaonds). The statistical uncertainties are shown but may be smaller than the size of the markers.}
\label{fig:recoObjects:D2}
\end{figure}

Variables commonly used in jet-tagging algorithms include the energy-correlation-function ratio, $D_2$, for two-body decays and the $n$-subjettiness ratio, $\tau_{32}$, for three-body decays~\cite{Thaler_2011,d2}. Figure~\ref{fig:recoObjects:D2} shows the $D_2$ variable reconstructed using the UFO algorithm with \GEANT, \AFII, and \AFIII on a $W^\prime$ sample. \AFIII significantly improves the modelling of $D_2$, particularly at lower values.
 
\begin{figure}[h]
\centering
\subfloat[]{\label{fig:recoObjects:tau32a}
\includegraphics[width=0.49\textwidth]{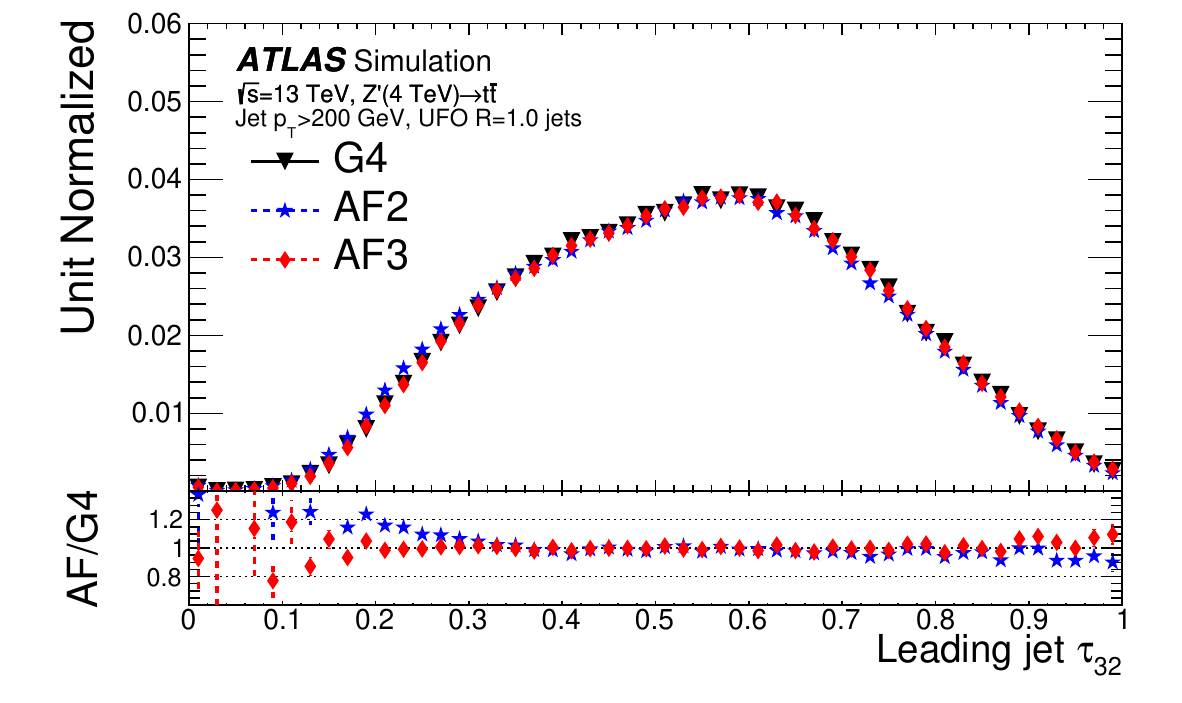}}
\subfloat[]{\label{fig:recoObjects:tau32b}\includegraphics[width=0.49\textwidth]{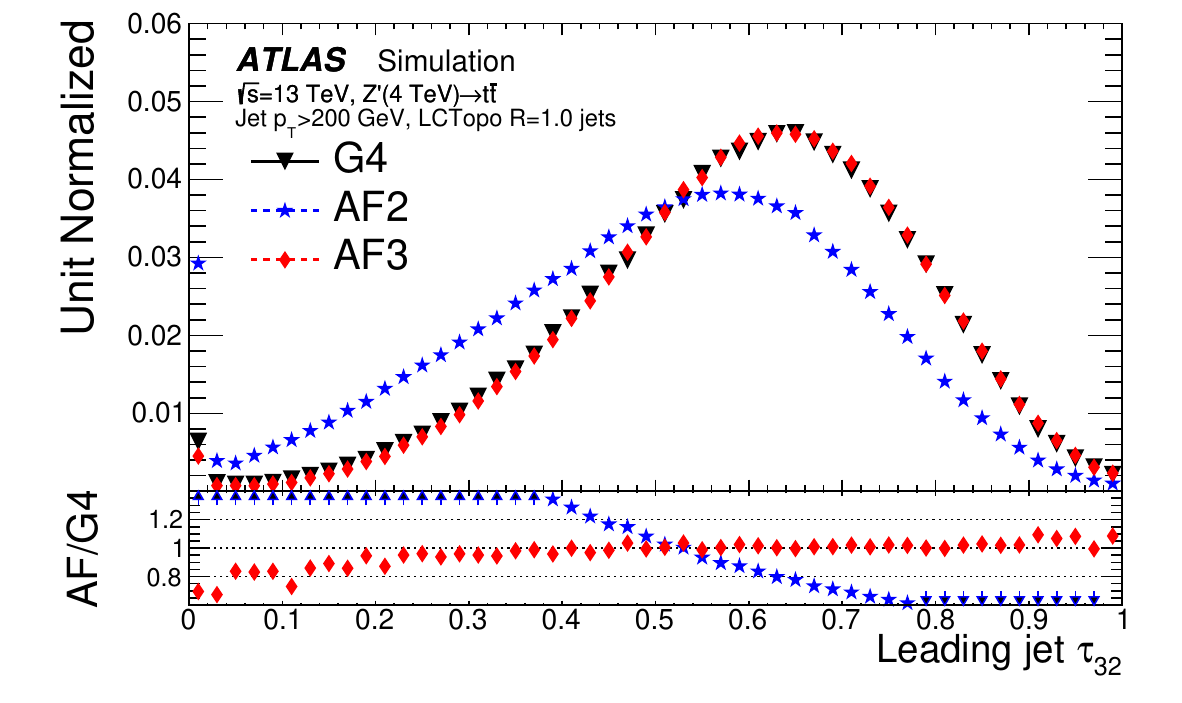}}
\caption{The $\tau_{32}$ variable for the leading jets in a $Z^\prime$ sample reconstructed using the UFO algorithm with radius parameter $R=1.0$ (a) and the LCTopo algorithm (b) with \GEANT (black triangles), \AFII (blue stars), and \AFIII (red diamonds). The statistical uncertainties are shown but may be smaller than the size of the markers, and the dark blue arrows indicate that a point is beyond the $y$-axis range.}
\label{fig:recoObjects:tau32}
\end{figure}

Figure~\ref{fig:recoObjects:tau32} shows $\tau_{32}$ for different large-radius jet algorithms. For the UFO jets in Figure~\ref{fig:recoObjects:tau32a}, \AFII reproduces the distribution of \GEANT to within 20\% and \AFIII improves this further to within 10\%.  For the LCTopo jets shown in
Figure~\ref{fig:recoObjects:tau32b} the modelling from \AFII is poor but is significantly improved with \AFIII, which obtains agreement to within 20\%. The improvement for LCTopo is expected to be larger than for UFO because UFO includes tracking information.
 
\subsubsection{Reconstructed hadronic $\tau$-lepton decays}
\label{sec:recoTaus}
 
\begin{figure}[h]
\centering
\subfloat[]{	\label{fig:recoObjects:taudecaypta}
\includegraphics[width=0.49\textwidth]{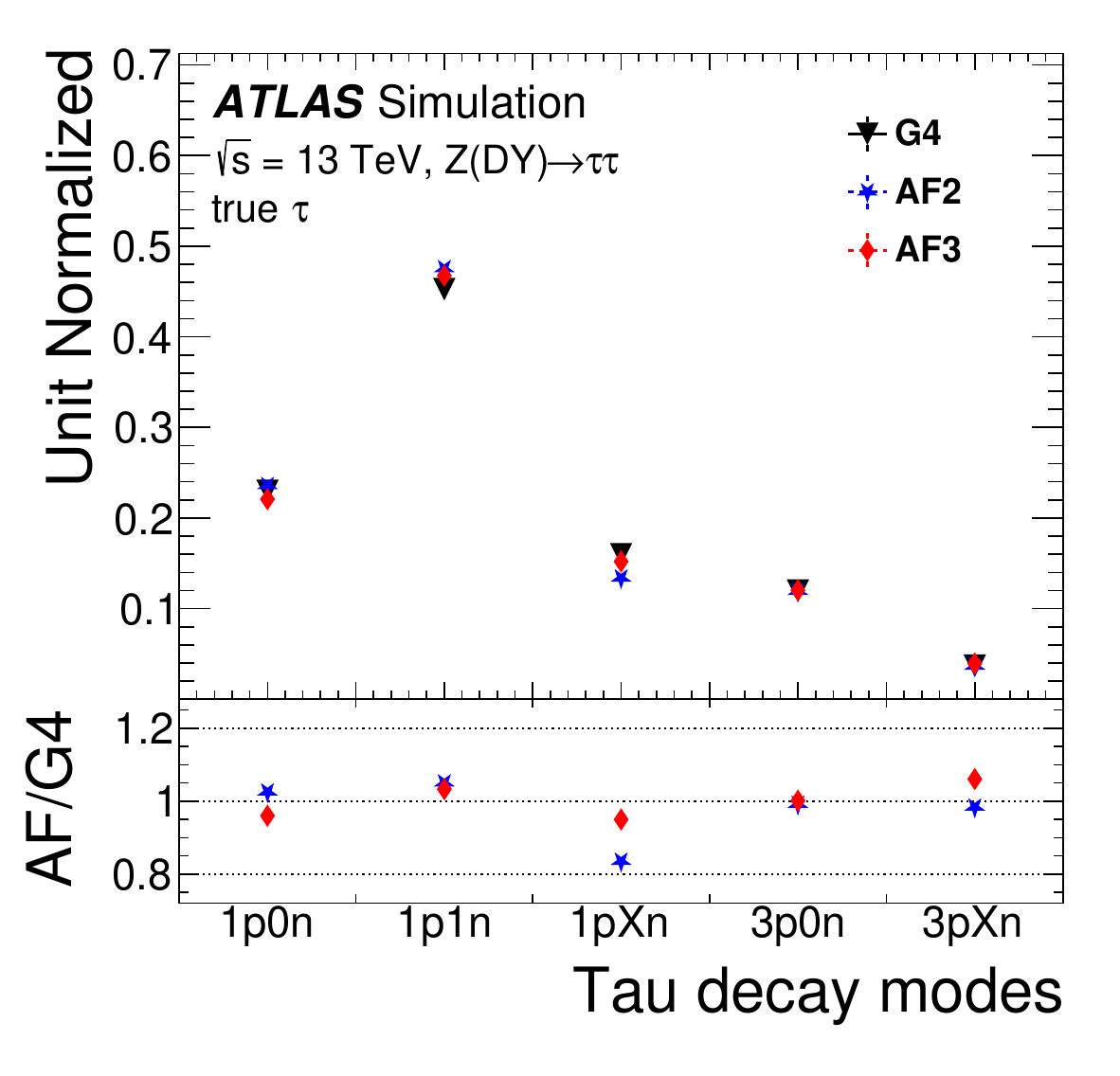}}
\subfloat[]{	\label{fig:recoObjects:taudecayptb}\includegraphics[width=0.49\textwidth]{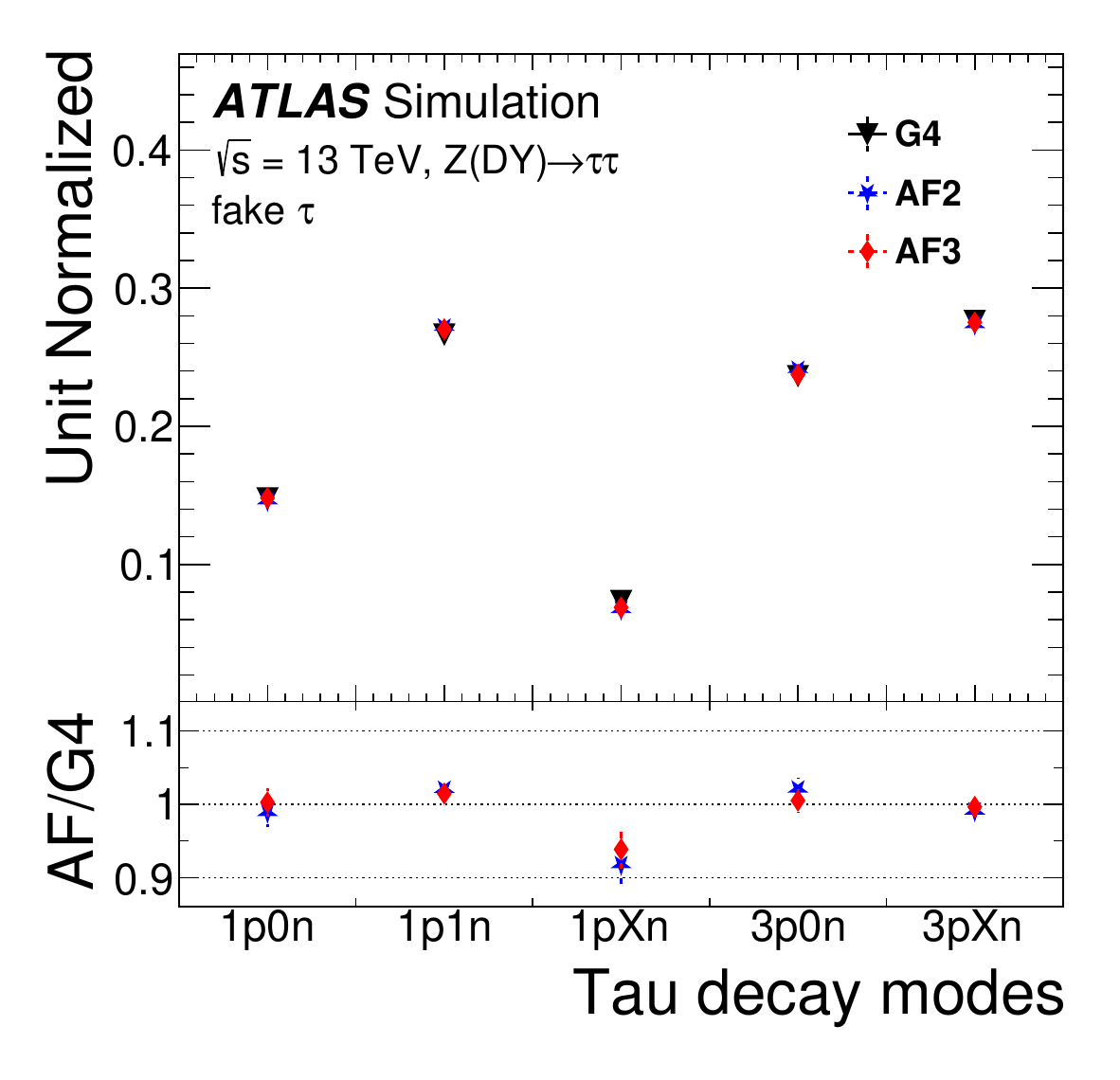}}\\
\caption{Hadronic $\tau$-lepton decay modes for reconstructed $\tau$-leptons matched to true $\tau$-leptons (a) and reconstructed $\tau$-leptons not matched to true $\tau$-leptons (b) in a $Z^\star/\gamma^\star \rightarrow \tau\tau$ Drell--Yan sample filtered for an off-shell mass of 2.0--2.25~\TeV{}. The decays with one or three charged-particle tracks are denoted by 1p and 3p respectively. X$(=1,2,3)$ denotes the number of neutral particles. The statistical uncertainties are shown but may be smaller than the size of the markers.}
\label{fig:recoObjects:taudecaypt}
\end{figure}

Hadronically decaying $\tau$-leptons are reconstructed in the ATLAS detector using their decays to one or three charged hadrons along with neutral particles~\cite{PERF-2014-06,ATL-PHYS-PUB-2015-045,ATLAS-CONF-2017-029}. The decays are labelled by the number (Y) of charged particles and the number (X) of neutral particles, YpXn. The $\tau$ reconstruction algorithm is seeded by the presence of a reconstructed jet. Figure~\ref{fig:recoObjects:taudecaypt} compares the number of events in different $\tau$ decay topologies identified in a $Z^\star/\gamma^\star \rightarrow \tau\tau$ Drell--Yan (DY) sample, filtered for an off-shell mass of 2.0--2.25~\TeV,  for \GEANT, \AFII and \AFIII using $\tau$-candidates with $\pt{}>10~\GeV{}$ and $|\eta|<2.5$. For all cases, except 1pXn, both \AFII and \AFIII agree with \GEANT to better than 10\% for reconstructed $\tau$ matched to a true $\tau$ and better than 5\% for reconstructed $\tau$ ummatched to a true $\tau$ (i.e.\ for fake $\tau$-leptons). The 1pXn case has more neutral calorimeter clusters, and the improved lateral correlations of calorimeter clusters resulted in the better agreement of AF3 with G4. The performance of \AFII and \AFIII is similar, with slightly better performance in \AFII for true $\tau$-leptons and slightly better performance in \AFIII for fake $\tau$-leptons.
 
\begin{figure}[h]
\centering
\includegraphics[width=0.49\textwidth]{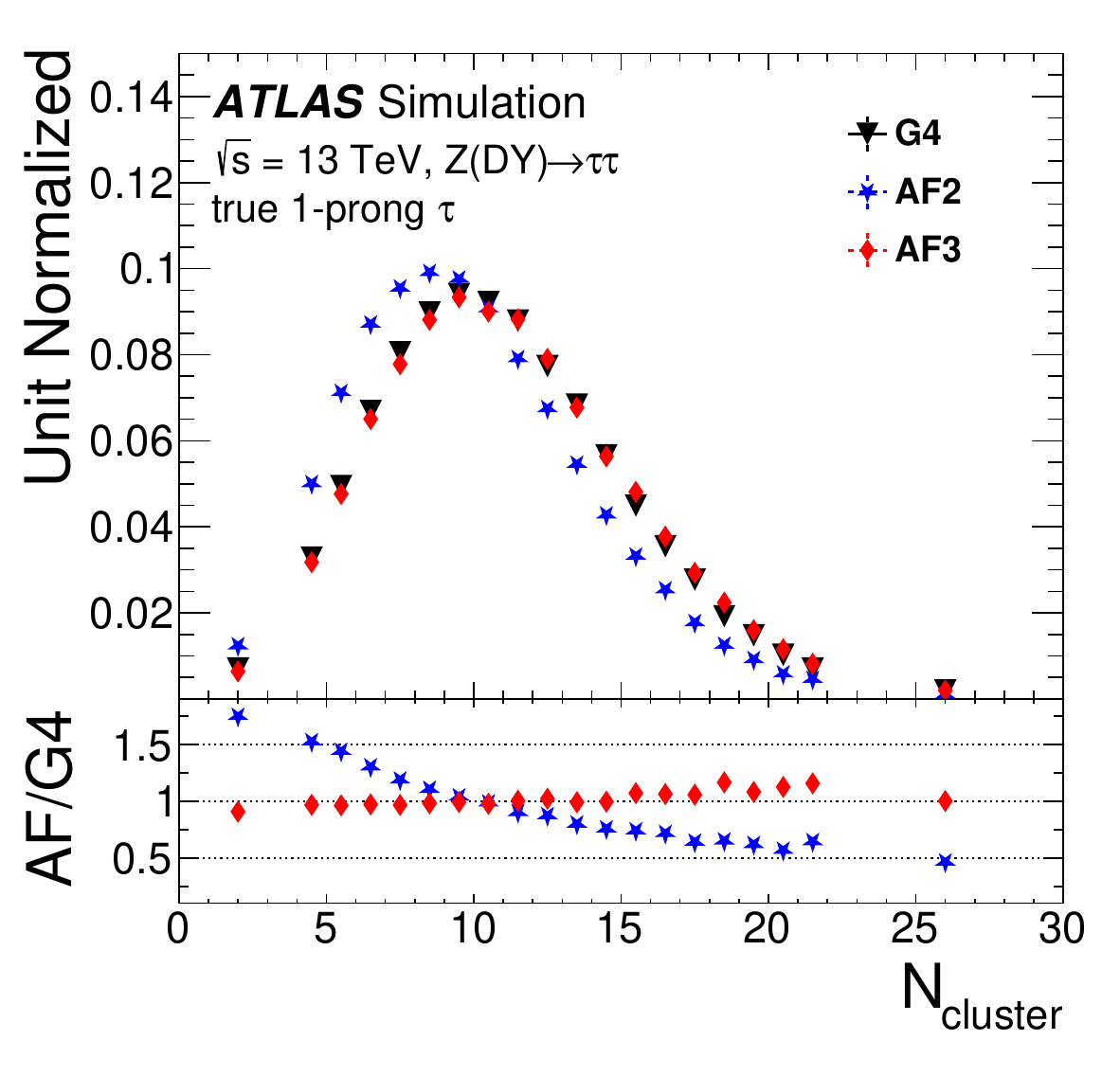}
\includegraphics[width=0.49\textwidth]{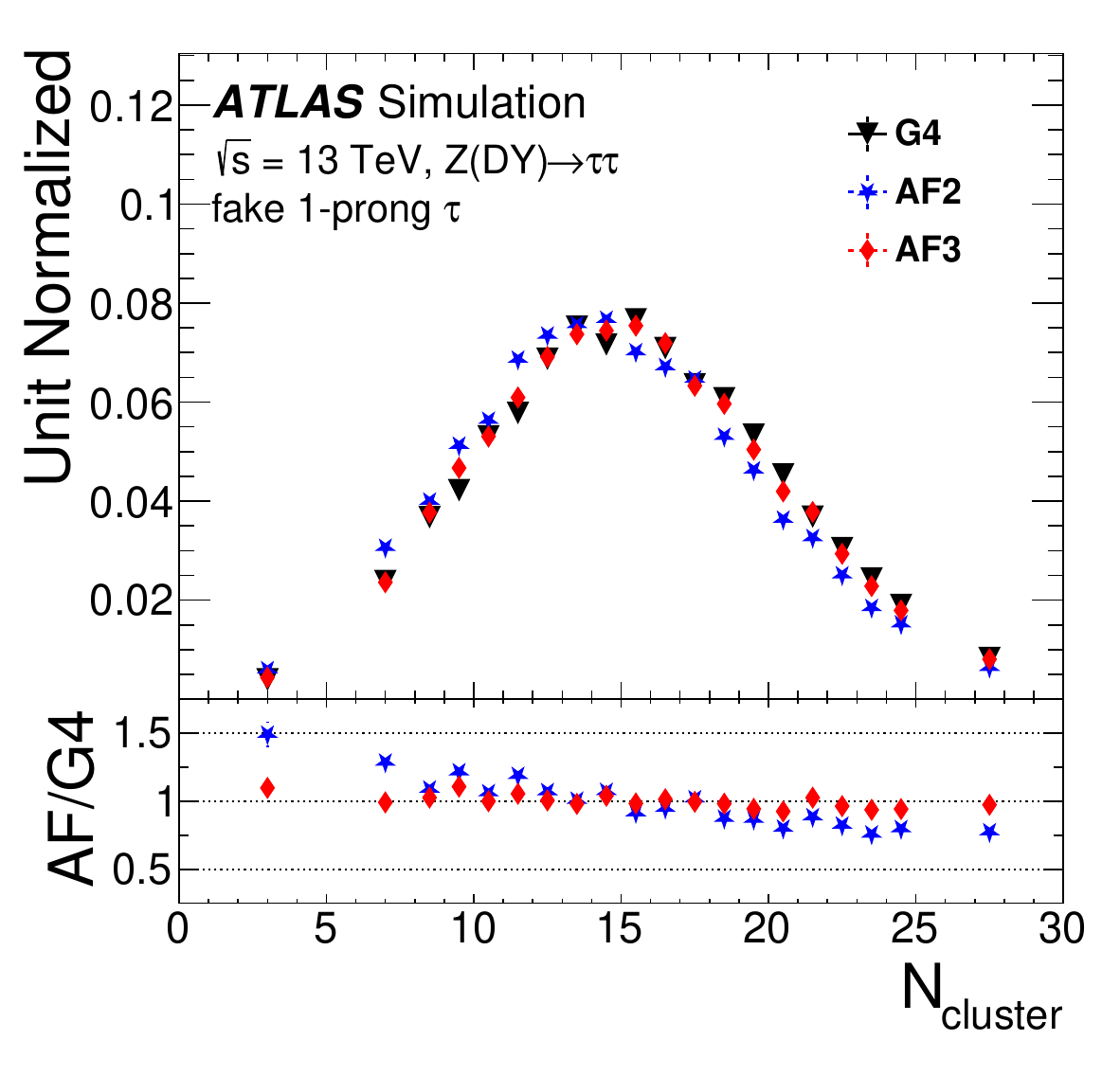}\\
\caption{Number of clusters in hadronic $\tau$-decay candidates reconstructed with one charged track (1p) and either matched (a) or not matched (b) to a true $\tau$-lepton in an $Z^\star/\gamma^\star \rightarrow \tau\tau$ Drell--Yan sample filtered for an off-shell mass of 2.0--2.25~\TeV{}. The statistical uncertainties are shown but may be smaller than the size of the markers.}
\label{fig:recoObjects:taudecayclusters}
\end{figure}

Accurate modelling of the structure of the constituents within $\tau$-jets can be challenging for fast simulation but is crucial in obtaining an accurate simulation of $\tau$ candidates. Figure~\ref{fig:recoObjects:taudecayclusters} compares the numbers of simulated clusters within true (left) and fake (right) $\tau$ candidates. In both cases, \AFII significantly underestimates the number of clusters, while \AFIII is consistent with \GEANT within statistical uncertainties.
 
\subsubsection{Reconstructed muons}
\label{sec:recoMuons}
 
\begin{figure}[h]
\centering
\subfloat[]{\label{fig:recoObjects:promptmuona}
\includegraphics[width=0.49\textwidth]{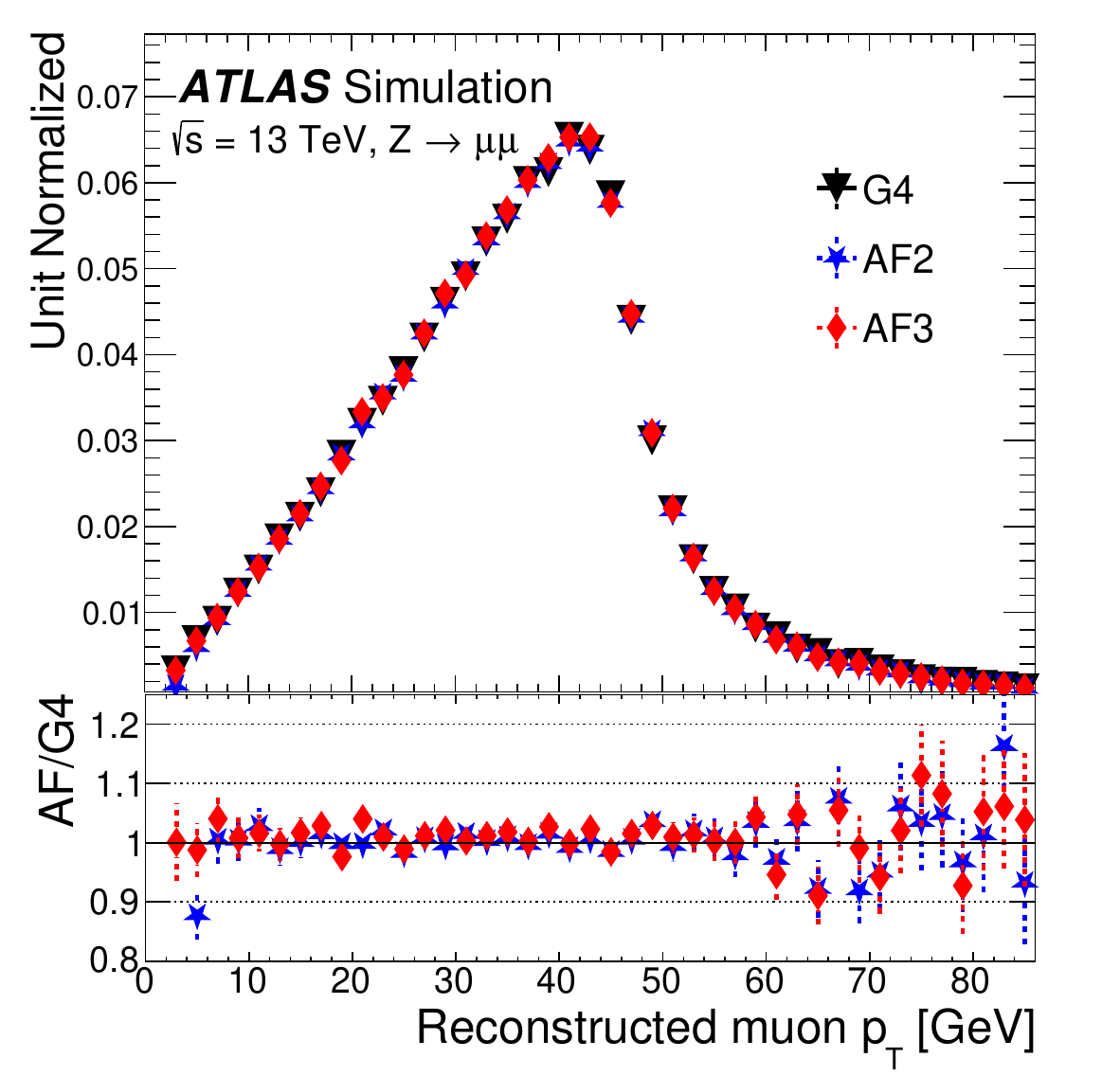}}
\subfloat[]{\label{fig:recoObjects:promptmuonb}
\includegraphics[width=0.49\textwidth]{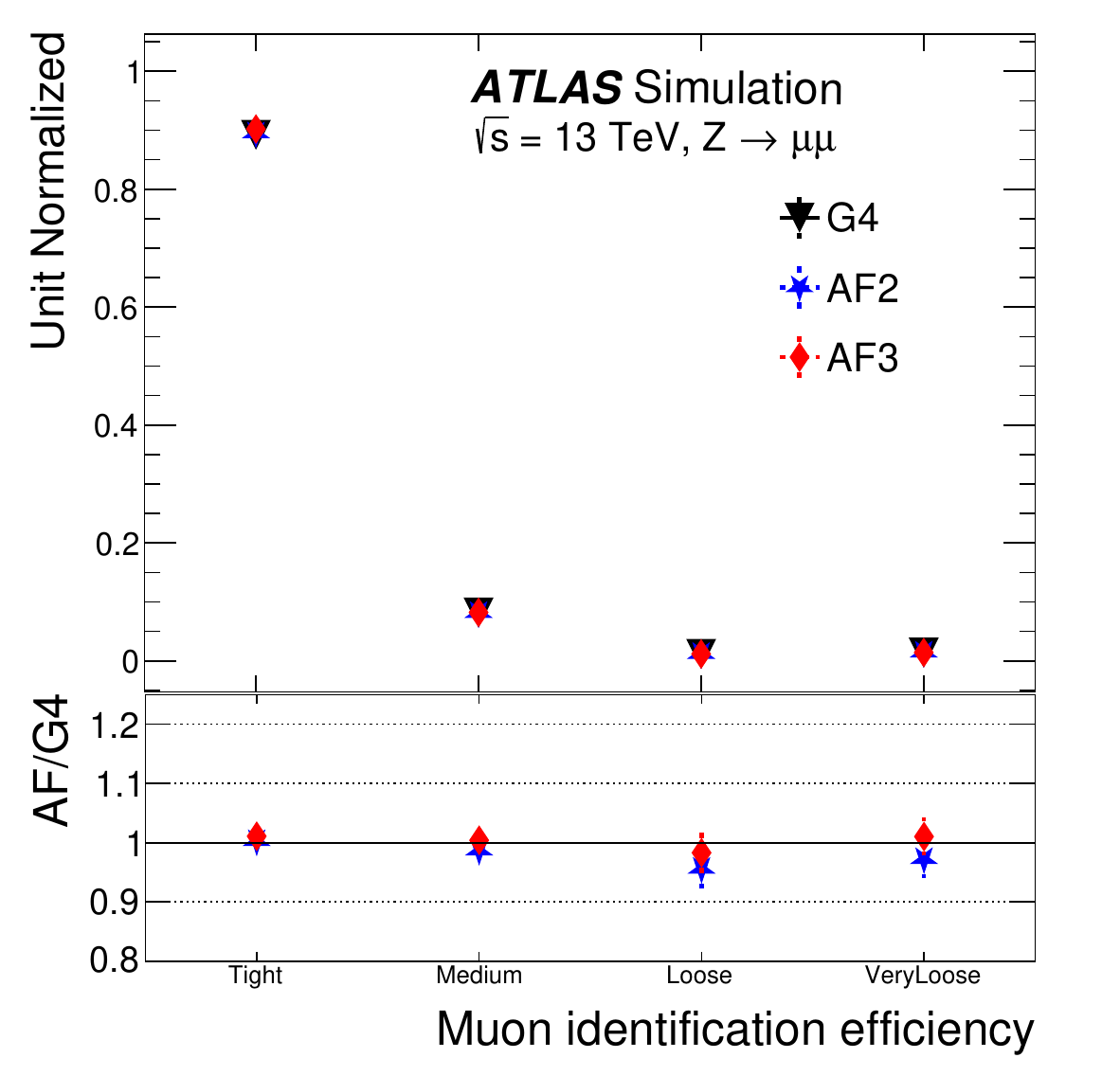}}\\
\caption{The (a) reconstructed muon transverse momentum distribution and (b) identification efficiency for different muon working points for a $Z \rightarrow \mu\mu$ sample generated with $\pT(Z) = 0$ for \GEANT, \AFII, and \AFIII. The statistical uncertainties are shown but may be smaller than the size of the markers.}
\label{fig:recoObjects:promptmuon}
\end{figure}

Muons are reconstructed from tracks in the muon spectrometer matched to tracks in the inner detector. The \pT distributions of all reconstructed muons from \GEANT, \AFII and \AFIII $Z \rightarrow \mu\mu$ samples are compared in Figure~\ref{fig:recoObjects:promptmuona}. Both \AFII and \AFIII reproduce the \pT spectrum from \GEANT. Figure~\ref{fig:recoObjects:promptmuonb} compares the number of muon candidates passing the different muon reconstruction working points. Both \AFII and \AFIII agree with \GEANT within uncertainties as expected because prompt muons are almost exclusively simulated with \GEANT for all three samples.
 
\begin{figure}[htbp]
\centering
\includegraphics[width=0.45\textwidth]{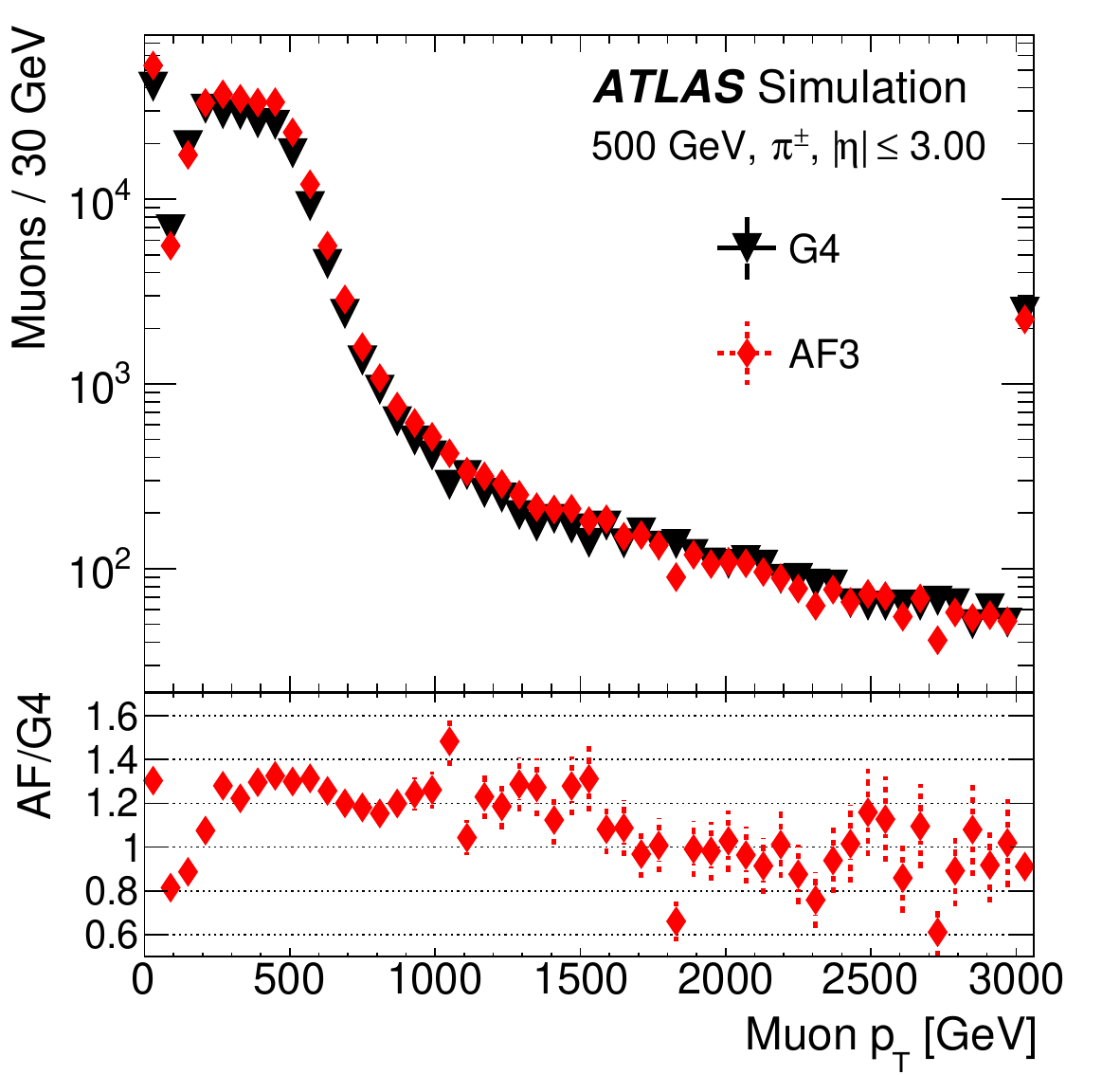}
\caption{Comparison of muon punch-through simulation in \AFIII and \GEANT as a function of the \pT of misidentified muons from 500~\GeV\ single-pion events. The statistical uncertainties are shown but may be smaller than the size of the markers.}
\label{fig:mpt_valid_500GeV_Pt}
\end{figure}

The performance of the muon punch-through simulation is validated by comparing misidentified muon candidates from hadronic activity produced in fully simulated \GEANT events with those produced by \AFIII.  Figure~\ref{fig:mpt_valid_500GeV_Pt} compares the reconstructed \pT of fake muons created by 500 \GeV\ single pions (inclusive in $\eta$) between \GEANT and \AFIII. As muon punch through is not simulated in \AFII, only \GEANT and \AFIII are shown. Agreement to better than 20\% is observed in most parts of the distributions.
 
\begin{figure}[h]
\centering
\includegraphics[width=0.7\textwidth]{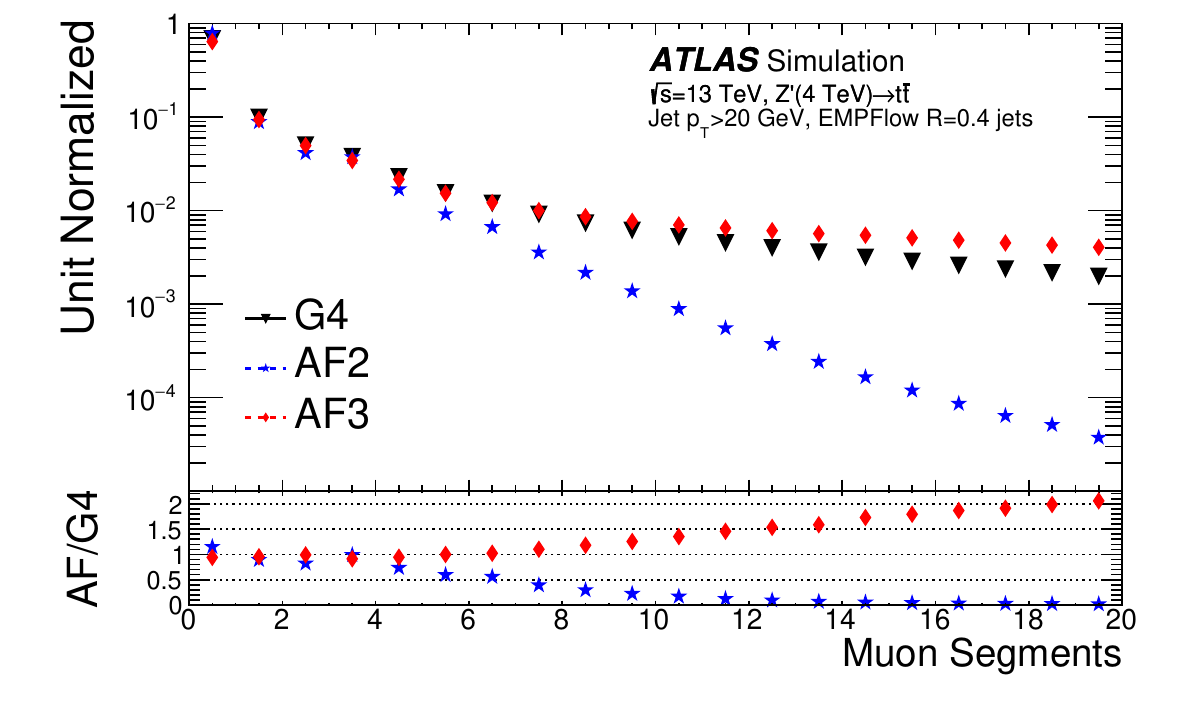}
\caption{Comparison of muon segments in jets reconstructed with a radius parameter of 0.4 using the EMPFlow algorithm in a  $Z^\prime\rightarrow\ttbar$ sample with a $Z^\prime$ mass of 4~\TeV\ in \GEANT (black triangles), \AFII (blue stars), and \AFIII (red diamonds). The statistical uncertainties are shown but may be smaller than the size of the markers.}
\label{fig:mpt_Zprime}
\end{figure}

The number of muon segments in jets reconstructed in the muon spectrometer is shown in Figure~\ref{fig:mpt_Zprime}. A $Z^\prime\rightarrow\ttbar$ sample is used because it includes prompt muons from the (anti-)top quark decays and particles produced by jets punching through the calorimeter. \AFII underestimates the number of muon segments, while \AFIII shows better agreement with \GEANT. In particular, \AFII reproduces the number of muon segments only up to three while \AFIII reproduces the number of muon segments up to seven.
 
\subsubsection{Reconstructed \MET{}}
\label{sec:recoMET}
 
\begin{figure}[h]
\centering
\includegraphics[width=0.49\textwidth]{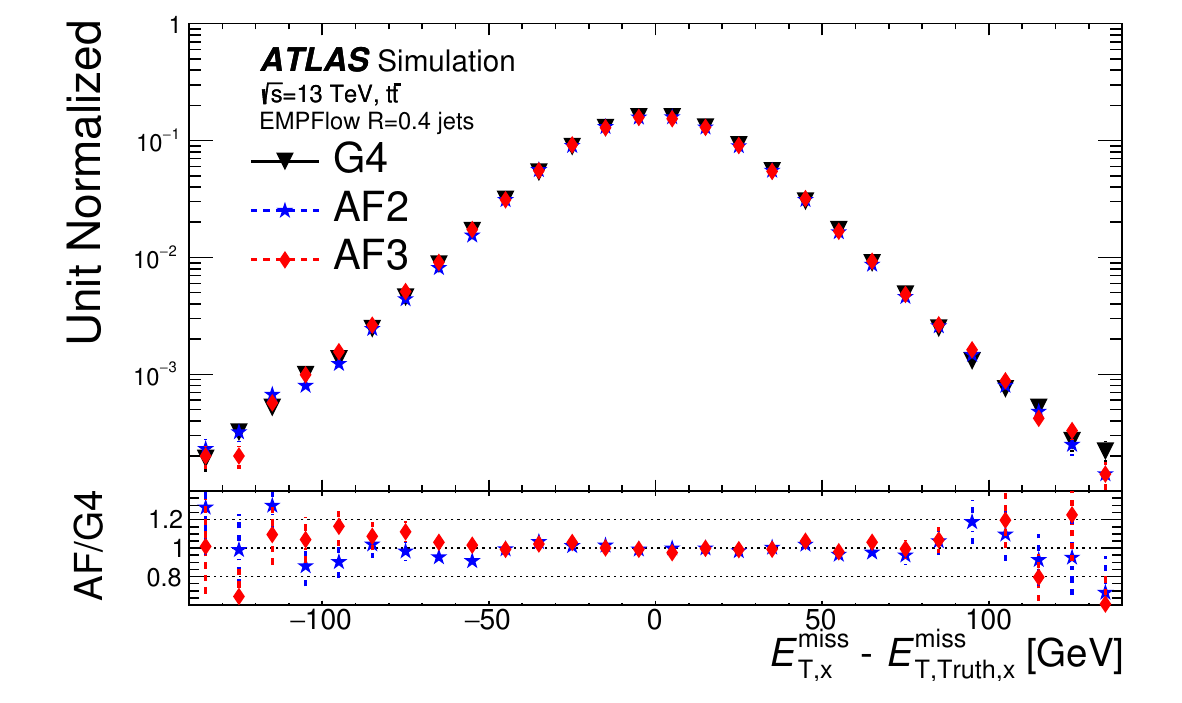} \includegraphics[width=0.49\textwidth]{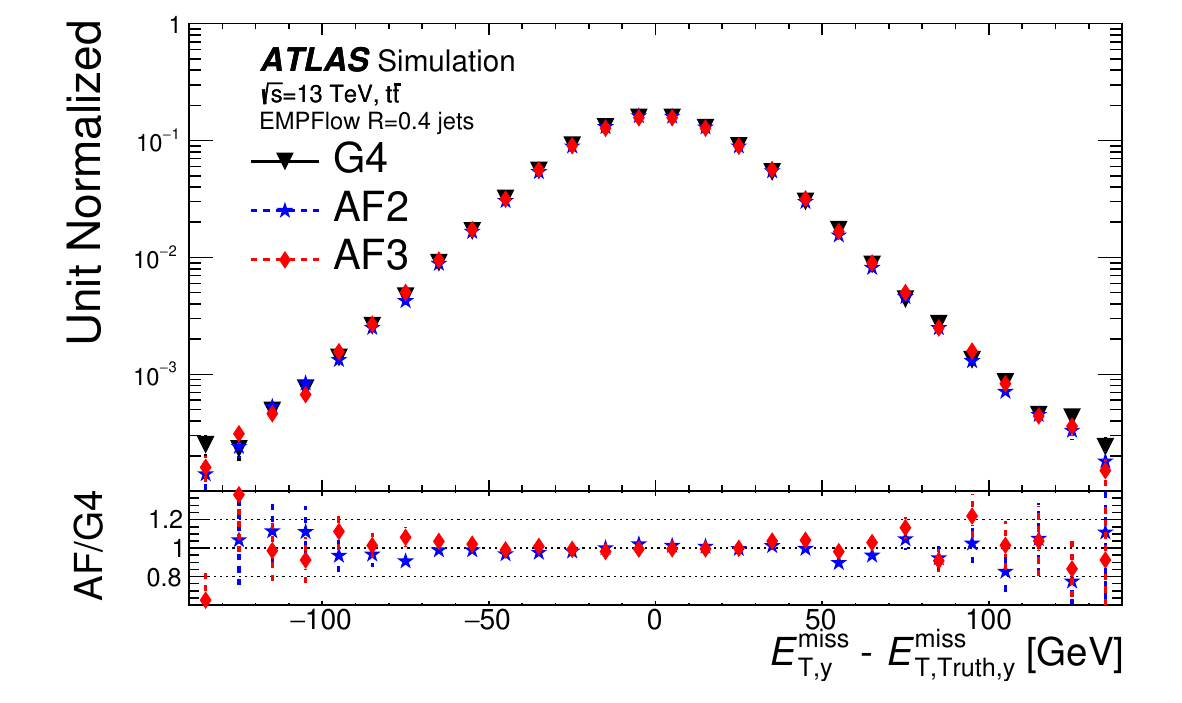}\\
\caption{The difference between the true \MET{} and the reconstructed \MET{} in the $x$ (a) and $y$ (b) directions for a \ttbar{} sample for \GEANT (black triangles), \AFII (blue stars), and \AFIII (red diamonds). The statistical uncertainties are shown but may be smaller than the size of the markers.}
\label{fig:recoObjects:met}
\end{figure}

The missing transverse momentum (\MET)~\cite{PERF-2016-07} is the negative vector sum of the reconstructed momenta of EMPFlow jets, electrons, photons, $\tau$-leptons, and muons, plus any other tracks associated with the hard-scatter primary vertex, and is used to look for transverse momentum imbalance in $pp$ collisions. The performance of the \MET reconstruction is therefore sensitive to the modelling of all reconstructed objects. Figure~\ref{fig:recoObjects:met} shows the difference between the true \MET{} and the reconstructed \MET{} in the $x$ and $y$ directions in \ttbar{} events. Both \AFII and \AFIII reproduce the \MET{} distribution from \GEANT within the statistical uncertainties. Moreover, no significant differences between \AFII and \AFIII are observed, and this is attributed to their good agreement in the jet \pT{} shown in Figure~\ref{fig:recoObjects:jetsa}.
 
\FloatBarrier
 
\subsection{Performance of \AFIII in physics analysis}
\label{sec:physana}
The performance of \AFIII for physics analysis is studied using reconstructed particle masses from selected physics analyses.
 
\begin{figure}[t]
\centering
\includegraphics[width=0.48\textwidth]{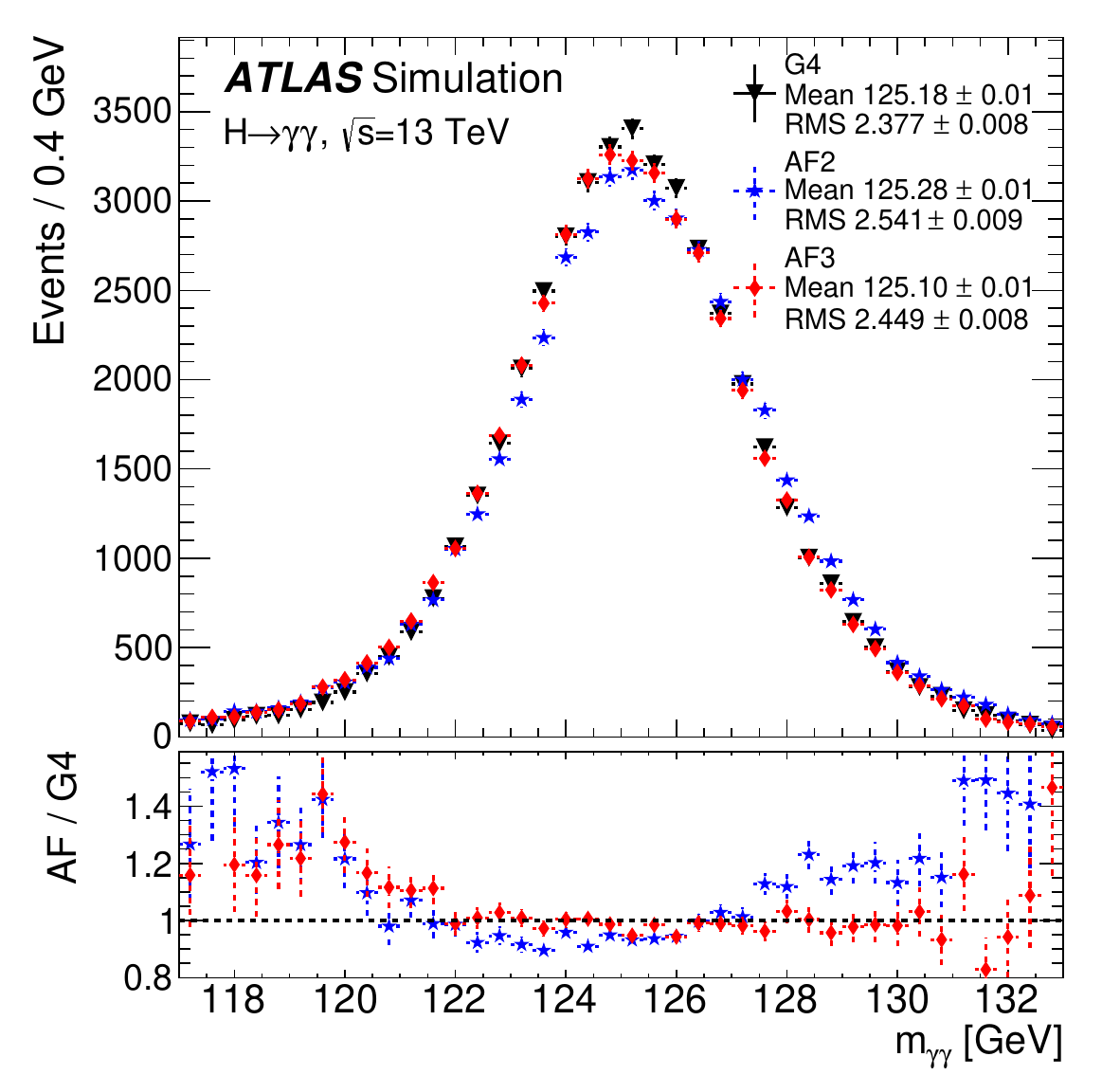}
\caption{ \label{fig:reco:egamma_mgg}
The reconstructed diphoton invariant mass distribution from a selection targeting events with Higgs boson decays into two photons. Events are selected by requiring two photons with $\pT > 0.35 m_{\gamma\gamma}$ and $\pT > 0.25 m_{\gamma\gamma}$, and with $|\eta| < 1.37$ or $1.52 < |\eta| < 2.47$. The statistical uncertainties are shown but may be smaller than the size of the markers.
}
\end{figure}

The reconstructed mass of Higgs bosons decaying into two photons is used to further evaluate the performance for photons. Events are selected by requiring two photons with $\pT > 0.35 m_{\gamma\gamma}$ and $\pT > 0.25 m_{\gamma\gamma}$, and with $|\eta| < 1.37$ or $1.52 < |\eta| < 2.47$. A comparison of \GEANT, \AFII and \AFIII is shown in Figure~\ref{fig:reco:egamma_mgg}. Both \AFII and \AFIII reproduce the mean of the distribution with high accuracy. \AFIII has better modelling of the width and agrees with \GEANT to within 5\%, while \AFII overestimates the width of the distribution by 10\%.
 
\begin{figure}[t]
\centering
\subfloat[]{\includegraphics[width=0.48\textwidth]{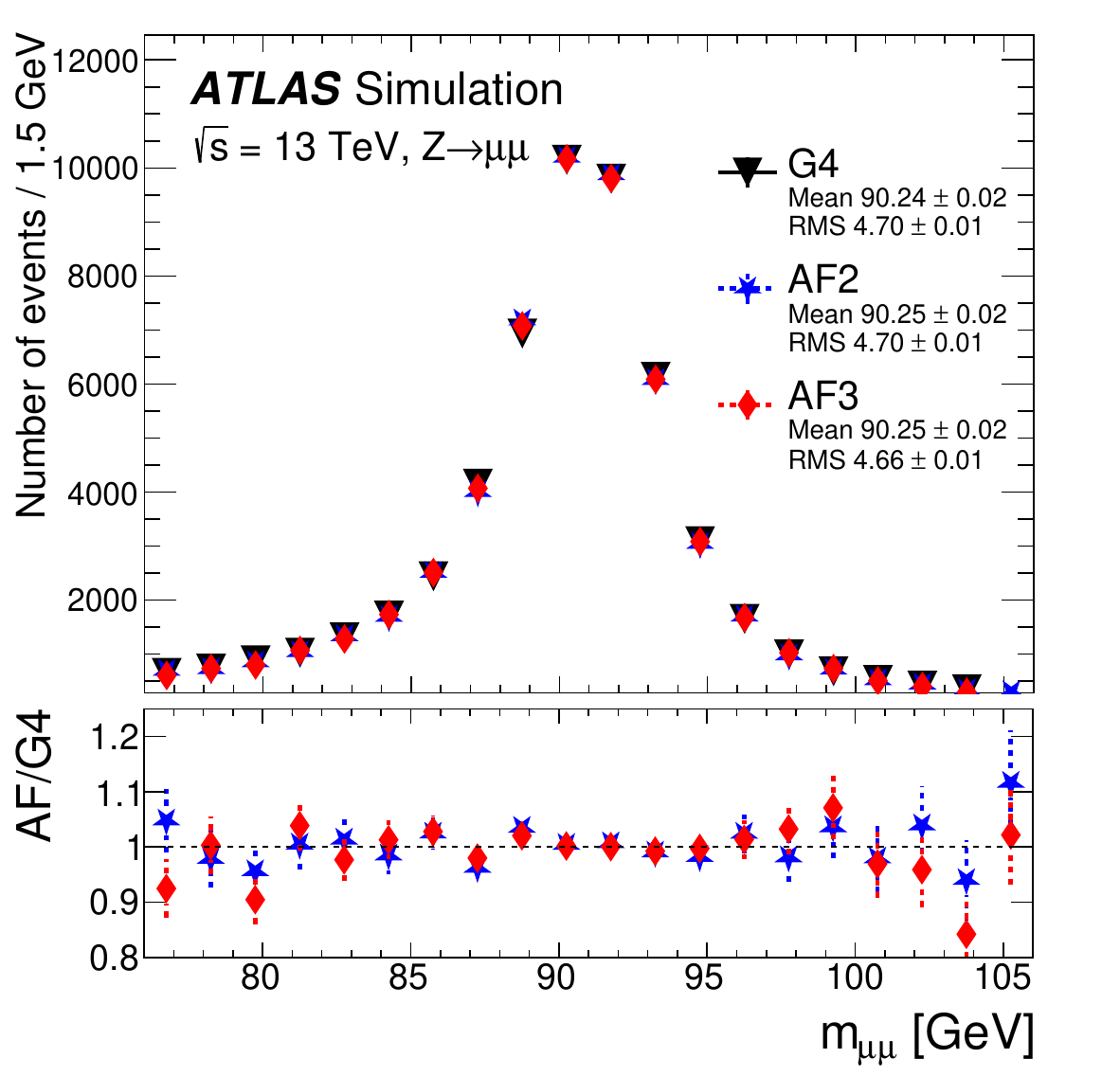}\label{fig:phys:Zmumu}}
\subfloat[]{\includegraphics[width=0.48\textwidth]{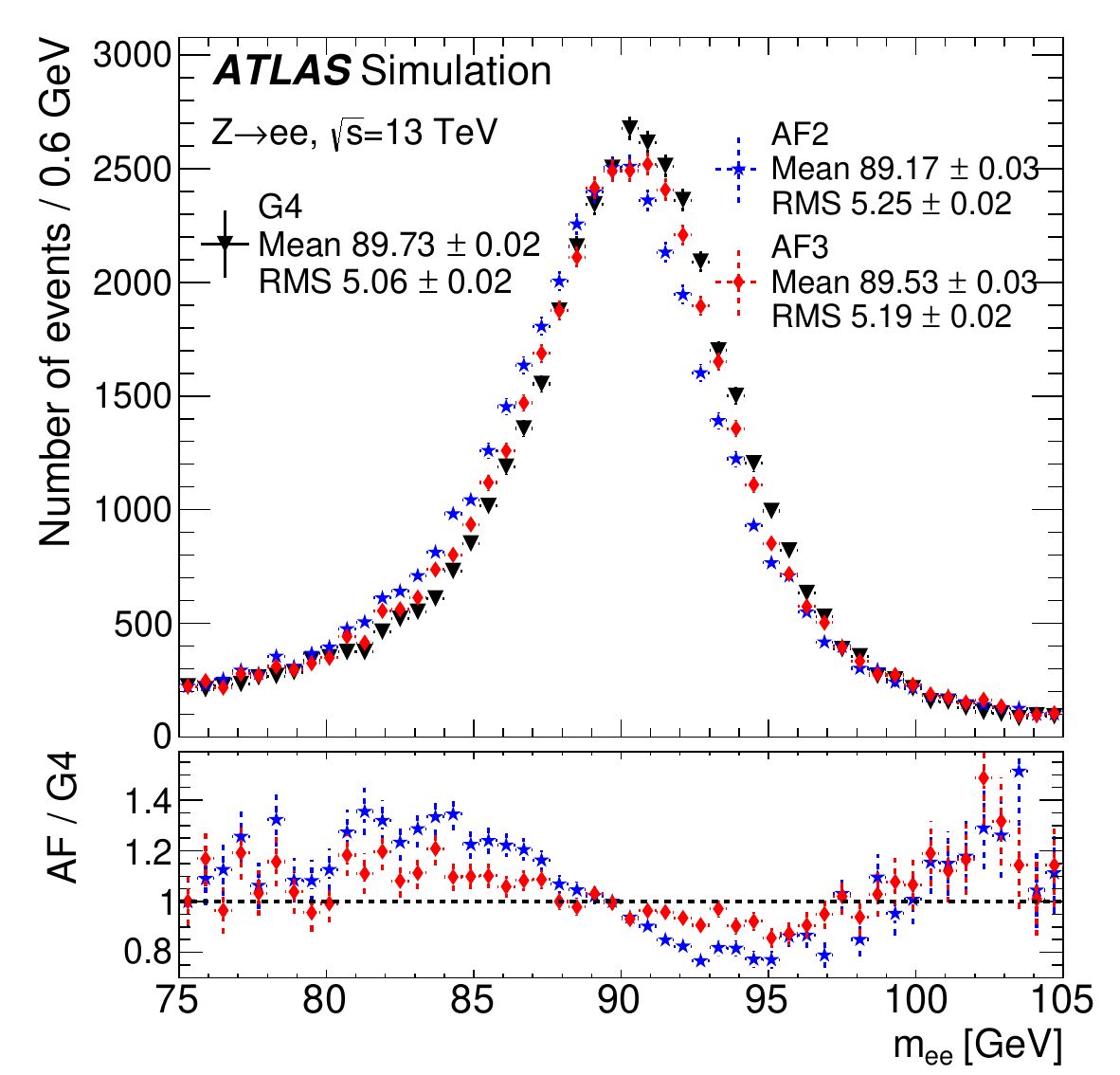}\label{fig:phys:Zee}} \\
\subfloat[]{\includegraphics[width=0.48\textwidth]{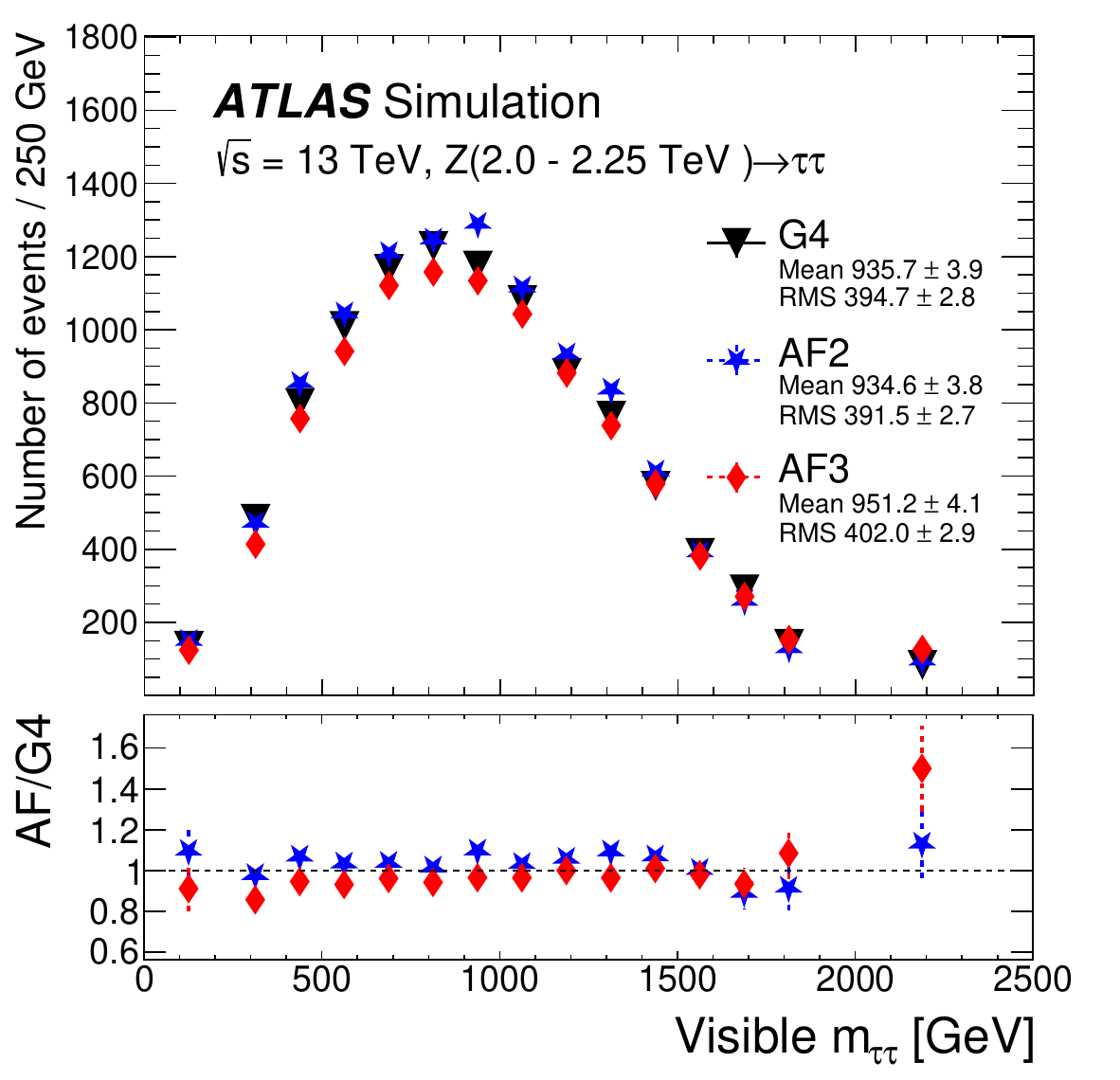}\label{fig:phys:Ztautau}}
\caption{ \label{fig:phys:resonance}
Invariant mass distribution from a selection targeting events with a $Z$ boson decaying into (a) two muons or (b) two electrons with $\pT > 25$~\GeV\ and $|\eta| < 1.37$ or $1.52 < |\eta| < 2.47$, and (c) the visible part of the invariant mass of two hadronically decaying $\tau$-leptons in Drell--Yan $Z^\star/\gamma^\star \rightarrow \tau\tau$ events filtered for an off-shell mass of 2.0--2.25~\TeV{}. The statistical uncertainties are shown but may be smaller than the size of the markers.
}
\end{figure}

Events containing Drell--Yan processes are used extensively in physics performance studies as well as to probe the Standard Model. The invariant mass of the $Z$ boson obtained when using \GEANT, \AFII, and \AFIII is shown in Figures~\ref{fig:phys:Zmumu} and~\ref{fig:phys:Zee}, with the $Z$ boson reconstructed from either a pair of muons or a pair of electrons. The visible invariant mass~\cite{ATLAS:2015boj} of off-shell 2.0--2.25~\TeV{} $Z^\star/\gamma^\star$ bosons reconstructed using the hadronic decay modes of two $\tau$-leptons is shown in Figure~\ref{fig:phys:Ztautau}, where the width is slightly overestimated by \AFIII. Otherwise, no significant differences are observed between the three distributions, which validates the simulation of prompt muons in \AFIII.
 
\begin{figure}[h]
\centering
\subfloat[]{\includegraphics[width=0.49\textwidth]{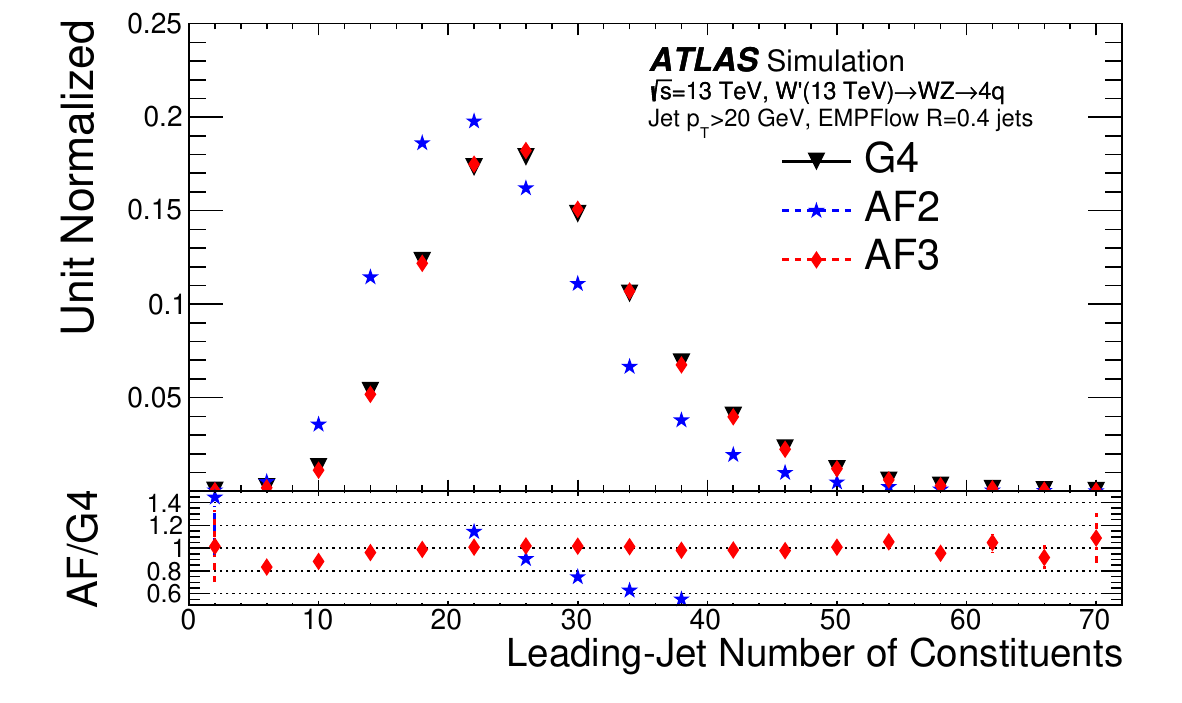}}
\subfloat[]{\includegraphics[width=0.49\textwidth]{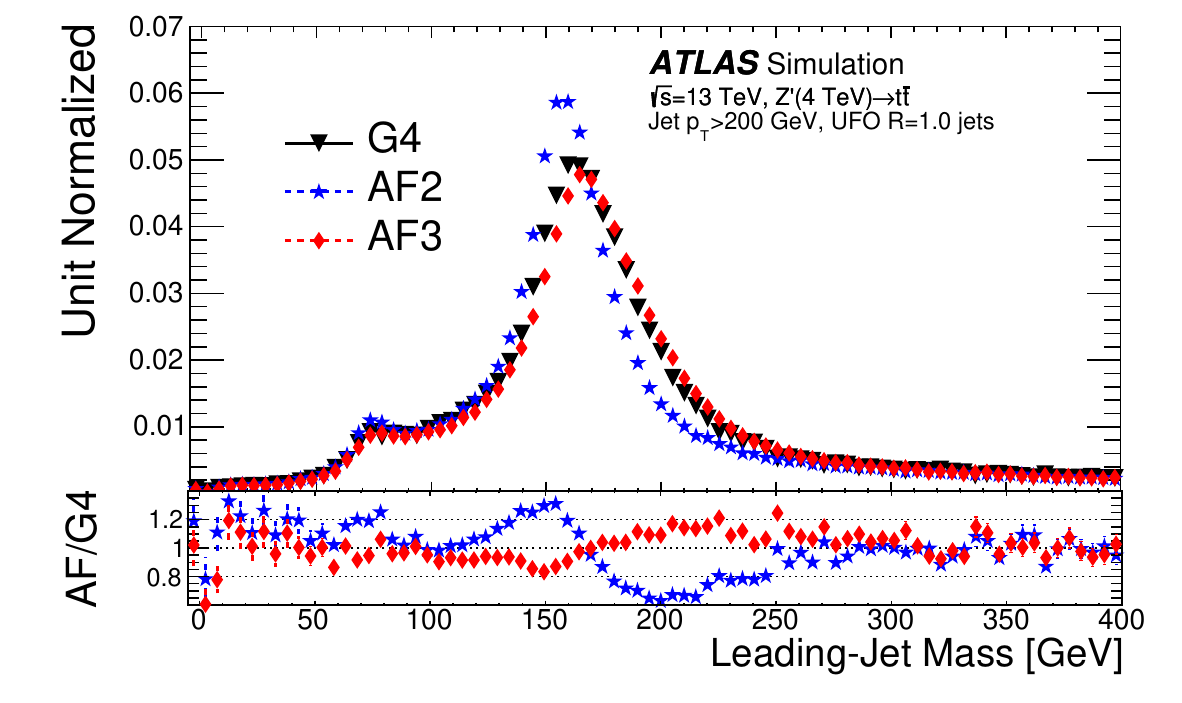}}
\caption{Distribution of the (left) number of constituents in the leading $R=0.4$ EMPFlow jets in the $W^\prime$ sample and (right) the mass of trimmed $R=1.0$ UFO jets in the $Z^\prime$ sample in \GEANT (black triangles), \AFII (blue stars), and \AFIII (red diamonds). The statistical uncertainties are shown but may be smaller than the size of the markers.}
\label{fig:recoObjects:massd2}
\end{figure}

The performance of the simulation for jet substructure is evaluated with the $Z^\prime\rightarrow\ttbar$ events as the benchmark for `beyond the Standard Model' signatures with boosted high-\pT objects in the calorimeter. The leading jet's mass  with its constituents calibrated to the EM scale is shown in Figure~\ref{fig:recoObjects:massd2}; the distribution has a large peak near the mass of the top quark, and a small peak at the $W$ boson mass. \AFII significantly underestimates the mass and the width of both peaks compared to \GEANT, while \AFIII is in better agreement with the \GEANT distribution than \AFII{}.
 
\FloatBarrier
 
\subsection{Computing performance with \AFIII}
\label{sec:techperf}
 
\begin{figure}
\centering
\includegraphics[width=0.7\textwidth]{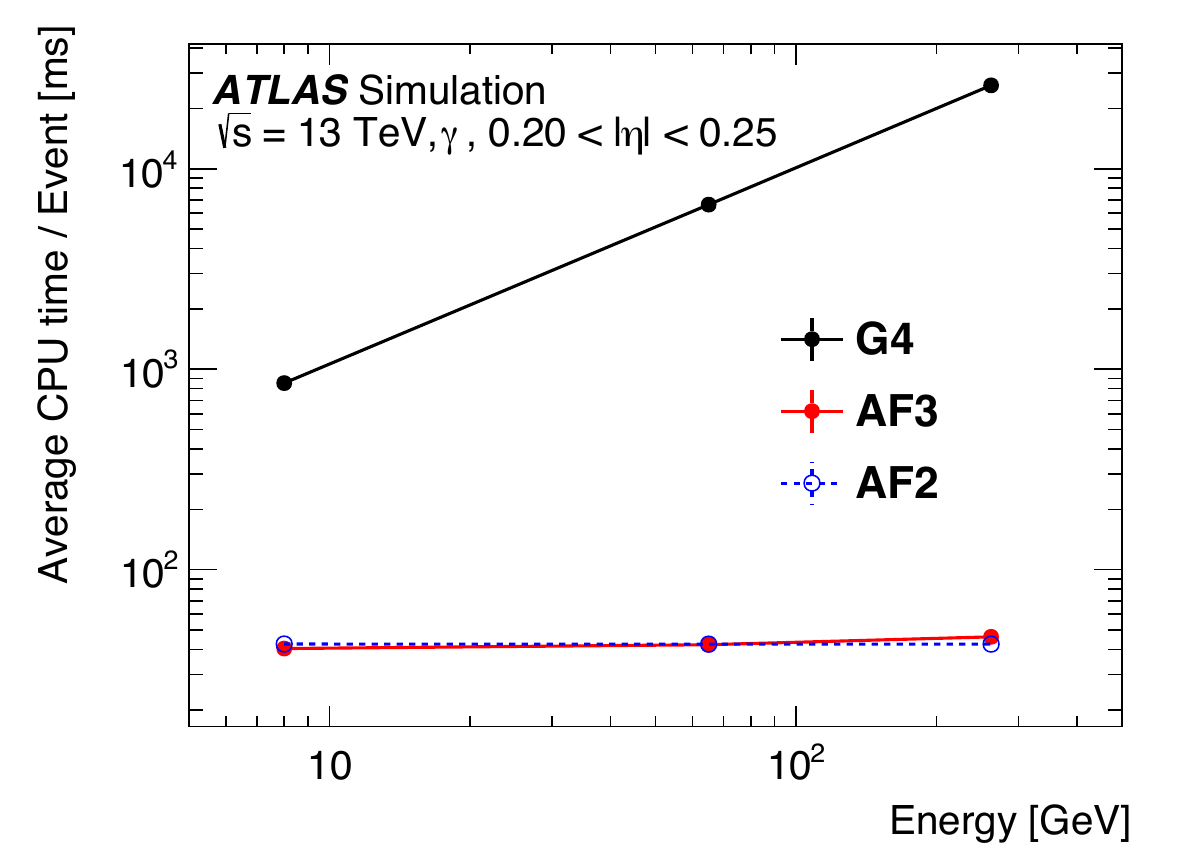}
\caption{Comparison of the CPU performance of \AFIII with \GEANT and \AFII. The average CPU time to simulate an event is estimated using 10\,000 single photons at $0.20 < |\eta| < 0.25$  for three different energies: 8~\GeV{}, 65~\GeV{}, and 256~\GeV{}. These photons are generated on the calorimeter surface and provide a comparison for calorimeter-only simulation time. \label{fig:tech:cpu}}
\end{figure}

The time required to simulate a particle in \GEANT increases with energy due to increasing shower depth and complexity, whereas in \AFII and \AFIII the time is independent of the particle energy because it requires a single lookup in the parameterization file. To illustrate this, the average CPU time, calculated with a 4-core Intel i7-3770 CPU at 3.40 GHz, required to simulate a single photon produced on the calorimeter surface at $0.20 < |\eta| <0.25$ is shown in Figure~\ref{fig:tech:cpu} as a function of energy. For an 8~\GeV{} photon produced on the calorimeter surface, \AFIII is approximately 20 times faster than \GEANT, while for a 256~\GeV{} photon, \AFIII is approximately 600 times faster.
 
For the full detector simulation, the computing performance of \AFIII is compared with that of \GEANT by simulating the same 1000 \ttbar{} events; this is a complex process ideal for a variety of benchmarking needs and is used extensively by the ATLAS experiment for this purpose. Each simulation algorithm is executed on a 8-core Intel Xeon E5 CPU at 3.20 GHz. On average, \GEANT requires 167 seconds to simulate a single event, while \AFIII only requires 32 seconds, thereby obtaining a speed-up of the simulation by a factor of five. If the calorimeter simulation alone is considered, \AFIII is $O(500)$ times faster than \GEANT. This means that the simulation time is dominated by the simulation of the inner detector performed by \GEANT. Therefore, further gains in the simulation speed of physics samples would require the use of fast simulation techniques in the tracking detector. Due to the size of the parameterization file, \AFIII requires 7\,GB of proportional set size (PSS) memory, while the full simulation requires 2.7\,GB in total when using eight separate cores. The parameterization requires 5\,GB of PSS memory and this is shared by the cores and is within the PSS memory budget available. The PSS memory required by \AFIII can be reduced in the future through the use of compression algorithms.
 
\FloatBarrier
 
\section{Conclusion}
\label{sec:conclusion}
An updated version of the fast simulation for the ATLAS experiment, \AFIII, is introduced in this paper. \AFIII significantly improves the modelling of reconstructed objects for physics analyses beyond that obtained by \AFII. In most cases, \AFIII and \GEANT agree to within a few percent. Key improvements include the modelling of the response in the forward calorimeters and of shower substructure within jets. Moreover, \AFIII requires only 20\% as much CPU as \GEANT to simulate an event. The version of \AFIII described in this paper is currently being used by ATLAS to simulate 7~billion events for physics analyses of the Run~2 data. Further updates and improvements to the modelling are anticipated for Run~3 and beyond.

\section*{Acknowledgements}
 

We thank CERN for the very successful operation of the LHC, as well as the
support staff from our institutions without whom ATLAS could not be
operated efficiently.
 
We acknowledge the support of
ANPCyT, Argentina;
YerPhI, Armenia;
ARC, Australia;
BMWFW and FWF, Austria;
ANAS, Azerbaijan;
SSTC, Belarus;
CNPq and FAPESP, Brazil;
NSERC, NRC and CFI, Canada;
CERN;
ANID, Chile;
CAS, MOST and NSFC, China;
Minciencias, Colombia;
MSMT CR, MPO CR and VSC CR, Czech Republic;
DNRF and DNSRC, Denmark;
IN2P3-CNRS and CEA-DRF/IRFU, France;
SRNSFG, Georgia;
BMBF, HGF and MPG, Germany;
GSRI, Greece;
RGC and Hong Kong SAR, China;
ISF and Benoziyo Center, Israel;
INFN, Italy;
MEXT and JSPS, Japan;
CNRST, Morocco;
NWO, Netherlands;
RCN, Norway;
MNiSW and NCN, Poland;
FCT, Portugal;
MNE/IFA, Romania;
JINR;
MES of Russia and NRC KI, Russian Federation;
MESTD, Serbia;
MSSR, Slovakia;
ARRS and MIZ\v{S}, Slovenia;
DSI/NRF, South Africa;
MICINN, Spain;
SRC and Wallenberg Foundation, Sweden;
SERI, SNSF and Cantons of Bern and Geneva, Switzerland;
MOST, Taiwan;
TAEK, Turkey;
STFC, United Kingdom;
DOE and NSF, United States of America.
In addition, individual groups and members have received support from
BCKDF, CANARIE, Compute Canada and CRC, Canada;
COST, ERC, ERDF, Horizon 2020 and Marie Sk{\l}odowska-Curie Actions, European Union;
Investissements d'Avenir Labex, Investissements d'Avenir Idex and ANR, France;
DFG and AvH Foundation, Germany;
Herakleitos, Thales and Aristeia programmes co-financed by EU-ESF and the Greek NSRF, Greece;
BSF-NSF and GIF, Israel;
Norwegian Financial Mechanism 2014-2021, Norway;
La Caixa Banking Foundation, CERCA Programme Generalitat de Catalunya and PROMETEO and GenT Programmes Generalitat Valenciana, Spain;
G\"{o}ran Gustafssons Stiftelse, Sweden;
The Royal Society and Leverhulme Trust, United Kingdom.
 
The crucial computing support from all WLCG partners is acknowledged gratefully, in particular from CERN, the ATLAS Tier-1 facilities at TRIUMF (Canada), NDGF (Denmark, Norway, Sweden), CC-IN2P3 (France), KIT/GridKA (Germany), INFN-CNAF (Italy), NL-T1 (Netherlands), PIC (Spain), ASGC (Taiwan), RAL (UK) and BNL (USA), the Tier-2 facilities worldwide and large non-WLCG resource providers. Major contributors of computing resources are listed in Ref.~\cite{ATL-SOFT-PUB-2021-003}.
 

\clearpage
 
\FloatBarrier

\printbibliography
 
\clearpage \input{atlas_authlist}

\end{document}

%% file: atlas_authlist.tex
 
\begin{flushleft}
\hypersetup{urlcolor=black}
{\Large The ATLAS Collaboration}

\bigskip

\AtlasOrcid[0000-0002-6665-4934]{G.~Aad}$^\textrm{\scriptsize 98}$,    
\AtlasOrcid[0000-0002-5888-2734]{B.~Abbott}$^\textrm{\scriptsize 124}$,    
\AtlasOrcid[0000-0002-7248-3203]{D.C.~Abbott}$^\textrm{\scriptsize 99}$,    
\AtlasOrcid[0000-0002-2788-3822]{A.~Abed~Abud}$^\textrm{\scriptsize 34}$,    
\AtlasOrcid[0000-0002-1002-1652]{K.~Abeling}$^\textrm{\scriptsize 51}$,    
\AtlasOrcid[0000-0002-2987-4006]{D.K.~Abhayasinghe}$^\textrm{\scriptsize 91}$,    
\AtlasOrcid[0000-0002-8496-9294]{S.H.~Abidi}$^\textrm{\scriptsize 27}$,    
\AtlasOrcid[0000-0002-9987-2292]{A.~Aboulhorma}$^\textrm{\scriptsize 33e}$,    
\AtlasOrcid[0000-0001-5329-6640]{H.~Abramowicz}$^\textrm{\scriptsize 157}$,    
\AtlasOrcid[0000-0002-1599-2896]{H.~Abreu}$^\textrm{\scriptsize 156}$
% ...
\bigskip
\\

% ...
$^{27}$Physics Department, Brookhaven National Laboratory, Upton NY; United States of America.\\
% ...
$^{33}$$^{(a)}$Facult\'e des Sciences Ain Chock, R\'eseau Universitaire de Physique des Hautes Energies - Universit\'e Hassan II, Casablanca;$^{(b)}$Facult\'{e} des Sciences, Universit\'{e} Ibn-Tofail, K\'{e}nitra;$^{(c)}$Facult\'e des Sciences Semlalia, Universit\'e Cadi Ayyad, LPHEA-Marrakech;$^{(d)}$LPMR, Facult\'e des Sciences, Universit\'e Mohamed Premier, Oujda;$^{(e)}$Facult\'e des sciences, Universit\'e Mohammed V, Rabat;$^{(f)}$Mohammed VI Polytechnic University, Ben Guerir; Morocco.\\
$^{34}$CERN, Geneva; Switzerland.\\
% ...
$^{51}$II. Physikalisches Institut, Georg-August-Universit\"{a}t G\"ottingen, G\"ottingen; Germany.\\
% ...
$^{91}$Department of Physics, Royal Holloway University of London, Egham; United Kingdom.\\
% ...
$^{98}$CPPM, Aix-Marseille Universit\'e, CNRS/IN2P3, Marseille; France.\\
$^{99}$Department of Physics, University of Massachusetts, Amherst MA; United States of America.\\
% ...
$^{124}$Homer L. Dodge Department of Physics and Astronomy, University of Oklahoma, Norman OK; United States of America.\\
% ...
$^{156}$Department of Physics, Technion, Israel Institute of Technology, Haifa; Israel.\\
$^{157}$Raymond and Beverly Sackler School of Physics and Astronomy, Tel Aviv University, Tel Aviv; Israel.\\
% ...
$^{a}$ Also at Borough of Manhattan Community College, City University of New York, New York NY; United States of America.\\
$^{b}$ Also at Bruno Kessler Foundation, Trento; Italy.\\
$^{c}$ Also at Center for High Energy Physics, Peking University; China.\\
$^{d}$ Also at Centro Studi e Ricerche Enrico Fermi; Italy.\\
$^{e}$ Also at CERN, Geneva; Switzerland.\\
$^{f}$ Also at D\'epartement de Physique Nucl\'eaire et Corpusculaire, Universit\'e de Gen\`eve, Gen\`eve; Switzerland.\\
% ...

\end{flushleft}
